\documentclass[aps,prd,nofootinbib,twocolumn,showpacs]{revtex4}
\usepackage[hyperindex,breaklinks]{hyperref}
\expandafter\let\csname equation*\endcsname\relax
\expandafter\let\csname endequation*\endcsname\relax
\usepackage[caption=false]{subfig}
\usepackage{amsmath}
\usepackage{amssymb}
\usepackage{graphicx}
\usepackage{tensor}
\usepackage{setspace}
\usepackage{pzccal}
\usepackage[sort&compress]{natbib}
\usepackage{graphicx}
\usepackage{epstopdf}
\usepackage{natbib}
\usepackage{appendix}
\usepackage{alphalph}
\setcitestyle{square,comma}
\def\inner#1{\left< #1 \right>}

\def\b#1{\textbf{#1}}
\def\c#1{\overline{#1}}

\def\t#1#2{\tensor{#1}{#2}}
\begin{document}
\title{Quasilocal energy exchange and the null cone}
\author{Nezihe Uzun}
\address{Department of Physics and Astronomy, University of Canterbury,
Private Bag 4800, Christchurch 8140, New Zealand}
\date{\today}
%
\begin{abstract}
Energy is at best defined quasilocally in general relativity. Quasilocal energy definitions depend on the conditions one imposes on the boundary Hamiltonian, i.e., how a finite region of spacetime is ``isolated.'' Here, we propose a method to define and investigate systems in terms of their matter plus gravitational energy content. We adopt a generic construction, that involves embedding of an arbitrary dimensional world sheet into an arbitrary dimensional spacetime, to a $2+2$ picture. In our case, the closed 2-dimensional spacelike surface $\mathbb{S}$, that is orthogonal to the 2-dimensional timelike world sheet $\mathbb{T}$ at every point, encloses the system in question. The integrability conditions of $\mathbb{T}$ and $\mathbb{S}$ correspond to three null tetrad gauge conditions once we transform our notation to the one of the null cone observables. We interpret the Raychaudhuri equation of $\mathbb{T}$ as a work-energy relation for systems that are not in equilibrium with their surroundings. We achieve this by identifying the quasilocal charge densities corresponding to rotational and nonrotational degrees of freedom, in addition to a relative work density associated with tidal fields. We define the corresponding quasilocal charges that appear in our work-energy relation and which can potentially be exchanged with the surroundings. These charges and our tetrad conditions are invariant under type-III Lorentz transformations, i.e., the boosting of the observers in the directions orthogonal to $\mathbb{S}$. We apply our construction to a radiating Vaidya spacetime, a $C$-metric and the interior of a Lanczos-van Stockum dust metric. The delicate issues related to the axially symmetric stationary spacetimes and possible extensions to the Kerr geometry are also discussed.
\end{abstract}
\bigskip
\maketitle

\section{Introduction}
In general relativity, there is no unique definition of matter plus gravitational energy exchange definition for a system. For the case of pure gravity, for example, gravitational radiation and the energy loss associated with it, can be identified unambiguously only at null infinity, $\mathfrak{I}^+$, of an isolated body \citep{Bondietal:1962}. Essentially it is assumed that observers are sufficiently far away from the body in question so that the asymptotic metric is flat and the perturbations around it correspond to the gravitational radiation. Also it is assumed that the spacetime admits the peeling property, i.e., the Weyl scalars behave asymptotically and outgoing null hypersurfaces are assumed to intersect $\mathfrak{I}^+$ through closed spacelike 2-surfaces whose departure from the unit sphere is small \citep{Lehner_Moreschi:2007}. It is known that the wave extraction and the interpretation of the physically meaningful quantities are often challenging for numerical relativity simulations based on those asymptotic regions.

On the other hand, for astrophysical and larger scale investigations, we would like to know how systems behave in the strong field regime. We would like to understand the behavior of binary black hole or neutron star mergers and how those objects affect their close environment. Considering the fact that gravitational energy cannot be localized due to the equivalence principle, there have been a considerable number of attempts to understand the energy exchange mechanisms of arbitrary gravitating systems quasilocally (see \citep{Szabados:2004} for a detailed review), on top of the earlier global investigations \citep{Thorne_Hartle:1984, Purdue:1999, Favata:2000}. However, not all of the quasilocal energy investigations are constructed on, or translated into, the formalism that the numerical relativity community uses. 
In the present paper, we aim to present a method with which one can investigate the quasilocal energy exchange of a system. This involves the observables of timelike congruences, however, we present the corresponding null cone observables as well once we perform a transformation between the two formalisms.

In \citep{Capovilla_Guven:1994} Capovilla and Guven (CG) generalize the Raychaudhuri equation which gives the focusing of an arbitrary dimensional timelike world sheet that is embedded in an arbitrary dimensional spacetime.
Previously, in \citep{Uzun_Wiltshire:2015}, we applied their formalism to a 2-dimensional timelike world sheet, $\mathbb{T}$, embedded in a 4-dimensional spherically symmetric spacetime. This allowed us to define quasilocal thermodynamic equilibrium conditions and the corresponding quasilocal thermodynamic potentials in a natural way.

In the present paper, we will consider more generic systems, which are not in equilibrium with their surroundings. Also the systems we consider here are not necessarily spherically symmetric. Our main aim is to present a method for the calculation of the energylike quantities of these systems which can be exchanged quasilocally. While doing so, we will switch from Capovilla and Guven's notation to the notation of Newman-Penrose (NP) formalism \citep{Newman_Penrose:1961}. Firstly, this will ease our calculations. Secondly, the transformation of the original formalism of CG to NP poses basic questions about the null tetrad gauge invariance of numerical relativity in terms of quasilocal concerns. Namely, if one wants to investigate a system quasilocally one needs to define it consistently throughout its evolution by keeping the boost invariance of the quasilocal observers. This fixes a gauge for the complex null tetrad constructed through their local double dyad in our 2+2 approach.  

The construction of the paper is as follows. In Sec.~\ref{Mass-energy exchange}, we survey some of the local, global and quasilocal approaches in the literature to investigate matter plus gravitational mass-energy exchange. We will show just how broad the literature is in terms of energy exchange investigations. In Sec.~\ref{Null tetrad gauge} we start to question how to best define a $ quasilocal \, system$ and introduce our choice of system definition. Section \ref{RaychaudhuriCG} gives a concise summary of Capovilla and Guven's formalism which is used to derive the Raychaudhuri equation of a world sheet \citep{Capovilla_Guven:1994}. In Sec.~\ref{RaychaudhuriNP} we present the contracted Raychaudhuri equation in the NP formalism and demonstrate how our gauge conditions affect it.
Later, in Sec.~\ref{Work-energy}, we give physical interpretations to the variables of the contracted Raychaudhuri equation in terms of the quasilocal charge densities. We define the associated quasilocal charges and end up with a work-energy relation. According to our interpretation, the contracted Raychaudhuri equation of the world sheet of the quasilocal observers gives information about how much rotational and nonrotational quasilocal energy the system possesses, in addition to the work that should be done by the tidal fields to create such a system. In Sec.~\ref{Applications} we present applications of our method to a radiating Vaidya spacetime, $C$-metric and interior of a Lanczos-van Stockum dust source. We present the delicate issues related to our construction in Sec.~\ref{Delicate} and give a summary and a discussion in Sec.~\ref{Discussion}. Our derivations, together with the relevant equations of the NP formalism, are presented in Appendices~\ref{Appendix:A}, \ref{Appendix:B} and \ref{Otherderivations}.

We use $\left(-,+,+,+\right)$ signature for our spacetime metric. Therefore one has to be careful about the definitions of the spin coefficients and curvature scalars when comparing them to Newman and Penrose's original construction in \citep{Newman_Penrose:1961}. However, that is not a complication for our contracted Raychaudhuri equation as it is independent of the metric signature. Also note that we use natural units through out the paper so that $c,G,h,k_B$ are set to 1.
\section{Mass-energy exchange: local, global and quasilocal}\label{Mass-energy exchange}
\subsection{Local approaches}
For local investigations of the gravitational energy flux, the Weyl tensor plays the central role. Newman and Penrose introduce five complex Weyl curvature scalars which incorporate all of the information of the Weyl tensor by \citep{Newman_Penrose:1961}
\begin{eqnarray}
\psi _0&=&\t {C}{_\mu _\nu _\alpha _\beta}l^\mu m^\nu l^\alpha m^\beta \label{Psi0},\\
\psi _1&=&\t {C}{_\mu _\nu _\alpha _\beta}l^\mu n^\nu l^\alpha m^\beta \label{Psi1},\\
\psi _2&=&\t {C}{_\mu _\nu _\alpha _\beta}l^\mu m^\nu \c{m}^\alpha n^\beta \label{Psi2},\\
\psi _3&=&\t {C}{_\mu _\nu _\alpha _\beta}l^\mu n^\nu \c{m}^\alpha n^\beta \label{Psi3},\\
\psi _4&=&\t {C}{_\mu _\nu _\alpha _\beta}n^\mu \c{m}^\nu n^\alpha \c{m}^\beta \label{Psi4},
\end{eqnarray}
where $C_{\mu \nu \alpha \beta}$ is the Weyl tensor of the spacetime, $\{l_{\mu}, n_{\mu}, m_{\mu}, \c{m}_{\mu}\}$ is the NP complex null tetrad and the only surviving inner products of the null vectors with each other are $\inner{\b{l},\b{n}}=-1$ and $\inner{\b{m},\c{\b{m}}}=1$.

The dynamics of timelike observers, who live in different 
Petrov-type spacetimes, was investigated by Szekeres previously \citep{Szekeres:1965}. In this method, one can assign physical meanings to the Weyl scalars. However, we note that this is only possible once we adapt our NP tetrad to the principal null direction(s) of the spacetime in question. Once we relax this condition, Weyl curvature scalars cannot be interpreted as the way it was done in Szekeres' work.

Let us decompose the Weyl tensor into its electric and magnetic parts. One can define a super-Poynting vector through them via \citep{Maartens_Bassett:1997}
$
\mathcal{P}_{\mu}=\t {\epsilon} {_{\mu}_{\alpha}_{\beta}}\t {\mathcal{E}}{^{\alpha}_{\nu}}\t {\mathcal{B}}{^{\beta}^{\nu}},
$ where $\t {\mathcal{E}}{_{\mu}_{\nu}}=\t {h}{^{\alpha}_{\mu}}\t {h}{^{\beta}_{\nu}}\t {C}{_{\alpha}_{\sigma}_{\beta}_{\gamma}}t^{\sigma}t^{\gamma}$ is its electric part, $\t {\mathcal{B}}{_{\mu}_{\nu}}=-\frac{1}{2}\t {h}{^{\alpha}_{\mu}}\t {h}{^{\beta}_{\nu}}\t {\epsilon}{_{\alpha}_{\sigma}_{\gamma}_{\kappa}}\t {C}{^{\gamma}^{\kappa}_{\beta}_{\rho}}t^{\sigma}t^{\rho}$ is the magnetic part, $t^{\mu}$ is the timelike vector orthogonal to the 3-dimensional spacelike hypersurfaces, $\t {h}{^{\mu}_{\nu}}$ is the corresponding projection operator and $\epsilon _{{\mu}{\nu}{\alpha}{\beta}}$ is the Levi-Civita tensor. The super-Poynting vector represents the gravitational energy flux density following its electromagnetic analogy. 
In \citep{Zhangetal:2012} it is shown that choosing a transverse tetrad, rather than a principal tetrad, aligns the gravitational wave propagation direction with the super-Poynting vector. Authors indicate that if we have a device which in principle works like Szekeres' ``gravitational compass'' \citep{Szekeres:1965} we can detect the gravitational waves locally.\footnote{In fact, recently, it has been announced that the gravitational waves have been detected by local measurements of the two LIGO interferometers \citep{Abbottetal:2016}.} This is of course applicable for a purely gravitational case.
\subsection{Global approaches}
For gravitational waves, Bondi mass loss \cite{Bondietal:1962} is one of the most widely used expressions to determine the energy lost by the system via gravitational radiation at null infinity. For an asymptotically flat spacetime, with NP variables, the Bondi mass reads as \cite{Szabados:2004}
\begin{eqnarray}
M_B = -\frac{1}{4\pi}\int_{\mathcal{S}}\left(\psi ^{\left(0\right)}_2 +\sigma ^{\left(0\right)} \dot{\c{\sigma}}^{\left(0\right)} \right)d\mathcal{S},
\end{eqnarray}
where $\mathcal{S}$ is the closed spacelike surface located at null infinity, $\sigma = -\inner{\b{m},D_\b{m}\b{l}}$ is one of the NP spin coefficients and the superscript ``$\left(0 \right)$'' represents the leading order part of the object with respect to a radial expansion.
The mass loss associated with the gravitational waves is determined once the ``time'' derivative, denoted by the overdot, of the Bondi mass is calculated in Bondi coordinates. Note that in the tetrad formalism approach of Bondi, the null tetrad is required to satisfy certain conditions. In the Bondi-Metzner-Sachs gauge one has
\begin{equation}
\kappa=\pi=\varepsilon=0, \qquad \rho=\overline{\rho}, \qquad \tau=\overline{\alpha}+\beta,
\end{equation}  
which gives the symmetry group of the conformal boundary at null infinity.

In terms of other global investigations, the energy loss of a relativistic body through its interaction with the external field can be traced back to Misner, Thorne and Wheeler's mass definition \cite{MTW:1974} constructed via an effective energy-momentum pseudotensor. Developed by many, including \cite{Thorne_Hartle:1984, Thorne:1980, Zhang:1986, Flanagan:1997}, the methodology for calculation of the mass-energy loss of an isolated relativistic body via its interaction with an external field is in fact very similar to the Newtonian analysis \cite{Purdue:1999}.

One can calculate the mass-energy loss via \cite{Thorne_Hartle:1984, Purdue:1999}
\begin{equation}
-\frac{d\mathcal{M}_\mathcal{S}}{dt}=\int_{\partial \mathcal{S}} \left(-g\right)t^{0J}n_J r^2 d\Omega , 
\end{equation}
where $M_{\mathcal{S}}$ is the mass inside the 3-sphere $\mathcal{S}$ which gives the mass of the isolated object, $M$, to leading order under the slow rotation assumption; $\partial \mathcal{S}$ is the 2-dimensional boundary of $\mathcal{S}$, $-g$ is the square of the 4-metric density, $t^{\alpha \beta}$ is the Landau-Lifshitz pseudotensor \cite{Landau_Lifshitz:1975}, $n^J =x^J/r$ are the radial vector components and $d\Omega$ is the 2-dimensional volume element. If one keeps only the $\mathcal{E} \mathcal{I}$ cross terms, where $\mathcal{E} _{JK}=\t {R}{_{J}_{0}_{K}_{0}}$, $\t {R}{_{\mu} _{\nu} _{\alpha} _{\beta}}$ is the Riemann tensor of the external field and $\mathcal{I} _{JK}$ is the mass quadrupole moment of the isolated body, one gets
\begin{equation}
-\frac{d\mathcal{M}_\mathcal{S}}{dt}=\frac{d}{dt}\left(\frac{1}{10}\mathcal{E} ^{JK}I_{JK}\right)+\frac{1}{2}\mathcal{E} ^{JK}\frac{d \mathcal{I}_{JK}}{dt},
\end{equation}
in which only the zeroth and first order time derivatives and the leading order term in the perturbative expansion are considered. In this approach, the first term on the right-hand side is interpreted as the rate of change of the interaction energy of the body and the external field, whereas the second term is interpreted as the rate of work done by the external field on the body. Therefore,
\begin{equation}\label{dWdt_global}
\frac{dW}{dt}= -\frac{1}{2}\mathcal{E} ^{JK}\frac{d \mathcal{I} _{JK}}{dt}
\end{equation}
is sometimes referred to as $\it {tidal \, heating}$ even though the energy loss/gain is not solely via the cooling/heating of the body in question \citep{Purdue:1999}.

There have been debates about whether or not the total mass of the body, which is taken as the sum of the self energy and the interaction energy, is ambiguous in this picture \cite{Thorne_Hartle:1984, Purdue:1999, Favata:2000} 
\footnote{The discussion began with Thorne and Hartle's statement that there exists an ambiguity in the total mass energy of the body \cite{Thorne_Hartle:1984}. Later, Purdue concluded that there is no ambiguity at least in the rate of work done on the system up to leading order \cite{Purdue:1999}. Furthermore, Favata considered different ``localisations'' of gravitational energy and concluded that the total mass-energy of the system does not depend on the choice of the energy-momentum pseudotensor and is thus unambiguous \cite{Favata:2000}.}.
For the time being, let us bear in mind that results obtained in this approach are true up to the leading order of the energy calculations of an external field and of an asymptotically flat spacetime which models a slowly rotating body at null infinity. Also, in general, one should be careful about using energy-momentum pseudotensors to calculate the mass-energy of a system since not all of them satisfy the conservation law with correct weight \cite{Szabados:2004}.\footnote{Let $t^{\mu \nu}_{(2k)}$ be a gravitational stress-energy pseudotensor with $k \in \mathbb{R}$. Some of the well known pseudotensors in general relativity can be defined via $2|g|^{k+1}\left(8\pi G\,t^{\alpha \beta}_{(2k)}-G^{\alpha \beta}\right)$ $:=\partial _\mu \partial _\nu \left(|g|^{k+1}\left[g^{\alpha \beta}g^{\mu \nu} -g^{\alpha \nu}g^{\beta \mu}\right]\right)$. Then Einstein field equations imply that $\partial _\alpha \left(|g|^{k+1}\left[t^{\alpha \beta}_{(2k)} +T^{\alpha \beta}\right]\right)=0$ where $T^{\alpha \beta}$ is the matter stress energy tensor. This shows that there is only one pseudotensor, $t^{\mu \nu}_{(-2)}$, which satisfies the conservation of the ``total'' stress-energy tensor with the correct weight.}

\subsection{Quasilocal approaches}
When quasilocal calculations of the mass-energy exchange of generic systems are considered, it is seen that the effective matter plus gravitational energy, momentum and stress energy densities can be attributed to the extrinsic or intrinsic geometry of a closed, spacelike, 2-dimensional surface in many applications. These spacelike 2-surfaces can be considered as the $t$-constant surfaces of the (2+1) timelike boundary of the spacetime. Alternatively, they can be considered as the embedded surfaces of spacelike 3-hypersurfaces or embedded surfaces of the spacetime itself \citep{Hawking:1968, Hayward:1993Q, Brown_York:1992, Kijowski:1997, Booth_Mann:1998, Epp:2000, Liu_Yau:2003}.

For example, suppose $\mathcal{B}$ is a (2+1) dimensional timelike boundary of a finite spacetime domain. Brown and York \cite{Brown_York:1992} define $\tau _{\mu \nu}=\left(\Theta \gamma _{\mu \nu}-\Theta _{\mu \nu}\right)/\left(8 \pi \right)$ as the object that carries information about the matter plus gravitational energy content of a given system by following a Hamiltonian approach. Here $\Theta _{\mu \nu}$ is the extrinsic curvature of the world tube and $\gamma _{\mu \nu}$ is the 3-metric induced on it that is fixed. 
Then the matter plus gravitational energy flux density, $f_{BY}$, follows from the world tube derivative of the matter plus gravitational energy tensor, i.e.,
\begin{equation}\label{flux_BY}
f_{BY}=\t {\gamma} {_\mu ^\alpha} D_\alpha \left(\tau ^{\mu \nu} t_\nu \right),
\end{equation} 
where $t^\mu$ is a timelike vector field which is not necessarily orthogonal to the $t$-constant spacelike surfaces $\mathcal{S}_t$, $\t {\gamma} {_\mu ^\alpha}$ is the projection operator on to the world tube and $D_\alpha $ is the spacetime covariant derivative.

In \citep{Booth_Creighton:2000}, the authors define the rate of work done on a quasilocal system via Eq.~(\ref{flux_BY}) by specifically choosing $t^\mu$ not to be a timelike Killing vector field of the world tube metric. According to Booth and Creighton, in vacuum, the rate of work done on the system by its environment is given by
\begin{equation}\label{dWdt_Booth&Creighton}
\frac{dW}{dt}=-\frac{1}{2}\int _{\mathcal{S}_t} d^2x \sqrt{-\gamma}\tau ^{\mu \nu} \$_t \gamma _{\mu \nu},
\end{equation}
where $\$_t$ is the operator that is obtained by projecting the covariant derivative operator defined by the induced metric of $\mathcal{B}$ on the spacelike 2-surface. Equation (\ref{dWdt_Booth&Creighton}) is used  to calculate the tidal heating quasilocally in the weak field limit, which serves as an excellent example to compare the quasilocal formalisms with the global ones. Their results show that the leading terms of the rate of work done is not exactly equal to the one given by the global method, eq.~(\ref{dWdt_global}). It is only the so-called \emph{irreversible} part, the portion that is expended to deform the body, that is equal to $\frac{1}{2}\mathcal{E}^{JK}d\mathcal{I}_{JK}/dt$ and hence attributed to tidal heating. However, there exists an additional portion which is stored as the potential energy in the system, called the \emph{reversible part}, which differs from the results of the global method.

In \cite{Eppetal:2008}, Epp \emph{et al.} take one step further and come up with a more concrete definition of  matter plus gravitational energy flux between the initial, $\mathcal{S}_i$, and final, $\mathcal{S}_f$, slices of a world tube. This approach is more concrete in the sense that the 2-surfaces have certain conditions on them. The authors define a rigid quasilocal frame by demanding the 2-surfaces to have zero expansion and shear when they are considered to be embedded in the world tube. In this approach, the energy flux density in vacuum is calculated as $\alpha_ \mu \mathcal{P}^\mu$. Here $\alpha _\mu$ is the proper acceleration of the observers projected on the 2-surface, $\mathcal{P}^\mu$ are constructed via the normal and tangential projections of $\tau _{\mu \nu}$, as defined by Brown and York \cite{Brown_York:1992}. On the spacelike 2-surfaces $\mathcal{P}^\mu = \sigma ^{\mu \nu} u^\rho \tau _{\nu \rho}$ and $\sigma _{\mu \nu}$ is the metric induced on the 2-surfaces. This is a coordinate approach. However, the conditions they impose on the spacelike 2-surface can be translated into null tetrad gauge conditions once a change of  formalism is applied. In the next section, we will see that our definition of a system is not as restrictive as the one of Epp \emph{et al.}.
\section{Null tetrad gauge conditions and the quasilocal calculations}\label{Null tetrad gauge}

In the present paper, we have no intention to discuss the advantages and disadvantages of numerical relativity calculations at finite distances.\footnote{For example see G{\'{o}}mez and Winicour's discussion on this issue \citep{Gomez_Winicour:1992}. Also see \citep{Friedrich:1981} for a construction of a conformal method and see \citep{Husa:2002} for a pedagogical review of conformal methods in numerical relativity.} However, we would like to keep track of the quasilocal observables and the null cone observables simultaneously as they are not always investigated in tandem in numerical relativity simulations.

Consider the case of a perturbed rotating black hole. In real astrophysical cases, our ultimate goal is to get information about the properties --such as the mass, angular momentum and their dissipation rates-- of this black hole via the gravitational radiation we detect. In such a case, we have the freedom to choose a null tetrad for gravitational radiation calculations and a corresponding orthonormal tetrad for the quasilocal energy calculations. One of our aims, in this paper, is to check whether or not those tetrad choices are consistent with each other when the different formalisms are considered. 

For example, there is a geometrically motivated transverse tetrad, the so-called quasi-Kinnersley tetrad \citep{Kinnersley:1969}, which is considered to be one of the best choices to study the gravitational wave extraction from a perturbed Kerr black hole \citep{Nerozzietal:2005qK, Nerozzietal:2005BS, Nerozzi:2006aj}. In \citep{Zhangetal:2012}, Zhang \emph{et al.} investigate the directions of energy flow using the super-Poynting vector and show that the wave fronts of passing radiation are aligned with the quasi-Kinnersley tetrad. However, in the current section, we introduce certain null tetrad gauge conditions for a quasilocal system which are not satisfied by the quasi-Kinnersley tetrad. This might mean that even though one can measure the gravitational radiation emitted from a region properly, one might not be able to extract the quasilocal properties of its source consistently. What we mean by this sentence will be more clear once we introduce our formalism and give a detailed discussion of this specific issue in Sec.~\ref{Delicate}.

When the quasilocal properties are taken into consideration, one has to start the investigation with a proper definition of a \emph{system}. This is the missing ingredient in many quasilocal approaches in the literature. In the present paper, we use a purely geometrical method to define our system. We will mainly consider a 2-dimensional timelike world sheet embedded into a 4-dimensional spacetime. The instantaneously defined 2-dimensional spacelike surface orthogonal to the world sheet at every point, encloses the system in question. 

The motivation behind the choice of such a geometric construction comes from the fact that the well-defined quasilocal energy definitions, which are made by following a Hamiltonian approach, rely on the $\it {mean \,extrinsic\,curvature}$ of a spacelike 2-surface.
It is a measure of boost-invariant matter plus gravitational energy density of the system \citep{Lau:1996, Kijowski:1997, Epp:2000, Liu_Yau:2003}. Hence the extrinsic geometry of this 2-surface, when it is embedded directly into a generic spacetime for example, is thought to have a more fundamental importance in terms of the quasilocal energy and energy exchange calculations.

In order to see how we define a system in the present paper, let us follow \citep{Capovilla_Guven:1994} and consider an embedding of an oriented world sheet with an induced metric, $\t {\eta}{_a_b}$, written in terms of orthonormal basis tangent vectors, $\{ \t {E}{_a} \}$,
\begin{equation}
g({\t {E}{_a}},{\t {E}{_b}})={\t {\eta}{_a_b}},
\end{equation}
where $g_{\mu \nu}$ is the 4-dimensional spacetime metric.
Now consider the two unit normal vectors, $\{ \t {N}{_i} \}$, of the world sheet which are defined up to a local rotation by
\begin{eqnarray}
g({\t {N}{_i}},{\t {N}{_j}})&=&{\t {\delta}{_i_j}},\\
g({\t {N}{^i}},{\t {E}{_a}})&=&0,
\end{eqnarray}
where $\{a,b\}=\{\hat{0},\hat{1}\}$ and $\{i,j\}=\{\hat{2},\hat{3}\}$ are the dyad indices and the Greek indices will refer to 4-dimensional spacetime coordinates.
Also note that to raise (or lower) the indices of tangential and normal dyad indices of an object, one should use ${\t {\eta}{^a^b}}$ (or ${\t {\eta}{_a_b}}$) and $\t {\delta}{^i^j}$ (or $\t {\delta}{_i_j}$) respectively, where in an orthonormal basis $\eta _{ab}=\left(-1,1\right)$ and $\delta _{ij}=\left(1,1\right)$.

Let us call this embedded timelike world sheet $\mathbb{T}$, and the spacelike surface which is orthogonal to $\mathbb{T}$ at every point,  $\mathbb{S}$. For a physically meaningful construction, we want the tangent spaces of these embedded surfaces to be integrable \citep{Capovilla_Guven:1994}.

According to Frobenius theorem, involutivity is a sufficient condition for the existence of an integral manifold through each point \citep{Lee:2003}. In other words, let $D^k$ be a $k$-dimensional distribution on a manifold $M$, which is 
required to be $C^\infty$. $D^k$ is involutive if for the vector fields $\b{X}$,$\b{Y}\in D^k$ their Lie bracket satisfies $\left[\b{X},\b{Y}\right]\in D^k$ \citep{Szekeres:2004}.

Therefore our tangent basis vectors $\{E_a,N_i\}$ need to satisfy
\begin{eqnarray}
\left[E_a,E_b\right]=\t {f}{^c _a_b}E_c, \label{Frob_T}\\
\left[N_i,N_j\right]=\t {h}{^k _i_j}N_k.\label{Frob_S}
\end{eqnarray}
Note that one can construct a complex null tetrad, $\{\b{l},\b{n},\b{m},\c{\b{m}}\}$, via an orthonormal double dyad and vice versa according to
\begin{eqnarray}
\t {E}{^\mu _{\hat{0}}} &=& \frac{1}{\sqrt{2}}\left(l^{\mu}+n^{\mu}\right),
\label{eq:null_to_t-like1}\\
\t {E}{^\mu _{\hat{1}}} &=& \frac{1}{\sqrt{2}}\left(l^{\mu}-n^{\mu}\right),
\label{eq:null_to_t-like2}\\
\t {N}{^\mu _{\hat{2}}} &=& \frac{1}{\sqrt{2}} \left(m^{\mu}+\c{m}^{\mu}\right),
\label{eq:null_to_t-like3}\\
\t {N}{^\mu _{\hat{3}}} &=& -\frac{i}{\sqrt{2}}\left(m^{\mu}-\c{m}^{\mu}\right).
\label{eq:null_to_t-like4}
\end{eqnarray}
Now let us see the gauge conditions that the Frobenius theorem, when applied to the tangent spaces of $\mathbb{T}$ and $\mathbb{S}$, imposes on a null tetrad constructed via the tangent vectors of $\mathbb{T}$ and $\mathbb{S}$. We can rewrite Eq.~(\ref{Frob_T}) as
\begin{eqnarray}
\t {E}{^\mu _a}D_{\mu}\t {E}{^\nu _b}-\t {E}{^\mu _b}D_{\mu}\t {E}{^\nu _a} = \t {f}{^c _a_b}\t {E}{^\nu _c} := \t {F}{^{\nu} _a _b}.
\end{eqnarray}
Considering the only nonzero component of $ F_{ab}$, i.e.,  $ F_{\hat{0}\hat{1}}=-F_{\hat{1}\hat{0}}$ and expressions (\ref{eq:null_to_t-like1})--(\ref{eq:null_to_t-like2}) we can write
\begin{eqnarray}
\t {F}{^{\nu}_{\hat{0}}_{\hat{1}}} &=& \t {E}{^\mu _{\hat{0}}}D_{\mu}\t {E}{^\nu _{\hat{1}}}-\t {E}{^\mu _{\hat{1}}}D_{\mu}\t {E}{^\nu _{\hat{0}}} \nonumber \\
&=&\t {f}{^{\hat{0}}_{\hat{0}}_{\hat{1}}}\t {E}{^\nu _{\hat{0}}} +
\t {f}{^{\hat{1}}_{\hat{0}}_{\hat{1}}}\t {E}{^\nu _{\hat{1}}}\nonumber \\
&=&\frac{1}{2}\left[\left(l^{\mu}+n^{\mu}\right) D_{\mu}\left(l^{\nu}-n^{\nu}\right) \right. \nonumber \\
&& \qquad {} \left.
- \left(l^{\mu}-n^{\mu}\right) D_{\mu}\left(l^{\nu}+n^{\nu}\right)\right]\nonumber \\
&=&\frac{1}{\sqrt{2}}\left[
\t {f}{^{\hat{0}}_{\hat{0}}_{\hat{1}}}\left(l^{\nu}+n^{\nu}\right)  +
\t {f}{^{\hat{1}}_{\hat{0}}_{\hat{1}}}\left(l^{\nu}-n^{\nu}\right)\right].
\end{eqnarray}
Thus,
\begin{eqnarray}
\left(D_{\b{l}}n^{\nu}-D_{\b{n}}l^{\nu}\right)
&=&
-\frac{1}{\sqrt{2}}
\left[
\left(\t {f}{^{\hat{0}}_{\hat{0}}_{\hat{1}}}+
\t {f}{^{\hat{1}}_{\hat{0}}_{\hat{1}}}\right)l^{\nu}
\right. \nonumber \\
&&\qquad \left.
+\left(\t {f}{^{\hat{0}}_{\hat{0}}_{\hat{1}}}-
\t {f}{^{\hat{1}}_{\hat{0}}_{\hat{1}}}\right)n^{\nu} 
\right] \label{F^nu_01}.
\end{eqnarray}
Now if we take the inner product of both sides of Eq.~(\ref{F^nu_01}) with the null vector $\b{m}$ we get
\begin{equation}
\inner{\b{m},D_{\b{l}}\b{n}}-\inner{\b{m},D_{\b{n}}\b{l}}=\c{\pi}-
\left(-\tau \right)= 0,
\end{equation}
which follows from the propagation equations (\ref{Dnl}) and (\ref{Dln}) of the spin coefficients of the Newman-Penrose formalism \cite{Newman_Penrose:1961}. 

Likewise when we rewrite Eq.~(\ref{Frob_S}) we get
\begin{eqnarray}
\t {N}{^\mu _i}D_{\mu}\t {N}{^\nu _j}-\t {N}{^\mu _j}D_{\mu}\t {N}{^\nu _i}=\t {h}{^k_i_j}\t {N}{^\nu _k} := \t {H}{^\nu _i_j}.
\end{eqnarray} 
If we consider the nonvanishing component $H_{\hat{2}\hat{3}}$ with the expressions (\ref{eq:null_to_t-like3})-(\ref{eq:null_to_t-like4}) we can write 
\begin{align}
\t {H}{^{\nu}_{\hat{2}}_{\hat{3}}} &= \t {N}{^\mu _{\hat{2}}}D_{\mu}\t {N}{^\nu _{\hat{3}}}-\t {N}{^\mu _{\hat{3}}}D_{\mu}\t {N}{^\nu _{\hat{2}}}\nonumber \\
&=\t {h}{^{\hat{2}}_{\hat{2}}_{\hat{3}}}\t {N}{^\nu _{\hat{2}}}
+\t {h}{^{\hat{3}}_{\hat{2}}_{\hat{3}}}\t {N}{^\nu _{\hat{3}}}\nonumber \\
&= -\frac{i}{2}\left(m^\mu +\c{m}^\mu \right)D_{\mu}\left(m^\nu -\c{m}^\nu \right)
\nonumber \\
&\qquad {} 
+\frac{i}{2}\left(m^\mu - \c{m}^\mu \right)D_{\mu}\left(m^\nu +\c{m}^\nu \right)\nonumber \\
&= \frac{1}{\sqrt{2}}\left[\t {h}{^{\hat{2}}_{\hat{2}}_{\hat{3}}} \left(m^\nu +\c{m}^\nu \right)-
i \t {h}{^{\hat{3}}_{\hat{2}}_{\hat{3}}}\left(m^\nu - \c{m}^\nu \right)\right].\nonumber \\
\end{align} 
Hence,
\begin{eqnarray}
\left(D_{\b{m}}\c{m}^\nu - D_{\c{\b{m}}} m^\nu\right)
&=& -\frac{1}{\sqrt{2}}\left[m^\nu \left(\t {h}{^{\hat{3}}_{\hat{2}}_{\hat{3}}}+i \t {h}{^{\hat{2}}_{\hat{2}}_{\hat{3}}} \right)
\right. \nonumber \\
&&\qquad \left.
-\c{m}^\nu \left(\t {h}{^{\hat{3}}_{\hat{2}}_{\hat{3}}}-i \t {h}{^{\hat{2}}_{\hat{2}}_{\hat{3}}} \right)\right]\label{H^nu_ij}.
\end{eqnarray}
Taking the inner product of both sides of Eq.~(\ref{H^nu_ij}) with the null vectors $\b{l}$ and $\b{n}$ respectively gives, 
\begin{eqnarray}
\inner{\b{l},D_{\b{m}}\c{\b{m}}}-\inner{\b{l},D_{\c{\b{m}}}\b{m}}&=&\c{\rho}-
\rho= 0,\\
\inner{\b{n},D_{\b{m}}\c{\b{m}}}-\inner{\b{n},D_{\c{\b{m}}}\b{m}}&=&
\left(-\mu \right)-\left(-\c{\mu} \right) = 0,
\end{eqnarray}
which follow from the propagation equation (\ref{Dmm_}).

Therefore we will state that for quasilocal energy calculations in our 2+2 approach, the following three null gauge conditions must be satisfied,
\begin{eqnarray}\label{Gaugeconditions}
\tau + \c{\pi}=0, \qquad \rho = \c{\rho}, \qquad \mu = \c{\mu}.
\end{eqnarray}

It is easy to check that under a type-III Lorentz transformation of the complex null tetrad, i.e.,
\begin{eqnarray}
\textbf{l} & \rightarrow & a^2 \textbf{l},\\
\textbf{n} & \rightarrow & \frac{1}{a^2}\, \textbf{n},\\
\textbf{m} & \rightarrow & e^ {2i\theta} \textbf{m},\\
\overline{\textbf{m}} & \rightarrow & e^ {-2i\theta} \overline{\textbf{m}},
\end{eqnarray}
the gauge conditions (\ref{Gaugeconditions}) are preserved.
This is because transformation of the spin coefficients $\tau,~\pi, ~\rho,~\mu$ under type-III Lorentz transformation follows as \cite{ODonnell:2003}
\begin{eqnarray}
\tau & \rightarrow & e^ {2i\theta} \tau ,\\
\pi & \rightarrow & e^ {-2i\theta} \pi ,\\
\rho & \rightarrow & a^2 \rho ,\\
\mu & \rightarrow & \frac{1}{a^2} \mu ,
\end{eqnarray}
in which $a^2$ and $2\theta$ respectively refer to the $boost$ and $spin$ parameters in Newman-Penrose formalism. They are arbitrary real functions.
Note that this transformation corresponds to 
\begin{eqnarray}
\t {E}{^{\mu} _{\hat{0}}} & \rightarrow & \gamma \left(\t {E}{^{\mu} _{\hat{0}}}-\beta \t {E}{^{\mu} _{\hat{1}}}\right),\\
\t {E}{^{\mu} _{\hat{1}}} & \rightarrow & \gamma \left(\t {E}{^{\mu} _{\hat{1}}}-\beta \t {E}{^{\mu} _{\hat{0}}}\right),
\end{eqnarray}
where 
\begin{equation}
\beta =\frac{a^4-1}{a^4+1}\qquad and \qquad \gamma=\frac{1}{\sqrt{1-\beta ^2}},
\end{equation}
meaning that a type-III Lorentz transformation of the null tetrad corresponds to the boosting of the timelike observers along $\t {E}{^{\mu}_{\hat{1}}}$ on $\mathbb{T}$. This is the property we want to preserve in the definition and the investigation of our quasilocal system.
\section{Raychaudhuri equation of a timelike world sheet}\label{RaychaudhuriCG}
In \citep{Capovilla_Guven:1994}, Capovilla and Guven construct a formalism to investigate the extrinsic geometry of an arbitrary dimensional timelike world sheet embedded in an arbitrary dimensional spacetime. We use their formalism to investigate the properties of a 2-dimensional world sheet, $\mathbb{T}$, embedded in a 4-dimensional spacetime as introduced in the previous section. Note that the Raychaudhuri equation of $\mathbb{T}$ carries information about how much the congruence of timelike \emph{world sheets} --- rather than world lines --- expands, shears or rotates. In their construction, Capovilla and Guven define three types of covariant derivatives, whose distinction we now introduce.

Let the torsionless covariant derivative defined by the spacetime coordinate metric be ${\t {D}{_\mu}}$ and its projection onto the world sheet be denoted by ${\t {D}{_a}}={\t {E}{^\mu_a}}{\t {D}{_\mu}}$.
On the world sheet $\mathbb{T}$, ${\t {\nabla}{_a}}$ is defined with respect to the intrinsic metric and ${\t {\tilde{\nabla}}{_a}}$ is defined on tensors under rotations of the normal frame, i.e., on $\mathbb{S}$. Likewise the projection of the spacetime covariant derivative on the instantaneous 2-surface $\mathbb{S}$ is  ${\t {D}{_i}}={\t {N}{^\mu_i}}{\t {D}{_\mu}}$. On $\mathbb{S}$, ${\t {\nabla}{_i}}$ is defined with respect to the intrinsic metric and ${\t {\tilde{\nabla}}{_i}}$ is defined on tensors under rotations of the normal frame of $\mathbb{S}$.

To study the deformations of $\mathbb{T}$ and $\mathbb{S}$, the following  extrinsic variables are introduced \citep{Capovilla_Guven:1994}.
The extrinsic curvature, Ricci rotation coefficients and extrinsic twist of
$\mathbb{T}$ are respectively defined by
\begin{eqnarray}
\t {K}{_a_b^i}&=&-\t {g}{_\mu _\nu}\left(\t {D}{_a}\t {E}{^\mu _b}\right)  \t {N}{^\nu ^i}=\t {K}{_b_a^i},\label{eq:K_a_b^i} \\
\t {\gamma}{_a_b_c}&=&\t {g}{_\mu _\nu}\left(\t {D}{_a}\t {E}{^\mu _b}\right)  \t {E}{^\nu _c}=-\t {\gamma}{_a_c_b},\label{eq:gamma_a_b_c} \\
\t {w}{_a^i^j}&=&\t {g}{_\mu _\nu}\left(\t {D}{_a}\t {N}{^\mu ^i}\right)  \t {N}{^\nu^j}=-\t {w}{_a^j^i}, \label{eq:w_a^i^j}
\end{eqnarray}
while the extrinsic curvature, Ricci rotation coefficients and extrinsic twist of $\mathbb{S}$ are respectively defined by
\begin{eqnarray}
\t {J}{_a^i^j}&=&\t {g}{_\mu _\nu}\left(\t {D}{^i}\t {E}{^\mu _a}\right)  \t {N}{^\nu ^j}, \label{eq:J_a^i^j} \\
\t {\gamma}{_i_j_k}&=&\t {g}{_\mu _\nu}\left(\t {D}{_i}\t {N}{^\mu _j}\right)  \t {N}{^\nu _k}=-\t {\gamma}{_i_k_j},\label{gamma_i_j_k} \\
\t {S}{_a_b^i}&=&\t {g}{_\mu _\nu}\left(\t {D}{^i}\t {E}{^\mu _a}\right)  \t {E}{^\nu _b}=-\t {S}{_b_a^i}. \label{eq:S_a_b^i}
\end{eqnarray}
By using those extrinsic variables one can investigate how the orthonormal basis $\{ \t {E}{_a},\t {N}{^i} \}$  varies when perturbed on $\mathbb{T}$ according to
\begin{eqnarray}
\t {D}{_a}\t {E}{_b}&=&\t {\gamma}{_a_b^c}\t {E}{_c}-\t {K}{_a_b^i}\t {N}{_i},\\
\t {D}{_a}\t {N}{^i}&=&\t {K}{_a_b^i}\t {E}{^b}+\t {w}{_a^i^j}\t {N}{_j},
\end{eqnarray}
or perturbed on $\mathbb{S}$ according to
\begin{eqnarray}
\t {D}{_i}\t {E}{_a}=\t {S}{_a_b_i}\t {E}{^b}+\t {J}{_a_i_j}\t {N}{^j},\\
\t {D}{_i}\t {N}{_j}=-\t {J}{_a_i_j}\t {E}{^a}+\t {\gamma}{_i_j^k}\t {N}{_k}.
\end{eqnarray}
Then the generalized Raychaudhuri equation, after being contracted with the orthogonal basis metrics ${\t {\eta}{^a^b}}$ and ${\t {\delta}{_i_j}}$ is given by
\begin{eqnarray}\label{eq:Raych_CapGuv}
\left(\t {\tilde{\nabla}}{_b}\t {J}{_a^i^j}\right){\t {\eta}{^a^b}}\t {\delta}{_i_j} &=-\left(\t {\tilde{\nabla}}{^i}\t {K}{_a_b^j}\right){\t {\eta}{^a^b}}\t {\delta}{_i_j}-\t {J}{_b^i_k}\t {J}{_a^k^j}{\t {\eta}{^a^b}}\t {\delta}{_i_j}
\nonumber \\
&\qquad {}
+g(R(\t {E}{_b},\t {N}{^i})\t {E}{_a},\t {N}{^j}){\t {\eta}{^a^b}}\t {\delta}{_i_j}
\nonumber \\
&\qquad {}
-\t {K}{_b_c^i}\t {K}{_a^c^j}{\t {\eta}{^a^b}}\t {\delta}{_i_j},
\end{eqnarray}
where $\t {R}{^\alpha_\beta_\mu_\nu}$ is the Riemann tensor of the 4-dimensional spacetime \cite{Capovilla_Guven:1994}, and
\begin{equation}
g(R(\t {E}{_a},\t {N}{_i})\t {E}{_b},\t {N}{_j})=\t {R}{_\alpha_\beta_\mu_\nu}\t {E}{^\mu_a}\t {N}{^\nu_i}\t {E}{^\beta_b}\t {N}{^\alpha_j}.
\label{Rproj}
\end{equation}
Note that $\t {w}{_b_i^k}$ transforms as a connection under the rotation of $\mathbb{S}$ and 
\begin{equation}\label{eq:CurlyCovJ}
\t {\tilde{\nabla}}{_b}\t {J}{_a_i_j}=\underbrace{\t {\nabla}{_b}\t {J}{_a_i_j}}_\text{$\t {D}{_b}\t {J}{_a_i_j}-\t {\gamma}{_b_a^c}\t {J}{_c_i_j}$}-\,\, \t {w}{_b_i^k}\t {J}{_a_k_j}-\,\, \t {w}{_b_j^k}\t {J}{_a_i_k}.
\end{equation}
Likewise, $\t {S}{_a_b^i}$ transforms as a connection under the rotation of $\mathbb{T}$ such that
\begin{equation}\label{eq:CurlyCovK}
\t {\tilde{\nabla}}{_i}\t {K}{_a_b^j}=\underbrace{\t {\nabla}{_i}\t {K}{_a_b^j}}_\text{$\t {D}{_i}\t {K}{_a_b^j}-\t {\gamma}{_i^j_k}\t {K}{_a_b^k}$}-\,\, \t {S}{_a_c_i}\t {K}{_b^c^j}-\,\, \t {S}{_b_c_i}\t {K}{_a^c^j}.
\end{equation}
Previously, in \citep{Uzun_Wiltshire:2015}, we interpreted Eq.~(\ref{eq:Raych_CapGuv}) for spherically symmetric systems by defining a quasilocal thermodynamic equilibrium state and the associated quasilocal thermodynamic potentials. To define quasilocal thermodynamic equilibrium, we minimized the quasilocal Helmholtz free energy density which was defined via the mean extrinsic curvature of $\mathbb{S}$. This showed us that the equilibrium takes place when the system is defined by the set of quasilocal observers who are located at the apparent horizon. For further details and the natural outcomes of this interpretation one can refer to \citep{Uzun_Wiltshire:2015}. In the following sections we will investigate more general systems which are in nonequilibrium with their surroundings. Moreover, we will relax the condition of spherical symmetry.
\section{Raychaudhuri equation with the Newman-Penrose formalism}\label{RaychaudhuriNP}
We use the relations (\ref{eq:null_to_t-like1})-(\ref{eq:null_to_t-like4}) in order to rewrite the contracted Raychaudhuri equation of our 2-dimensional timelike world sheet, Eq.~(\ref{eq:Raych_CapGuv}), in the language of the NP formalism. This will allow us to compare the results of the investigations of the energy exchange mechanisms built on null cone variables and the notation that is used in quasilocal energy calculations. 

Note that Eq.~(\ref{eq:Raych_CapGuv}) is built on the extrinsic geometry of $\mathbb{T}$ and $\mathbb{S}$. Those extrinsic objects, like curvature, rotation and twist, are all measures of how much the dyad vectors change when they are propagated along each other.
Likewise in the NP formalism, spin coefficients are defined via the changes of null vectors when they are propagated along each other with the relevant projections. A short summary of the NP formalism and the detailed calculations of our formalism transformation can be found in Appendices~\ref{Appendix:A} and \ref{Appendix:B} respectively.

When the formalism transformation is applied, the contracted Raychaudhuri equation, (\ref{eq:Raych_CapGuv}), of $\mathbb{T}$ can be conveniently written as
\begin{equation}\label{Raych_simple}
\tilde{\nabla} _{\mathbb{T}}\mathcal{J}=-\tilde{\nabla} _{\mathbb{S}}\mathcal{K}-\mathcal{J}^2-\mathcal{K}^2+\mathcal{R_{\,W}},
\end{equation}
where
\begin{align}
\tilde{\nabla} _{\mathbb{T}}\mathcal{J} &:= \t {\eta}{^a^b}\t {\delta}{^i^j}\t {\tilde{\nabla}}{_b}\t {J}{_a_i_j} \nonumber \\
&= \left[D_{\b{n}}\left(\rho +\c{\rho }\right)-D_{\b{l}}\left(\mu +\c{\mu }\right)\right]\nonumber
\\
&\qquad {}
-\left[\left(\varepsilon +\c{\varepsilon}\right)\left(\mu +\c{\mu}\right)+\left(\gamma +\c{\gamma}\right)\left(\rho +\c{\rho}\right)\right]
\nonumber \\
&\qquad {}
+2\left[\left(\varepsilon - \c{\varepsilon}\right)\left(\mu -\c{\mu} \right)+\left(\gamma - \c{\gamma}\right)\left(\rho -\c{\rho}\right)\right],\label{DeltaJNPgen}\\
\tilde{\nabla} _{\mathbb{S}}\mathcal{K}
&:=\t {\eta}{^a^b}\t {\delta}{^i^j}\t {\tilde{\nabla}}{_i}\t {K}{_a_b_j}\nonumber \\ &=D_{\b{m}}\left(\pi -\c{\tau} \right)+D_{\c{\b{m}}}\left(\c{\pi} -\tau \right)
\\
&\qquad {}
-\left[\left(\c{\alpha}-\beta \right)\left( \pi -\c{\tau} \right)+ \left(\alpha -\c{\beta} \right)\left( \c{\pi} -\tau \right)\right]
\nonumber \\
&\qquad {}
+ 2\left[ \left(\c{\alpha}+\beta \right)\left( \pi + \c{\tau} \right) +\left(\alpha + \c{\beta} \right)\left( \c{\pi} + \tau \right)\right],\label{DeltaKNPgen}
\end{align}
\begin{align}
\mathcal{J}^2
&:=\t {J}{_b_i_k}\t {J}{_a_l_j}{\t {\eta}{^a^b}}\t {\delta}{^i^j}\t {\delta}{^l^k}
\nonumber \\
&= 2\left(\mu \c{\rho} + \c{\mu} \rho + \sigma \lambda + \c{\sigma} \c{\lambda} \right),\label{J2NPgen}\\
\mathcal{K}^2
&:=\t {K}{_b_c_i}\t {K}{_a_d_j}\t {\eta}{^a^b}\t {\eta}{^c^d} \t {\delta}{^i^j}
\nonumber \\
&= -2\left(\kappa \nu + \c{\kappa} \c{\nu} + \pi \tau + \c{\pi} \c{\tau} \right),\label{K2NPgen} \\
\mathcal{R_{\,W}}
&:=g(R(\t {E}{_b},\t {N}{_i})\t {E}{_a},\t {N}{_j}){\t {\eta}{^a^b}}\t {\delta}{^i^j}
\nonumber \\
&=D_{\b{n}}\left(\rho +\c{\rho }\right)-D_{\b{l}}\left(\mu +\c{\mu }\right)
\nonumber \\
&\qquad {}
+ D_{\b{m}}\left(\pi -\c{\tau} \right)+D_{\c{\b{m}}}\left(\c{\pi} -\tau \right)
\nonumber \\
&\qquad {}
-\left[\left(\alpha - \c{\beta} \right)\left( \c{\pi} - \tau \right)+\left(\c{\alpha}-\beta \right)\left( \pi - \c{\tau} \right)\right]
\nonumber \\
&\qquad {}
-\left[\left(\varepsilon +\c{\varepsilon}\right)\left(\mu +\c{\mu}\right)+\left(\gamma +\c{\gamma}\right)\left(\rho +\c{\rho}\right)\right]
\nonumber \\
&\qquad {}
-2\left(\kappa \nu + \c{\kappa} \c{\nu} \right) 
+ 2\left(\rho \c{\mu} +\c{\rho} \mu +\lambda \sigma + \c{\lambda} \c{\sigma} \right).\label{RWNPgen}
\end{align}
An alternative, more compact expression for $\mathcal{R_{\,W}}$ is
\begin{eqnarray}
\mathcal{R_{\,W}}&=& -2\left(\psi _2 + \c{\psi}_2 + 4\Lambda \right).\label{RWSimpleNPgen}
\end{eqnarray}
Now if we substitute the terms (\ref{DeltaJNPgen})--(\ref{RWSimpleNPgen}) back into Eq.~(\ref{Raych_simple}) we see that the Raychaudhuri equation is not yet satisfied. This is simply because Capovilla and Guven impose the integrability condition in their formalism to define the extrinsic objects\footnote{This can be seen by checking the symmetries of the extrinsic objects introduced at the previous section.} and we did not impose it after our change of formalism. We must further impose the null tetrad gauge conditions introduced in Sec.~\ref{Null tetrad gauge}. Thus, with $\tau +\c{\pi}=0$, $\rho =\c{\rho}$ and $\mu = \c{\mu}$ we get
\begin{align}
\tilde{\nabla} _{\mathbb{T}}\mathcal{J}
&= 2\left(D_{\b{n}}\rho-D_{\b{l}}\mu \right)
\nonumber \\
&\qquad {}
-2\left[\left(\varepsilon +\c{\varepsilon}\right)\mu +\left(\gamma +\c{\gamma}\right)\rho \right]\label{DeltaJNP},\\
\tilde{\nabla} _{\mathbb{S}}\mathcal{K}
&=2\left(D_{\b{m}}\pi - D_{\c{\b{m}}}\tau \right) 
\nonumber \\
&\qquad {}
- 2\left[\left(\c{\alpha}-\beta \right)\pi + \left(\alpha -\c{\beta} \right)\c{\pi}\right] \label{DeltaKNP},\\
\mathcal{J}^2
&= 4\mu \rho + 2\left( \sigma \lambda + \c{\sigma} \c{\lambda} \right)
\label{J2NP},\\
\mathcal{K}^2
&= -2\left(\kappa \nu + \c{\kappa} \c{\nu}\right) + 2\left(\pi \c{\pi} + \tau \c{\tau} \right)\label{K2NP},\\
\mathcal{R_{\,W}}
&=2 \left[D_{\b{n}}\rho - D_{\b{l}}\mu \right]
+ 2\left[D_{\b{m}}\pi - D_{\c{\b{m}}}\tau \right]
\nonumber \\
&\qquad {}
-2 \left[\left(\c{\alpha}-\beta \right)\pi + \left(\alpha -\c{\beta} \right)\c{\pi}\right] 
\nonumber \\
&\qquad {}
-2\left[\left(\varepsilon +\c{\varepsilon}\right)\mu +\left(\gamma +\c{\gamma}\right)\rho \right]- 2\left(\kappa \nu + \c{\kappa} \c{\nu} \right)
\nonumber \\
&\qquad {}
+2\left(\tau \c{\tau}+\pi \c{\pi}\right) + 4 \mu \rho 
+ 2\left(\sigma \lambda +  \c{\sigma} \c{\lambda} \right)\label{RiemNP} ,
\end{align}
and the alternative expression (\ref{RWSimpleNPgen}) is unchanged.
These variables now satisfy the Raychaudhuri equation as expected. 

We further note that since the Einstein field equations have not yet been applied, (\ref{DeltaJNP})--(\ref{RiemNP}) are purely geometrical results irrespective of the underlying gravitational theory that governs the dynamics of the quasilocal observers. In order to satisfy the Einstein equations, all 16 of the field equations of the spin coefficients should be satisfied. However, we need to emphasize that this version of the contracted Raychaudhuri equation contains all the information contained in two of the NP spin field equations. Let us consider the following NP spin field equations
\begin{align}
\tensor{D}{_{\textbf{l}}}\,\mu - \tensor{D}{_{\textbf{m}}}\,\pi &= \mu \, \overline{\rho}-\left( \varepsilon +\overline{\varepsilon } \right) \mu + \sigma \lambda + \pi \, \overline{\pi} 
\nonumber \\
&\qquad {}
- \left( \overline{\alpha }-\beta  \right) \pi
-\kappa \,\nu +\psi _2 +2\Lambda,\label{eq:NP field1}\\
\tensor{D}{_{\textbf{n}}}\,\rho -\tensor{D}{_{\overline{\textbf{m}}}}\,\tau &= -\overline{\mu}\,\rho+ \left( {\it \gamma}+\overline{{\it \gamma}}\right)\rho-\sigma\,\lambda-\tau\,\overline{\tau}
\nonumber \\
&\qquad {}
-\left(\alpha-\overline{\beta}\right)\tau
+\kappa \,\nu-\psi _2-2\Lambda.\label{eq:NP field2}
\end{align}
If we take $(\ref{eq:NP field1})+(\ref{eq:NP field1})^*-(\ref{eq:NP field2})-(\ref{eq:NP field2})^*$, where $*$ denotes the complex conjugate, then the result is the contracted Raychaudhuri equation of the world sheet under our gauge conditions.
We will not attempt to restrict the general set of equations (\ref{DeltaJNP})-(\ref{RiemNP}) by further imposing the Einstein equations. Rather, we will apply it to spacetimes that are already solutions of the Einstein field equations.
\section{A work-energy relation}\label{Work-energy}
In this section we are going to define quasilocal charges by using the terms that appear in the Raychaudhuri equation. Ultimately we will make definitions so as to end up with a work-energy relation that looks like the following
\begin{eqnarray}
E_{\rm Total}=E_{\rm Dilatational}+E_{\rm Rotational}+W_{\rm Tidal}.
\end{eqnarray}
In doing so, one of Kijowski's quasilocal energy definitions will be our anchor. Let us recall the two energy definitions made by Kijowski which are derived from a gravitational action \citep{Kijowski:1997},
\begin{equation}
E_{\rm K1}=-\frac{1}{16\pi}\oint_{\mathbb{S}}{d\mathbb{S} \left(\frac{H^2-k_0^2}{k_0}\right)},\label{eq:E_K1}
\end{equation}
\begin{equation}
E_{\rm K2}=-\frac{1}{8\pi}\oint_{\mathbb{S}}{d\mathbb{S} \left(\sqrt{H^2}-k_0\right)},\label{eq:E_K2}
\end{equation}
where the square of the mean extrinsic curvature, $H^2$, is the $k^2-l^2$ term that often appears in quasilocal energy definitions. The term $k_0$ is the extrinsic curvature of a spacelike 2-surface embedded into the 3-dimensional space of a reference spacetime which is chosen to be Minkowski, $\mathcal{M}^4$, in Kijowski's work. Previously, we identified Eq.~(\ref{eq:E_K1}) as internal energy \citep{Uzun_Wiltshire:2015} since it was associated with the quasilocal energy of a system in equilibrium which can potentially be used to do work, dissipate heat or exchange energy in other forms. The second expression (\ref{eq:E_K2}) is usually interpreted as the invariant mass energy of the system that is an analogue of a proper mass of a particle \cite{Epp:2000}. Therefore if we are after an expression which represents the energy that can be exchanged by the system, $H^2$ should be our central object.\footnote{The literature is divided into two camps in terms of the definition of the extrinsic curvature scalars $k$ and $l$. For example, let us consider $k := \sigma ^{\mu \nu}k_{\mu \nu}=\sigma ^{\mu \nu} \left(\t {\sigma}{^\alpha_\mu}\t {\sigma}{^\beta_\nu}D_\alpha \,n_\beta\right)$, where $\sigma _{\mu \nu}$ is the induced 2-metric on the closed spacelike surface, $\mathbb{S}$, and $\b{n}$ is the unit vector orthogonal to $\mathbb{S}$ when we consider its embedding in a spacelike 3-volume. For this definition, $k_0=+\frac{2}{r}$ for a round 2-sphere. This notation was used in Epp's \citep{Epp:2000}, Liu and Yau's \citep{Liu_Yau:2003} and in Szabados's review article \citep{Szabados:2004}. On the other hand, Brown and York \citep{Brown_York:1992} and Kijowski \citep{Kijowski:1997} follow the formal notation for the extrinsic curvature with an extra minus sign. Accordingly $k_0=-\frac{2}{r}$ for a round 2-sphere in their notation. In this paper, we follow the notation used by the first camp since the ``positivity'' theorem was first presented in this notation \citep{Liu_Yau:2003}. Moreover, we suspect most researchers refer to Szabados' review article to compare and contrast various quasilocal energy definitions. Therefore, in Kijowski's original paper \citep{Kijowski:1997}, $E_{\rm K1}$ and  $E_{\rm K2}$ are given in different forms than the ones presented in Eqs.~(\ref{eq:E_K1}) and (\ref{eq:E_K2}) respectively.}

The quasilocal energy definitions $E_{\rm K1}$ and $E_{\rm K2}$ of Kijowski both have the functional form $\left(H^2\right)^p$ with $p=1$ and $p=1/2$ respectively. This is due to Kijowski applying a Legendre transform on the boundary Hamiltonian with different boundary conditions. In the case of $E_{\rm K1}$, he controls the information on the boundary of the world tube by imposing conditions on the metric of the induced 2-surface and the associated curvature. He sets the components of the induced 2-metric of $\mathbb{S}$ to be time independent in this type of control, in order to avoid the extra volume inclusions. By contrast in $E_{\rm K2}$, the entire information of the world tube is controlled via imposing conditions on the 3-metric of the world tube. Those conditions require the world tube metric to have $g_{00}=1$ and  $g_{0A}=0$, where $A$ refers to the indices of the spacelike boundary of the world tube. Ultimately $E_{\rm K1}$ and $E_{\rm K2}$ might be used for situations where different boundary conditions apply. However, this does not cause any problem in terms of the dimensionality of the quasilocal energies as the so-called reference terms, which make sure that the energy definitions are boost invariant, do not appear in the same format.

Previously, in \citep{Uzun_Wiltshire:2015}, we defined quasilocal thermodynamic potentials at equilibrium for spherically symmetric spacetimes by using the terms that appear in the contracted Raychaudhuri equation, (\ref{eq:Raych_CapGuv}), of $\mathbb{T}$. We applied our formalism for metrics with boundary conditions $g_{00}=1$, $g_{0A}=0$ when the quasilocal observers are located at the apparent horizon. Therefore the quasilocal charges defined in \citep{Uzun_Wiltshire:2015} take the same form as $E_{\rm K2}$. Note that this refers to a very special state of the system in question. 

In the present paper, we would like to define quasilocal charges for nonequilibrium states and we would like to go beyond spherical symmetry. We will consider spacetimes with metrics that have time independent components for the induced 2-metric on $\mathbb{S}$ just as Kijowski did to define $E_{\rm K1}$. In order to define the quasilocal charges we will first multiply the contracted Raychaudhuri equation (\ref{Raych_simple}) by 2 \footnote{The reason behind this factor of 2 will be more clear in the following sections.}, and add the reference energy term, $k_0^2$, to each side. Since all of the terms that appear in Eq.~(\ref{Raych_simple}) have dimension $(length)^{-2}$ on account of their relationship to the Riemann tensor, to obtain a quasilocal energy expression we further divide by $k_0$ before integrating the equation on our closed 2-surface $\mathbb{S}$. Then we obtain the following quasilocal charges
\begin{eqnarray}
E_{\rm Tot}&=&-\frac{1}{16\pi}\oint_{\mathbb{S}}{d\mathbb{S} \left[\frac{-\left(2\tilde{\nabla} _{\mathbb{T}}\mathcal{J} +k_0^2\right)}{k_0}\right]},\label{ETotal}\\
E_{\rm Dil}&=&-\frac{1}{16\pi}\oint_{\mathbb{S}}{d\mathbb{S} \left[\frac{2\mathcal{J}^2-k_0^2}{k_0}\right]},\label{EDilatational}\\
E_{\rm Rot}&=&-\frac{1}{16\pi}\oint_{\mathbb{S}}{d\mathbb{S} \left[\frac{2\tilde{\nabla} _{\mathbb{S}}\mathcal{K} + 2\mathcal{K}^2}{k_0}\right]},\label{Erotational}\\
W_{\rm Tid}&=&-\frac{1}{16\pi}\oint_{\mathbb{S}}{d\mathbb{S} \left[\frac{-2\mathcal{R_{\,W}}}{k_0}\right]},\label{WTidal}
\end{eqnarray}
so that 
\begin{eqnarray}
E_{\rm Tot}&=&E_{\rm Dil}+E_{\rm Rot}+W_{\rm Tid}
\end{eqnarray}
is satisfied.

In the following sections, we will discuss our reasons for these quasilocal charge definitions.
The reasons behind naming our quasilocal charges like energy associated with dilatational or rotational degrees of freedom and work done by tidal fields of the system will be explained.
\subsection{Energy associated with dilatational degrees of freedom}
In spherical symmetry \citep{Uzun_Wiltshire:2015}, we were able to write $\mathcal{J}^2:=\t {J}{_b_i_k}\t {J}{_a_l_j}{\t {\eta}{^a^b}}\t {\delta}{^i^j}\t {\delta}{^l^k}$ in terms of the square of the mean extrinsic curvature, $H^2$, of $\mathbb{S}$ via $2\mathcal{J}^2=H^2$. Note that confining the quasilocal observers to radial world lines in a spherically symmetric system results in corresponding, purely radial, null congruences that are shear-free. Indeed, for the generic case,
\begin{eqnarray}
H^2 &:=&\t {J}{_a_i_k}\t {J}{_b_j_l}{\t {\eta}{^a^b}}\t {\delta}{^i^k}\t {\delta}{^j^l}=2\left(\rho + \c{\rho}\right)\left(\mu + \c{\mu}\right)\label{meanJ},\\
\mathcal{J}^2 &:=&\t {J}{_a_i_l}\t {J}{_b_j_k}{\t {\eta}{^a^b}}\t {\delta}{^i^k}\t {\delta}{^j^l}=2\left(\mu \c{\rho} + \c{\mu} \rho + \sigma \lambda + \c{\sigma} \c{\lambda} \right)\label{Raych_J}.
\end{eqnarray}
Therefore with two of our null tetrad gauge conditions, $\rho = \c{\rho},~\mu = \c{\mu}$ and the shear-free case, $\sigma=0$, 
\begin{eqnarray}
H^2=2\mathcal{J}^2=4\left(\mu \c{\rho} + \c{\mu} \rho + \sigma \lambda + \c{\sigma} \c{\lambda} \right)=8\mu \rho .
\end{eqnarray}
This is natural for radially moving observers of spherically symmetric systems. However, it is not clear which of the terms in (\ref{meanJ}) and (\ref{Raych_J}) carries more information about the generic system in question. 

According to the Goldberg-Sachs theorem, there exists a shear-free null congruence, $k^ \mu$, for a vacuum spacetime if \cite{Wald:1984}
\begin{eqnarray}
k _{[\mu} C _{\nu]\alpha \beta [\gamma} k_{\sigma ]}k^ \alpha k^\beta =0
\end{eqnarray} 
is satisfied. This means that if we wish to have the shear-free property, we need to pick a principal null tetrad for our systems in vacuum. However, there is no such $a\,priori$ necessity for our formalism to hold.

In \citep{Adamoetal:2012}, Adamo \emph{et al.} investigate the shear-free null geodesics of asymptotically flat spacetimes in detail. They note that the shear-free or asymptotically shear-free null congruences may provide information 
about the asymptotic center of mass or intrinsic magnetic dipole in certain cases. Also the importance of the twistor theory, which is solely constructed on shear-free null congruences, cannot be denied. At this point, we should also emphasize that the spacetimes we are interested in are not necessarily asymptotically flat.
 
In \citep{Ellis:2011}, Ellis investigated shear-free timelike and null congruences. He concluded that by imposing a shear-free condition on the null congruences, one puts a restriction on the way the distant matter can influence the local gravitational field. In that case, there is an information loss. Note that shear is also the central concept of Bondi's mass loss formulation. It is only if the null congruence has shear, that one can define a $news\,function$ which is solely responsible for the mass loss via gravitational radiation at null infinity \citep{Bondietal:1962}. Ellis also emphasized the fact that a nonrotating null congruence in vacuum cannot shear without expanding or contracting. Thus we cannot completely separate the effects of dilatation and shear for null congruences. We will combine them in the quasilocal charge constructed from the $\mathcal{J}^2$ term, (\ref{J2NP}), and write
\begin{eqnarray}
E_{\rm Dil}&=&-\frac{1}{16\pi}\oint_{\mathbb{S}}{d\mathbb{S} \left[\frac{2\mathcal{J}^2-k_0^2}{k_0}\right]}\nonumber \\
&=&-\frac{1}{16\pi}\oint_{\mathbb{S}}{d\mathbb{S} \left[\frac{8\mu \rho + 4 \left( \sigma \lambda + \c{\sigma} \c{\lambda} \right)-k_0^2}{k_0}\right]}.\label{Edil}
\end{eqnarray}

Since we claim that the Raychaudhuri equation of the world sheet incorporates the physically meaningful quasilocal energy densities, one might ask what the direct connection of our $\mathcal{J}^2$ term, (\ref{Raych_J}), to the boundary Hamiltonian --which is generically written in terms of the mean extrinsic curvature $H$, (\ref{meanJ}) --is. The link lies in the Gauss equation of the 2-surface $\mathbb{S}$ when it is embedded directly into spacetime \citep{Spivak:1979}, i.e.,
\begin{equation}\label{Gauss_S}
g(R(N_k, N_l)N_j,N_i)=\mathcal{R} _{ijkl} - \t {J}{_a_i_k}\t {J}{_b_j_l}{\t {\eta}{^a^b}} + \t {J}{_a_j_k}\t {J}{_b_i_l}{\t {\eta}{^a^b}},
\end{equation}  
where $\mathcal{R} _{ijkl}$ is the Riemann tensor associated with the 2-dimensional metric induced on $\mathbb{S}$. If we contract Eq.~(\ref{Gauss_S}) with $\t {\delta}{^i^k}\t {\delta}{^j^l}$ we find
\begin{eqnarray}\label{Gauss_S_simple}
\mathcal{J}^2=H^2-\mathcal{R}_{\, \mathbb{S}}+2\left(\Psi _2+\c{\Psi _2}-2\Lambda -2 \Phi _{11}\right),
\end{eqnarray}
in which $\mathcal{R}_{\, \mathbb{S}}:=\mathcal{R} _{ijkl}\t {\delta}{^i^k}\t {\delta}{^j^l}$ is the scalar intrinsic curvature of $\mathbb{S}$ and the derivation of $g(R(N_k, N_l)N^l,N^k)=-2\left(\Psi _2+\c{\Psi _2}-2\Lambda -2 \Phi _{11}\right)$ can be found in Appendix~\ref{Otherderivations}.
Equation (\ref{Gauss_S_simple}) not only allows us to connect our $\mathcal{J}^2$ term to the boundary Hamiltonian of general relativity, but it can also be used to relate different quasilocal energy definitions which are built on either the extrinsic or intrinsic curvature of $\mathbb{S}$.
\subsection{Energy associated with rotational degrees of freedom}
In the previous subsection we defined the quasilocal energy associated with the dilatational degrees of freedom by combining the real divergence and the possibly existing shear of the null congruence which is constructed from the timelike dyad that spans the timelike surface $\mathbb{T}$. Now we will distinguish which spin coefficients are most significant in defining the energy associated with the rotational degrees of freedom.

Recall that by imposing the integrability conditions on our local dyad we made sure that the tangent vectors of the spacelike surface $\mathbb{S}$ always stay within the surface. Later, we transformed our construction into the NP formalism and stated that these conditions imply that the null vectors $\{\b{m}, \c{\b{m}}\}$, constructed from the spacelike dyad of $\mathbb{S}$, should satisfy certain null gauge conditions throughout the evolution of the quasilocal system. Then, under such gauge conditions, the magnitude of the change of these null vectors should be related to how much the quasilocal system rotates. Note that this interpretation makes sense only when one forces  the spacelike dyad, constructed from  $\{\b{m}, \c{\b{m}}\}$, to stay on $\mathbb{S}$ throughout the evolution.

Now let us define the spacetime covariant derivative via the directional covariant derivatives of the null tetrad and write
\begin{eqnarray}
D_{\mu}&=-l_{\mu}D_{\b{n}}-n_{\mu}D_{\b{l}}+m_{\mu}D_{\c{\b{m}}}+\c{m}_{\mu}D_{\b{m}}.
\end{eqnarray} 
Then the change in components of $\{\b{m}, \c{\b{m}}\}$ follows as
\begin{align}
D_{\mu}m^{\mu}&=-\inner{\b{l},D_{\b{n}}\b{m}} -\inner{\b{n},D_{\b{l}}\b{m}} 
\nonumber \\
&\qquad 
+\inner{\b{m},D_{\c{\b{m}}}\b{m}}
+\inner{\c{\b{m}},D_{\b{m}}\b{m}},\nonumber \\
D_{\mu}\c{m}^{\mu}&=-\inner{\b{l},D_{\b{n}}\c{\b{m}}}-\inner{\b{n},D_{\b{l}}\c{\b{m}}}
\nonumber \\
&\qquad +\inner{\b{m},D_{\c{\b{m}}}\c{\b{m}}}
+\inner{\c{\b{m}},D_{\b{m}}\c{\b{m}}}\nonumber.
\end{align}
By using Eqs.~(\ref{Dlm})--(\ref{Dmm_}) we get
\begin{eqnarray}
D_{\mu}m^{\mu}&=\left(\c{\pi} -\tau \right)+\left(\beta - \c{\alpha}\right),\nonumber \\
D_{\mu}\c{m}^{\mu}&= \left(\pi -\c{\tau} \right)+\left(\c{\beta} - \alpha \right).\nonumber
\end{eqnarray}
Therefore, the spin coefficients $\{\pi,~\tau,~\alpha,~\beta \}$, their complex conjugates and their changes when one perturbs them on $\mathbb{S}$ can be used to define the energy associated with the rotational degrees of freedom. Since the terms $\tilde{\nabla} _{\mathbb{S}}\mathcal{K}$, (\ref{DeltaKNP}), and $\mathcal{K}^2$, (\ref{K2NP}), involve these spin coefficients and their changes we define
\begin{align}
E_{\rm Rot}
&=-\frac{1}{16\pi}\oint_{\mathbb{S}}d\mathbb{S} \left[\frac{2\tilde{\nabla} _{\mathbb{S}}\mathcal{K} + 2\mathcal{K}^2}{k_0}\right] \nonumber \\
&=-\frac{1}{16\pi}\oint_{\mathbb{S}}d\mathbb{S}\frac{4}{k_0} \left[D_{\b{m}}\pi - D_{\c{\b{m}}}\tau  - \pi \left(\c{\alpha}-\beta \right) 
\right.  \nonumber \\
&\qquad \qquad \qquad \qquad \left.
-\c{\pi} \left(\alpha -\c{\beta} \right)
+ \pi \c{\pi} + \tau \c{\tau}
\right. \nonumber \\
&\qquad \qquad \qquad \qquad \qquad \qquad \qquad \left.
-\kappa \nu - \c{\kappa} \c{\nu}\right].\label{Erot}
\end{align}
Note that the term $\left(\kappa \nu + \c{\kappa} \c{\nu}\right)$ vanishes if one picks the null vector $\b{l}$ or $\b{n}$, constructed from the timelike dyad that spans $\mathbb{T}$, to be a geodesic, i.e., $\kappa =0$ or $\nu =0$. In that case $E_{\rm Rot}$ can be written purely in terms of the spin coefficients $\{\pi,~\tau,~\alpha,~\beta \}$. However, there is no geometric or physical reason for us to demand our null congruences to be geodesic, and we will not impose the geodesic condition for the time being.  
\subsection{Work done by tidal distortions}
If we want to understand the properties of a system via its energy exchange mechanisms we need to account for the different types of associated energies,  especially in the nonequilibrium case. One needs to be careful about what is actually measured by the quasilocal observers. What is physical for any one observer is the tidal acceleration as measured by that observer's local ruler and clock. The work done by tidal distortions of the whole system, however, requires the quasilocal observers to be placed in such a geometric configuration that the observers all agree on the fact that they are measuring the properties of the same system. In the previous sections, we stated that this is guaranteed by our integrability conditions. 

In \cite{Hartle:1974}, Hartle investigates the changes in the shape of an instantaneous horizon of a rotating black hole through the intrinsic scalar curvature, $\mathcal{R}_{\, \mathbb{S}}$, of a spacelike 2-surface when it is embedded into a 4-dimensional spacetime. He $chooses$ a null tetrad gauge $so \,\,that$ $\mathcal{R}_{\, \mathbb{S}}$ can be written in terms of a simple combination of $\Psi _2$ and the spin coefficients in vacuum. In the end, he finds $\mathcal{R}_{\, \mathbb{S}}=4Re\left(-\Psi _2 +\rho \mu -\lambda \sigma \right)$.  In \cite{Hayward:1994}, Hayward provides a quasilocal version of the Bondi-Sachs mass via the Hawking mass \cite{Hawking:1968}, in which the central object is again the complex intrinsic scalar curvature given by $\mathcal{R}_{\, \mathbb{S}}^H= -\Psi _2 +\sigma \sigma ' -\rho \rho ' +\Phi _{11}+\Pi $, in the formalism of weighted spin coefficients.

We believe that the $\mathcal{R_{\,W}}$ term that appears in Eq.~(\ref{RiemNP}) has a more fundamental meaning than $\mathcal{R}_{\, \mathbb{S}}$ in terms of the tidal distortion. In order to show why this should be so, previously in \cite{Uzun_Wiltshire:2015}, we considered its analogue in the 3+1 picture.
In particular,
\begin{equation}\label{eq:relacc}
\frac{d^2\xi ^\mu}{d\tau ^2}=\tensor{R}{^\mu_\nu _\rho _\sigma}u^{\nu}u^{\rho}\xi ^{\sigma}
\end{equation}
is the relative tidal acceleration of the observers on neighboring timelike geodesics, where $\vec{\xi}$ is the spacelike separation 4-vector, $\tau$ is the proper time and $u^\mu$ are the 4-velocity vector field components.
Thus one can define an object which we named  \textit{relative work density}, that mimics $W=\vec{F}\cdot \vec{x}$ by
\begin{equation}\label{eq:relwork}
\left(\frac{d^2\xi ^\mu}{d\tau ^2}\right)\xi_{\mu}=\tensor{R}{^\gamma _\nu _\rho _\sigma}u^{\nu}u^{\rho}\xi^{\sigma}\xi_{\gamma},
\end{equation}
in which the separation vector was assumed to be residing on $\mathbb{S}$.
We also noted that, in the 3+1 picture, connecting the two world lines is essentially nonlocal. The reason for applying Eq.~(\ref{eq:relacc}) only for neighboring world lines is due to the fact that the observers are trying to approximate the value of a quantity, which is essentially quasilocal, locally \citep{Uzun_Wiltshire:2015}.  Therefore the quantity (\ref{eq:relwork}) in the 2+2 picture, i.e., $\mathcal{R_{\,W}}=g(R(\t {E}{_b},\t {N}{_i})\t {E}{_a},\t {N}{_j}){\t {\eta}{^a^b}}\t {\delta}{^i^j}=-2\left(\psi _2 + \c{\psi}_2 + 4\Lambda \right)$, should have a more fundamental importance, as it is an intrinsically quasilocal quantity. Therefore  by Eq.~(\ref{RWSimpleNPgen})
\begin{eqnarray}
W_{\rm Tid}&=&-\frac{1}{16\pi}\oint_{\mathbb{S}}{d\mathbb{S} \left[\frac{-2\mathcal{R_{\,W}}}{k_0}\right]}\nonumber \\
&=&-\frac{1}{16\pi}\oint_{\mathbb{S}}{d\mathbb{S} \left[\frac{4\left(\psi _2 + \c{\psi}_2 + 4\Lambda \right)}{k_0}\right]}.\label{WTid}
\end{eqnarray}
Note that the quasilocal tidal work of the system is written purely in terms of the Coulomb-like Weyl curvature scalar, $\psi _2$, and the Ricci scalar of the spacetime due to $\Lambda=R/24$. This interpretation does not contradict our intuition, since one would expect the quasilocal observers to measure greater magnitude of tidal distortion under higher Coulomb-like attraction and a higher Ricci curvature.
\subsection{Total energy}
In \cite{Uzun_Wiltshire:2015} we associated the $\sqrt{2\mathcal{J}^2}$ term with the Helmholtz free energy density for spherically symmetric systems in equilibrium. Likewise $\sqrt{2\left|\tilde{\nabla} _{\mathbb{T}}\mathcal{J}\right|}$ was interpreted as the Gibbs free energy density of the system that \emph{includes} the energy that is spontaneously exchanged with the surroundings to relax the system into its current state. However, in the present paper, we do not attempt to give a thermodynamic interpretation to the Raychaudhuri equation of Capovilla and Guven since systems far from equilibrium cannot be assigned unique thermodynamic relations even in classical thermodynamics \citep{Demirel:2007}. Therefore, by using the term $\tilde{\nabla} _{\mathbb{T}}\mathcal{J}$, (\ref{DeltaJNP}), the total energy is represented by 
\begin{align}
E_{\rm Tot}&=-\frac{1}{16\pi}\oint_{\mathbb{S}}d\mathbb{S} \left[\frac{-\left(2\tilde{\nabla} _{\mathbb{T}}\mathcal{J} +k_0^2\right)}{k_0}\right] \nonumber \\
&=-\frac{1}{16\pi}\oint_{\mathbb{S}}d\mathbb{S}\frac{1}{k_0} \left\{-4\left[D_{\b{n}}\rho -D_{\b{l}}\mu \right]
\right.  \nonumber \\
&\qquad \qquad \qquad \qquad \left.
+4\left[\left(\varepsilon +\c{\varepsilon}\right)\mu +\left(\gamma +\c{\gamma}\right)\rho \right]-k_0^2
\right\}.\label{Etot}
\end{align}
Here the total energy combines two types of terms: (i) the quasilocal energy the system possesses, (ii) the energy that is expended by the ``internal''(tidal) forces to bring the quasilocal observers in a geometric configuration to define $\mathbb{S}$. The first piece further splits into the energy associated with dilatational and rotational degrees of freedom. The second piece can be viewed as the energy that has already been expended by the system in order for it to create ``room'' for itself.
\subsection{On the boost invariance of the quasilocal charges}
Previously, in Sec.~\ref{Null tetrad gauge}, it was shown that our tetrad conditions, (\ref{Gaugeconditions}), are invariant under type-III Lorentz transformations which correspond to the boosting of physical observers in the only spacelike direction, $\t {E}{^{\mu} _{\hat{1}}}$, defined on $\mathbb{T}$. We also stated that for a well-defined construction, one would expect the matter plus gravitational energy of the system to be boost invariant. 

In Appendix~\ref{App:SpinBoost} we show that all of the terms, (\ref{DeltaJNP})-(\ref{RiemNP}), that appear in the contracted Raychaudhuri equation are invariant under such spin-boost transformations. Therefore all of the quasilocal charges we defined in the current section are invariant under the boosting of the observers along the spacelike direction orthogonal to $\mathbb{S}$.
\section{Applications}\label{Applications}
\subsection{Radiating Vaidya spacetime}
The Vaidya spacetime is used in investigations of radiating stars. It is associated with a spherically symmetric metric which reduces to the Schwarzschild metric when the mass function of the body is taken to be a constant. In standard coordinates with null coordinate, $u$, Vaidya metric is
\begin{equation}
ds^2=-\left(1-\frac{2M(u)}{r}\right)du^2-2du\,dr+r^2d\theta ^2 +r^2\sin ^2 \theta d\phi ^2.
\end{equation} 
Let us pick the following complex null tetrad, $\{\b{l},\b{n},\b{m},\c{\b{m}}\}$, with
\begin{eqnarray}
l^\mu &=& \partial _u-\left(\frac{1}{2}-\frac{M(u)}{r}\right)\partial _r,\\
n^\mu &=& \partial _r,\\
m^\mu &=& \frac{1}{\sqrt{2}}\left(\frac{1}{r}\partial _\theta +\frac{i}{r\sin{\theta}}\partial _\phi \right).
\end{eqnarray}
For such a complex null tetrad, $\kappa,~\nu,~\sigma,~\lambda,~\tau, ~\pi$ all vanish so that $\pi + \c{\tau}=0$ is trivially satisfied. Also $\rho = \c{\rho}$, $\mu = \c{\mu}$ as expected. Therefore all of our integrability conditions are satisfied. When we evaluate the spin coefficients, their relevant directional derivatives and the curvature scalars, then substitute them in Eq.~(\ref{Raych_simple}) we get
\begin{eqnarray}
\tilde{\nabla} _{\mathbb{T}}\mathcal{J}
&=& \frac{-2}{r^2}+\frac{8M(u)}{r^3}\label{DeltaJVaidya},\\
\tilde{\nabla} _{\mathbb{S}}\mathcal{K}
&=&0 \label{DeltaKVaidya},\\
\mathcal{J}^2
&=& \frac{2}{r^2}-\frac{4M(u)}{r^3}\label{JVaidya},\\
\mathcal{K}^2
&=&0\label{K^2Vaidya},\\
\mathcal{R_{\,W}}
&=&\frac{4M(u)}{r^3}\label{R_WVaidya}.
\end{eqnarray}
Here we immediately notice that the terms that have been associated with the rotational degrees of freedom, i.e., $\tilde{\nabla} _{\mathbb{S}}\mathcal{K}$ and $\mathcal{K}^2$, are zero. This is expected since Vaidya is a spherically symmetric spacetime. 

In order to calculate our quasilocal charges we need to first find the so-called reference curvature $k_0$. This requires the isometric embedding of the $u=$ constant, $r=$ constant surface to the $\mathcal{M}^4$, Minkowski spacetime, which is considered in the spherical coordinates $\{ \bar{r},\bar{\theta},\bar{\phi} \}$. For Vaidya, by setting $\{\bar{r}=r,\bar{\theta}=\theta,\bar{\phi}=\phi \}$ we see that the metric induced on $\mathbb{S}$ is trivially isometric to that of the 2-surface embedded in $\mathcal{M}^4$. Then $k_0$ is given by the scalar curvature of a 2-sphere, i.e.,  $k_0=2/\bar{r}=2/r$. From Eqs.~(\ref{Edil}), (\ref{Erot}), (\ref{WTid}) and (\ref{Etot}) we then have 
\begin{eqnarray}
E_{\rm Tot}
&=&\frac{-1}{16\pi}\int_{\mathbb{S}}d\mathbb{S}\frac{-\left[ 2\left(\frac{-2}{r^2}+\frac{8M(u)}{r^3}\right)+\frac{4}{r^2}\right]}{\frac{2}{r}}= 2M\left(u\right) \nonumber ,\\
E_{\rm Dil}
&=&\frac{-1}{16\pi}\int_{\mathbb{S}}{d\mathbb{S}\frac{\left[2\left(\frac{2}{r^2}-\frac{4M(u)}{r^3}\right)-\frac{4}{r^2}\right]}{\frac{2}{r}}}= M\left(u\right)\nonumber, \\
W_{\rm Tid}
&=&\frac{-1}{16\pi}\int_{\mathbb{S}}{d\mathbb{S}\frac{\left[-2\left(\frac{4M(u)}{r^3}\right)\right]}{\frac{2}{r}}}= M\left(u\right)\nonumber, \\
E_{\rm Rot}&=&0.
\end{eqnarray}

Note that we chose a null tetrad in order to satisfy our gauge conditions which turned out to be shear-free. Therefore $H^2=2\mathcal{J}^2$ holds in this case and thus $E_{\rm Dil}=E_{\rm K1}$. Also, the spacetime Ricci scalar, $24\Lambda$, vanishes. Therefore $\mathcal{R_{\,W}}=-2\left(\Psi _2 +\c{\Psi} _2 \right)=-4\Psi _2$ and the $\mathcal{R_{\,W}}$ term is solely determined by the Coulomb-like gravitational potential.

\begin{figure}[htp]

\subfloat[Vaidya at $u=0$, i.e, Schwarzschild geometry.]{\label{Fig:Schwenergy}
  \includegraphics[clip,width=0.8\columnwidth]{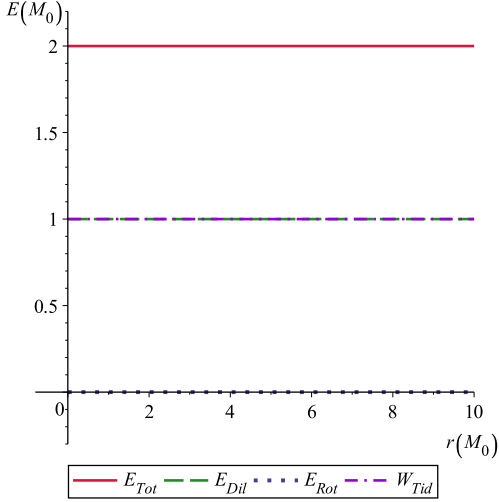}%
}

\subfloat[Vaidya at $u=1$, i.e, Minkowski geometry.]{\label{Fig:Minkenergy}
  \includegraphics[clip,width=0.8\columnwidth]{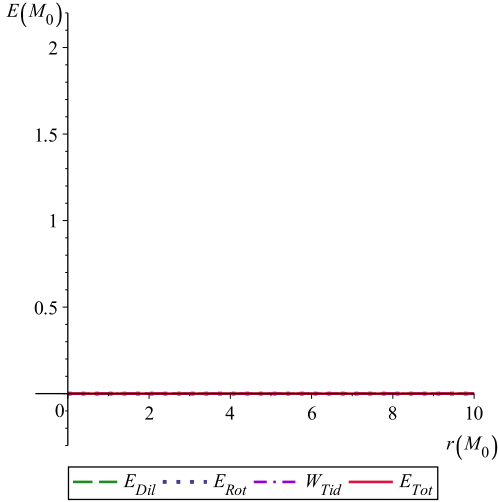}%
}

\subfloat[Time evolution of quasilocal charges.]{\label{Vaidyatimeevo}
  \includegraphics[clip,width=0.8\columnwidth]{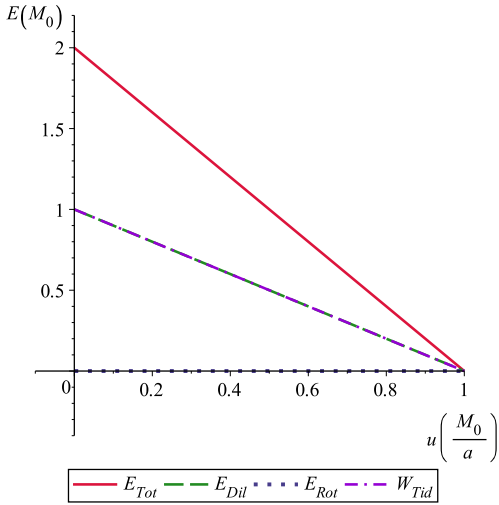}%
}
\caption{Our quasilocal charges give $E_{\rm K1}=E_{\rm Dil}=W_{\rm Tid}$ and $E_{\rm Rot}=0$ for each $u$ value. Charges are given in units of $M_0$ and the time parameter is in units of $M_0/(a\,c)$ where the speed of light, $c$, is 1 throughout the paper. }\label{fig:QLEdensities}

\end{figure}

To visualize a simple evolution, consider the mass function $M(u)=M_0-a\,u$, where $a$ is a positive constant. These kinds of linear mass functions have been used to investigate the black hole evaporation previously in the literature (cf. \citep{Hiscock:1980}, \cite{Waugh_Lake:1986}, \cite{Podolsky_Svitek:2005}). With this choice of mass function, at $u=0$ we have the case of a Schwarzschild black hole [see Fig.~\ref{Fig:Schwenergy}.] which, given enough time, eventually evaporates so that the spacetime becomes Minkowski [see Fig.~\ref{Fig:Minkenergy}.]. The quasilocal charges fall off linearly with the time parameter $u$ [see Fig~\ref{Vaidyatimeevo}.].
Now let us consider the $\tilde{\nabla} _{\hat{0}} E_{\rm Dil}=\tilde{\nabla} _{\hat{0}}\left(E_{\rm K1}\right)=\t {E}{^\mu _{\hat{0}}}\partial _\mu \left(E_{\rm K1}\right)$. Following relation (\ref{eq:null_to_t-like1}) and with the choices we have made here for $\b{l}$ and $\b{n}$,
\begin{eqnarray}
\tilde{\nabla} _{\hat{0}}E_{\rm Dil}&=&\frac{1}{\sqrt{2}}\left[\partial _u+\left(\frac{1}{2}+\frac{M(u)}{r}\right)\partial _r \right]M(u)\nonumber \\
&=&\frac{1}{\sqrt{2}}\frac{\partial M(u)}{\partial u}.\nonumber
\end{eqnarray}

According to the Einstein field equations, $-\frac{2}{r^2}\frac{\partial M(u)}{\partial u}=8\pi \tilde{\rho}$, where $\tilde{\rho}$ is the energy density of the null dust. This shows that the dilatational energy of  the system which could potentially be lost by work, heat or other forms is lost purely due to radiation, for the case of the Vaidya spacetime.
\subsection{The $C$-metric}
For our second application we want to consider a nonspherically symmetric spacetime. The $C$-metric is not spherically symmetric and it has many interpretations depending on its coordinate representation. We will consider the coordinate representation which was introduced by Hong and Teo \cite{Hong_Teo:2004},
\begin{equation}\label{HongTeometric}
ds ^2=\frac{1}{H}\left(-F\,d\tau ^2 +\frac{dy^2}{F}+\frac{dx^2}{G}+G\,d\phi ^2\right),
\end{equation}
with 
\begin{eqnarray}
H(x,y)&:=&A^2\left(x+y\right)^2 \nonumber,\\
G(x)&:=&\left(1-x^2\right)\left(1+2AMx\right)\nonumber,\\
F(y)&:=&-\left(1-y^2\right)\left(1-2AMy\right)\nonumber.
\end{eqnarray}
Griffiths \emph{et al.} \citep{Griffithsetal:2006} transformed this cylindrical form of the metric into spherical coordinates by applying the coordinate transformation $\{ \tau =At, x=\cos \theta , y= 1/(Ar) \}$ and gave physical interpretations to the $C$-metric.  The transformed metric is written as \citep{Griffithsetal:2006}
\begin{align}\label{Griffithsmetric}
ds ^2 =\frac{1}{\Delta}\left(-Q dt^2+\frac{dr^2}{Q}+\frac{r^2d\theta ^2}{P}+Pr^2\sin ^2\theta d\phi ^2 \right),
\end{align}
where 
\begin{eqnarray}
\Delta (r,\theta)&:= &\left(1+Ar\cos \theta \right)^2\nonumber,\\
Q(r)&:=&\left(1-\frac{2M}{r}\right)\left(1-A^2r^2\right)\nonumber,\\
P(\theta)&:=&1+2AM\cos \theta \nonumber,
\end{eqnarray}
with $A$ and $M$ being constants. Note that at $r=2M$ and at $r=1/A$ the metric has coordinate singularities and one needs to satisfy the  $A^2M^2<1/27$ condition in order to preserve the metric signature. Furthermore, Eq.~(\ref{Griffithsmetric}) reduces to the metric of the Schwarzschild black hole in standard curvature coordinates when one sets $A=0$. Because of this, following Griffiths \emph{et al.} \cite{Griffithsetal:2006}, we will interpret the $C$-metric as the metric of an accelerated black hole. At this point we note that the $C$-metric is sometimes interpreted as a metric representing two causally disconnected black holes that are joined by a strut and accelerating away from each other \citep{Bonnor:1983, Bonnor:1988, Cornish_Uttley:1995}. However, this interpretation is valid only when the metric is extended across each horizon, i.e., $r=2M$ and $r=1/A$ \citep{Griffithsetal:2006}. For the application of our quasilocal construction we will not consider such an extension of the metric, and the resulting quasilocal charges will correspond to the charges of a single accelerated black hole. 

Let us consider the following null tetrad that is generated by the double dyad of the quasilocal observers:
\begin{eqnarray}
l^\mu &=& \frac{1}{\sqrt{2}}\left[\frac{\Delta }{Q(r)}\right]^{1/2}\partial _t-\frac{1}{\sqrt{2}}\left[\Delta \,Q(r)\right]^{1/2}\partial _r ,\nonumber\\
n^\mu &=& \frac{1}{\sqrt{2}}\left[\frac{\Delta }{Q(r)}\right]^{1/2}\partial _t+\frac{1}{\sqrt{2}}\left[\Delta \,Q(r)\right]^{1/2}\partial _r ,\nonumber\\
m^\mu &=& \frac{1}{\sqrt{2}}\left[\frac{\Delta \, P(\theta )}{r^2}\right]^{1/2}\partial  _\theta +\frac{i}{\sqrt{2} \sin \theta}\left[\frac{\Delta}{r^2 P(\theta )}\right]^{1/2}\partial _\phi . \nonumber
\end{eqnarray}
For such a null tetrad, our integrability conditions $\{ \pi + \c{\tau}=0, \rho = \c{\rho}, \mu = \c{\mu} \}$ hold. The only vanishing spin coefficients are $\kappa,~\nu,~\lambda$ and  $\sigma$, meaning that our null congruences, constructed from the timelike dyads residing on the 2-surface $\mathbb{T}$, are composed of geodesics which are shear-free. As noted earlier this last property is not a necessary condition in our formalism. With the remaining nonvanishing spin coefficients and the variables of the contracted Raychaudhuri equation given in (\ref{DeltaJNP})-(\ref{RiemNP}) we get
\begin{eqnarray}
\tilde{\nabla} _{\mathbb{T}}\mathcal{J}
&=& \frac{1}{r^3}\left[P(\theta)\left(6r-2A^2r^3\right)-4A\cos \theta r^2
\right. \nonumber \\
&&\qquad \left. 
+8\left(M-r\right)\right],\label{C-DeltaJ}\\
\tilde{\nabla} _{\mathbb{S}}\mathcal{K}
&=&\frac{2A}{r}\left[2AM\cos ^2 \theta\left(2A\cos \theta r +3\right)
\right. \nonumber \\
&&\qquad \left. 
+\cos \theta \left(A\cos \theta r +2\right) 
\right. \nonumber \\
&&\qquad \qquad \left.
+A\left(r-2M\right)\right],\label{C-DeltaK} \\
\mathcal{J}^2
&=& \frac{2Q(r)}{r^2}, \label{C-J^2}\\
\mathcal{K}^2
&=& 2A^2P(\theta )\sin ^2\theta,\label{C-K^2} \\
\mathcal{R_{\,W}}
&=&4M\left(\frac{1}{r}+A\cos \theta\right)^3.\label{C-Rw}
\end{eqnarray}
In order to calculate the quasilocal charges we must first calculate the reference energy density, $k_0$. We isometrically embed $\mathbb{S}$ into $\mathcal{M}^4$, by setting
\begin{eqnarray}
\frac{r^2d\theta ^2}{\Delta \, P(\theta) }&=&\bar{r}^2d\bar{\theta}^2,\\
\frac{P(\theta)r^2\sin ^2\theta d\phi ^2}{\Delta}&=&\bar{r}^2\sin ^2\bar{\theta}d\bar{\phi}^2,
\end{eqnarray}
and demand that the observers measure the same solid angle in both coordinate systems.
This is satisfied by choosing $\bar{r}=r\Delta ^{-1/2}$ and then $k_0=2/\bar{r}$. Here we should note that for a generic $C$-metric the angular coordinates are defined within $\{ 0< \theta < \pi , -C\pi < \phi < C\pi \}$ where $C$ is the remaining parameter, other than $A$ and $M$, that parametrizes the spacetime. It is closely related to the ``deficit/excess angle'' that tells us how much $\mathbb{S}$ deviates from the spherical symmetry. For example, repeating Griffiths \emph{et al.}'s discussion
\begin{eqnarray}
\frac{\rm circumference}{\rm radius}=
\begin{cases}
\lim_{\theta \to 0}{\frac{2\pi CP(\theta)\sin \theta}{\theta}} =2\pi C\left(1+2AM\right)\nonumber\\
\nonumber\\
\lim_{\theta \to \pi}{\frac{2\pi CP(\theta)\sin \theta}{\pi - \theta}}=2\pi C\left(1-2AM\right)\nonumber\\
\end{cases}
\end{eqnarray}
shows us that setting $C=1$, as we choose to do here, will introduce excess and deficit angles on the spacelike surface $\mathbb{S}$ due to the conical singularities that are introduced. This, and our choices for coordinate functions of $\mathcal{M}^4$ will guarantee that the solid angle is the same for the quasilocal observers of the physical and the reference spacetimes.

\begin{figure}[htp]
  \includegraphics[clip,width=0.9\columnwidth]{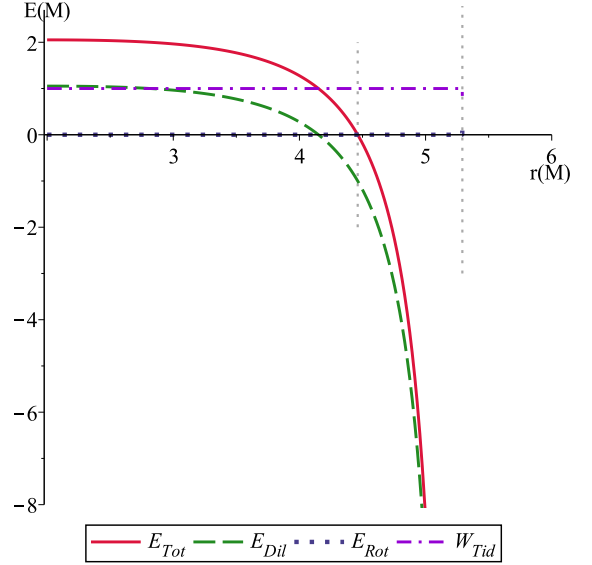}%
\caption{Quasilocal charges of the $C$-metric which is parametrized with $A=\frac{1}{\sqrt{28}M}$. Those quasilocal charges are meaningful only in the region $2M<r<\sqrt{28}M \approx 5.29M$ due to the coordinate singularities.}\label{fig:C_Energies}
\end{figure}

We obtain the quasilocal charges by substituting the quasilocal charge densities, in Eqs.~(\ref{C-DeltaJ})--(\ref{C-Rw}), into the definitions (\ref{ETotal})--(\ref{WTidal}) and numerically integrating them. The results are presented in Fig.~\ref{fig:C_Energies} for a specific choice of $A=1/(\sqrt{28}M)$ to perform the numerical integration. 

\begin{figure}[htp]
  \includegraphics[clip,width=0.9\columnwidth]{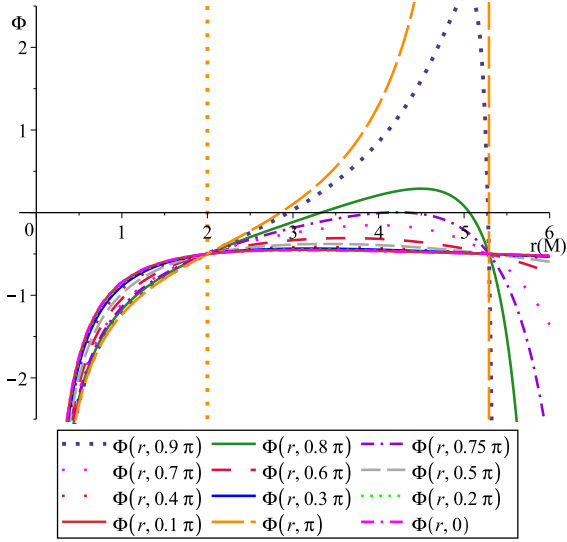}%
\caption{Radial behavior of the gravitational potential of the $C$-metric, which is parametrized with $A=\frac{1}{\sqrt{28}M}$, plotted for observers located at different polar angles. Those potentials are meaningful only in the region $2M<r<\sqrt{28}M \approx 5.29M$ due to the coordinate singularities.}\label{fig:C_gravpot28}
\end{figure}

From Fig.~\ref{fig:C_Energies} we immediately recognize that  $E_{\rm K1}=E_{\rm Dil}$ decreases as the size of the system increases. For the case of Schwarzschild, i.e., $A=0$, we expect this curve to be flat, as in Fig.~\ref{Fig:Schwenergy}. For lower values of acceleration, $E_{\rm Dil}$ gets flatter as expected. This shows that in order for the black hole to be accelerated more, more energy should be input to the system by an \emph{external} agent. In other words, the potential work that can be done \emph{by} the system is lower. Note that after a certain size of the system, $E_{\rm Dil}$ and $E_{\rm Tot}$ take negative values. It may seem counterintuitive that quasilocal observers could measure a ``negative energy.'' To better understand this result, consider the metric (\ref{Griffithsmetric}) and define  $g_{tt}=-\left(Q(r)/\Delta\right)=-\left[1+2\Phi (r,\theta)\right]$ where $\Phi(r,\theta)$ plays the role of the ``gravitational potential.'' In Fig.~\ref{fig:C_gravpot28} we plot $\Phi(r,\theta)$ for observers located at different polar angles.
We observe that for none of the observers, except the ones located at $\theta =\pi$, $\Phi(r,\theta)$ is monotonic. Moreover, for observers located at $\theta > 0.75 \pi$  the gravitational potential changes sign after a certain radial distance. This shows that the effect of the external agent on the system is repulsive. Then the positive total energy $E_{\rm Dil}+W_{\rm Tid}$, which corresponds to a system that has an otherwise attractive nature, cannot overcome the repulsive effect of the external agent which causes the black hole to accelerate. 
The $E_{\rm Tot}=0$ point can be viewed as the minimum energy state of the system, below which it cannot exist without the energy exchange provided by an external agent.  

\begin{figure}[htp]

\subfloat[$\Phi(r,\pi)$]{\label{Fig:PhiSp}
  \includegraphics[clip,width=0.8\columnwidth]{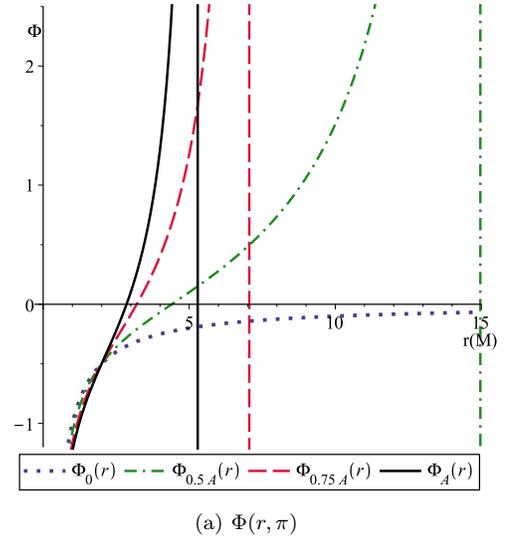}%
}

\subfloat[$\Phi(r,\pi /2)$]{\label{Fig:PhiEq}
  \includegraphics[clip,width=0.8\columnwidth]{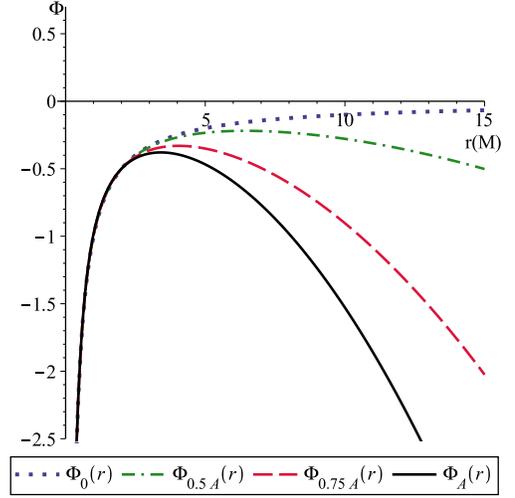}%
}

\subfloat[$\Phi(r,0)$]{\label{Fig:PhiNp}
  \includegraphics[clip,width=0.8\columnwidth]{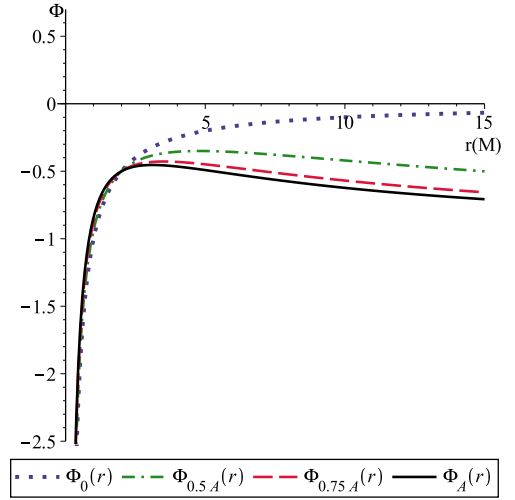}%
}
\caption{Acceleration parameter dependence of $\Phi(r,const.)$. For each acceleration parameter $A^*$, we consider only the region $2M<r<1/A^*$.}\label{fig:C_gravpots}
\end{figure}

Also recall that the $C$-metric is interpreted as two black holes which are accelerated \emph{away} from each other. This is a signature of the repulsive behavior we observe here. Note that here we are investigating one of the most extreme cases for an accelerated black hole, since as for acceleration parameters greater than $1/\left(\sqrt{27}M\right)$ the metric changes signature. Therefore the change in the behavior of the gravitational potential, and hence a change in the sign of the total energy of the system is not unexpected. We do not observe such behavior for the Schwarzschild geometry as the gravitational potential is monotonic with constant sign for a static black hole. In order to investigate how the acceleration parameter, $A$, affects the behavior of the gravitational potential, see Fig.~\ref{fig:C_gravpots}. We plot $\Phi(r,\theta)$ for observers located at $\theta =\pi$, $\theta =\pi /2$ and $\theta =0$ respectively in Figs.~\ref{Fig:PhiSp}, \ref{Fig:PhiEq} and \ref{Fig:PhiNp}. For each case, we investigate the effect of the acceleration parameter, $A$. We observe that only for $A=0$ case does the gravitational potential not change behavior.
For a more detailed investigation of the behavior of the gravitational potential of a $C$-metric, depending on the observer position and on the acceleration parameter, one can see \citep{Farhoosh_Zimmerman:1980}.

In order to understand what this means for the acceleration vector of an observer of the quasilocal system, let us set the 4-velocity of the observer to be $u^\mu = \t {E}{^\mu _{\hat{0}}} = \frac{1}{\sqrt{2}}\left(l^{\mu}+n^{\mu}\right)$. Then the acceleration vector is obtained by $a^\mu=D_{\t {E}{_{\hat{0}}}}\t {E}{^\mu _{\hat{0}}}=a_r \partial _r +a_\theta \partial 
_\theta$ with 
\begin{eqnarray}
a_r&=&-\frac{1}{r^2}\left[A^3r^4\cos \theta \left(AM\cos \theta +1\right)
\right. \nonumber \\
&&\qquad \left. 
+A^2r^2\cos ^2\theta  \left(r-3M\right)+A^2r^2\left(r-2M\right)
\right. \nonumber \\
&& \qquad \left. 
+Ar\cos \theta \left(r-4M\right)-M
\right],\label{C-a_r}\\
a_\theta &=& \frac{A\sin \theta}{r}P(\theta)\Delta ^{1/2} .
\end{eqnarray}
\begin{figure}[htp]

\subfloat[Radial component, $a_r$.]{\label{Fig:a_r}
  \includegraphics[clip,width=0.9\columnwidth]{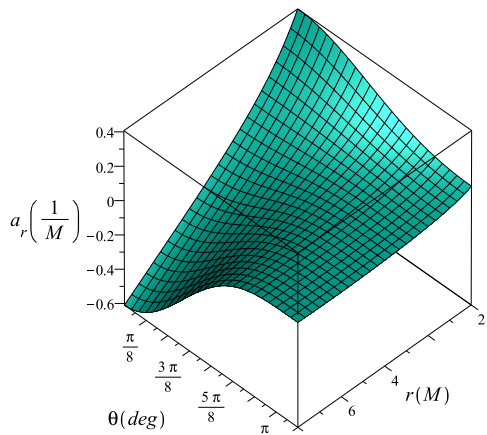}%
}

\subfloat[Tangential component, $a_\theta$.]{\label{Fig:a_theta}
  \includegraphics[clip,width=0.9\columnwidth]{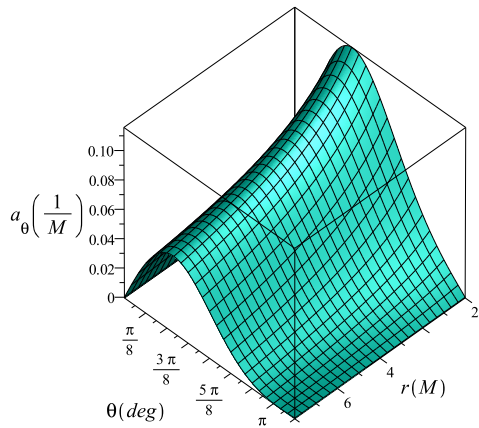}%
}

\caption{Radial and tangential dependence of the components of the acceleration vector. }\label{fig:acceleration}
\end{figure}
As it can be seen from Fig.~\ref{fig:acceleration} the sign of the radial component of the acceleration vector changes depending on the radial and angular position.
\begin{figure}[htp]
  \includegraphics[clip,width=0.9\columnwidth]{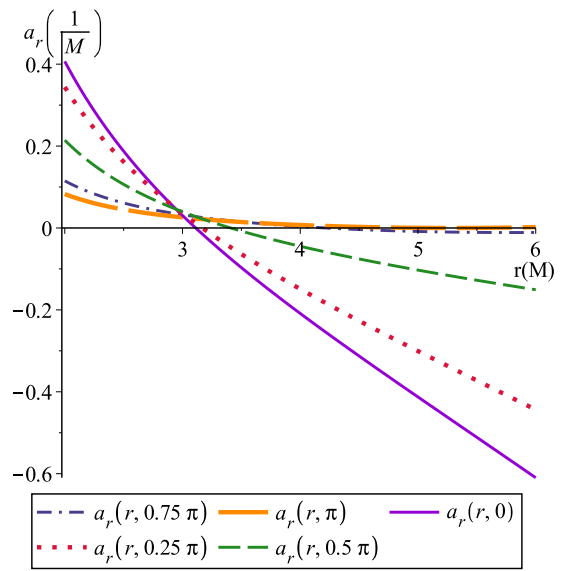}%
\caption{Radial behavior of $a_r$ for observers at different polar angles. We consider the acceleration vector only in the region $2M<r<\sqrt{28}M \approx 5.29M$.}\label{fig:a_r}
\end{figure}
In Fig.~\ref{fig:a_r} we plot the radial dependence of the radial component, $a_r$, for different observer positions. We observe that for all observers, except the one located at $\theta =\pi$, the direction of the radial acceleration flips. This is due to the change in the behavior of the gravitational potential and explains why $E_{\rm Dil}$ takes negative values after a critical point.

The reason that $E_{\rm Dil}$ and $E_{\rm Tot}$ diverge at $r = \sqrt{28}M$, in Fig.~\ref{fig:C_Energies}, results from this point being the second coordinate singularity of our $C$-metric, as we chose $A=1/(\sqrt{28}M)$ and the coordinate singularities occur at $\{ r=2M, r=1/A\}$. This result is expected since after this point, the nature of the spacetime geometry is different.

We also recognize that the system does not possess any energy which can be attributed to rotational degrees of freedom.  This is not immediately obvious since the densities (\ref{C-DeltaK}) and (\ref{C-K^2}) which appear in definition (\ref{Erot}) are nonzero. However, what is physical for the quasilocal observers are the quasilocal charges, not the quasilocal densities. Having zero energy associated with the rotational degrees of freedom is expected since the black hole in question is nonrotating.

Finally we observe that the work that has already been done by the tidal fields, $W_{\rm Tid}$, is positive for all system sizes and takes the same value as in the case of a static black hole. This means that although the individual observers could measure tidal squeezing and tidal stretching depending on their position, the overall effect on the system corresponds to a positive quasilocal charge.
\subsection{Lanczos-van Stockum dust}
For our next application we would like to consider a rotating spacetime. For this, we pick one of the simplest exact solutions of Einstein equations: a rigidly rotating dust cylinder. This solution was first found by Lanczos \citep{Lanczos:1924}, later rediscovered and matched to a vacuum exterior by van Stockum \citep{vanStockum:1937}. Its physicality and mathematical aspects have been investigated intensively in the literature \citep{Bonnor:1977, Bonnor:1980, daSilva:1996xt, deAraujoetal:2000, Bonnor:2005, Zinggetal:2006, Brateketal:2007, Gurlebeck:2009}. Also lately, rotating dust metrics have been used to model galaxies in attempts to understand the general relativistic effects on the galaxy rotation curves \citep{Cooperstock_Tieu:2006,Cooperstock_Tieu:2007,Balasin_Grumiller:2008}.

The original derivation of van Stockum does not end up with an asymptotically flat spacetime. The energy density of the dust, $\tilde{\rho}$, increases exponentially with increasing cylindrical radial coordinate, $x$, and it is given by $\tilde{\rho} =\omega ^2e^{\omega ^2x^2}/(2\pi)$. This is not realistic. Later investigations in the literature, naturally focus on creating more realistic models which are asymptotically flat. In such cases, components of the line element are given by series solutions \citep{deAraujoetal:2000, Cooperstock_Tieu:2006, Cooperstock_Tieu:2007, Brateketal:2007}. 

For our application in the current section, we want to focus on finding the quasilocal energy of the spacetime that is associated with the rotational degrees of freedom. We need to find an orthonormal dyad that satisfies the integrability conditions and this already is not an easy task for axially symmetric stationary spacetimes.\footnote{We discuss this in more detail in the next section.} Therefore we will consider the simplest interior solution given by van Stockum which has a line element
\begin{eqnarray}\label{vanStockcylmetric}
ds^2=-dt^2+a\,\left(dx^2+dz^2\right) +b\,d\psi ^2+c\,dt\,d\psi,
\end{eqnarray}
where
\begin{eqnarray}
a(x)&:=&e ^{-\omega ^2x^2}\nonumber,\\
b(x)&:=&\left(x^2-\omega ^2x^4\right) \nonumber,\\
c(x)&:=&2\omega x^2 \nonumber,
\end{eqnarray}
and $\omega$ is a constant that is associated with the angular velocity of the dust at $x=0$ with respect to ``distant stars''. Other than the singularity at $x=0$, the spacetime becomes singular at $x=1/\omega$ for the metric in (\ref{vanStockcylmetric}). Note that the $g_{\psi \psi}$ component of the metric changes sign when $x>1/\omega $. This introduces closed timelike curves into the spacetime that are not physical. Therefore we will consider systems within the $0<x<1/\omega $ range. 

It is possible to transform the metric into toroidal coordinates at this point and search for a double dyad which satisfies our gauge conditions (\ref{Gaugeconditions}). \footnote{The reason for choosing toroidal coordinates is that it simplifies the process of defining a smooth, closed, spacelike 2-surface in order to integrate the quasilocal densities. By using this toroidal surface, one can bypass the coordinate singularity at $x=0$. Note that without the existence of such a closed surface, quasilocal energies are not defined. This is closely related to the Stokes' Theorem which comes up in the derivation of the non-vanishing boundary Hamiltonian from an action principle of general relativity in a covariant formulation.} Eventually we would like to calculate our quasilocal charges. However, if we apply such a transformation, we lose the information about the actual symmetries of the system. Therefore, let us first consider a null tetrad in cylindrical coordinates which satisfies our gauge conditions, (\ref{Gaugeconditions}),
\begin{eqnarray}\label{vanStocktetradcyl}
l^\mu &=& \frac{1}{\sqrt{2}}\left[\partial _t+a(x)^{-1/2}\partial _x\right], \\
n^\mu &=& \frac{1}{\sqrt{2}}\left[\partial _t-a(x)^{-1/2}\partial _x\right],  \nonumber\\
m^\mu &=& \frac{i}{\sqrt{2}}\left[\omega x \partial _t+\frac{1}{x}\partial _\psi -i a(x)^{-1/2} \partial _z\right]. \nonumber
\end{eqnarray}
For such a tetrad $\{\pi=0, \tau=0 \}$ so that the condition $\pi+\c{\tau}=0$ is trivially satisfied. Also $\{\mu=\c{\mu}, \rho =\c{\rho} \}$ holds. Now let us perform two transformations on the spacelike coordinates. The first coordinate transformation, $Tr_1 :=\{ X= x\cos \psi , Y= x\sin \psi , Z=z\}$, relates the cylindrical coordinates to Cartesian coordinates, $\{ X, Y, Z \}$. The second one, $Tr_2 :=\{ X= \left( R_0+r\cos \theta \right)\cos \phi , Y= \left( R_0+r\cos \theta \right)\sin \phi , Z=r\sin \theta \}$, relates the toroidal coordinates, $\{r,\theta , \phi \}$, to the Cartesian coordinates. After applying $Tr_1$ and $Tr_2^{-1}$ successively on the metric and on the null tetrad we find
\begin{eqnarray}
ds ^2 = -dt^2+\zeta \left(dr^2+r^2d\theta ^2\right)+\chi d\phi ^2+ \xi dt d\phi ,
\end{eqnarray}
where
\begin{eqnarray}
R(r,\theta) &:=& R_0+r\cos \theta \nonumber ,\\
\zeta (r,\theta) &:=& e ^{-\omega ^2R^2} \nonumber ,\\
\chi (r,\theta)&:=& R ^2 \left(1-\omega ^2 R^2 \right) \nonumber ,\\
\xi (r,\theta)&:=& 2\omega R^2 \nonumber ,\\
\end{eqnarray}
and
\begin{eqnarray}\label{vanStocktetradtor}
l^\mu &=& \frac{1}{\sqrt{2}}\left[\partial _t+ \zeta ^{-1/2} \left(\cos \theta \partial _r-\frac{\sin \theta}{r}\partial _\theta \right)\right], \\
n^\mu &=& \frac{1}{\sqrt{2}}\left[\partial _t- \zeta ^{-1/2} \left(\cos \theta \partial _r-\frac{\sin \theta}{r}\partial _\theta \right) \right], \nonumber \\
m^\mu &=& \frac{i}{\sqrt{2}}\left[\omega R \partial _t-\frac{1}{R} \partial _\phi 
\right. \nonumber \\
&&\qquad \qquad \left. 
- i\zeta ^{-1/2}\left(\sin \theta  \partial _r + \frac{\cos \theta}{r} \partial _\theta \right)\right]. \nonumber
\end{eqnarray}
For this null tetrad, after calculating the spin coefficients and by following (\ref{DeltaJNP})--(\ref{RiemNP}), we find the following variables that appear in the contracted Raychaudhuri equation:
\begin{eqnarray}
\tilde{\nabla} _{\mathbb{T}}\mathcal{J}
&=& \frac{-\zeta ^{-1} \left(R^4\omega ^4+1\right)}{R^2} \label{DeltaJvanStock},\\
\tilde{\nabla} _{\mathbb{S}}\mathcal{K}
&=&0  \label{DeltaKvanStock}, \\
\mathcal{J}^2
&=& \frac{\zeta ^{-1} \left(R^4\omega ^4 +1\right)}{R^2} \label{JvanStock}, \\
\mathcal{K}^2
&=& -2\omega ^2 \zeta ^{-1}  \label{K^2vanStock},\\
\mathcal{R_{\,W}}
&=&-2\omega ^2 \zeta ^{-1} \label{R_WvanStock}.
\end{eqnarray}
In order to determine the reference energy density we isometrically embed $\mathbb{S}$ in $\mathcal{M}^4$ by setting
\begin{eqnarray}
\zeta r^2d\theta ^2&=&\bar{r}^2d\bar{\theta}^2,\\
\left(1 -\omega ^2R^2\right)R^2 d\phi ^2&=& \left(\bar{R}_0+\bar{r}\cos \bar{\theta}\right)^2 d\bar{\phi}^2,
\end{eqnarray}
so that the reference quasilocal observers are located at a flat 2-torus in Minkowski spacetime. In order to set the same surface area element both in the physical and in the reference spacetime, we choose
\begin{eqnarray}
\bar{r}&=&r \zeta ^{1/2},\\
d\bar{\theta}&=& d\theta ,\\
d\bar{\phi}&=&\frac{R\left(1-\omega ^2 R^2\right)^{1/2}}{r \zeta ^{1/2} \cos \theta + \bar{R}_0}d\phi ,
\end{eqnarray}
with $\bar{R}_0=R_0$. Then, when written in physical spacetime coordinates, the mean extrinsic curvature of the flat 2-torus, 
\begin{eqnarray}
k_0&=&\frac{\bar{R}_0+2\bar{r}\cos \bar{\theta}}{\bar{r}\left(\bar{R}_0+\bar{r}\cos \bar{\theta}\right)},
\end{eqnarray}
can be used as the reference energy density.

\begin{figure}[htp]
  \includegraphics[clip,width=0.9\columnwidth]{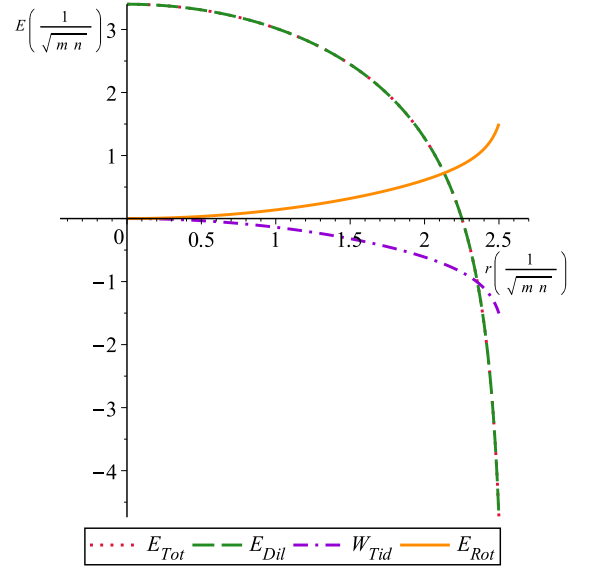}%
\caption{Quasilocal charges of the van Stockum dust. Charges are in length units which can be written as a function of individual mass of the dust particles, $m$, and the total number density, $n$.}
\label{fig:vanStockumEnergies}
\end{figure}

Now that the physical and the reference energy densities are determined, we can calculate the quasilocal charges via Eqs.~(\ref{ETotal})--(\ref{WTidal}). Recall that the spacetime is physically meaningful in the $0<R_0+r\cos \theta <1/\omega$ range. We choose  $R_0=5$ and $\omega=1/10$ for our numerical example, which introduces a coordinate singularity at $x=10$. In terms of the system size, we consider only the $0<r<2.5$ range for computational ease. The results are presented in Fig.~\ref{fig:vanStockumEnergies}, from which we immediately recognize that $E_{\rm Tot}=E_{\rm Dil}$, which is positive for a small sized system, diverges to $-\infty$ as the size of the system gets larger.

Let us try to understand what this result means. Previously, for asymptotically flat versions of the rotating dust, it has been argued by Bonnor that there has to be an infinitely large negative mass associated with the singularity, $x=0$, in order to cancel the effect of positive energy associated with the dust \citep{Bonnor:1977}. Later in \citep{Bonnor:2005} he argued that one can add an infinitely large negative mass layer into the spacetime to observe the same effect. Furthermore, Bratek \emph{et al.} \citep{Brateketal:2007} discussed the same issue and concluded that singularities of the asymptotically flat rotating dust are associated with the ``additional weird stresses'' of the negative active mass.

Here our spacetime is not asymptotically flat. However, we observe a similar behavior. Note that in our solution the energy density of the dust increases with increasing $x$. In such a case one would expect the system to get ever closer to a collapsed state as its size increases. Zingg \emph{et al.} \citep{Zinggetal:2006} and Gurlebeck \citep{Gurlebeck:2009} have argued that such a collapse is in fact expected for a Newtonian dust cylinder. We end up with a similar interpretation which agrees with their arguments. In our work, the fact that $E_{\rm Tot}=E_{\rm Dil}$ diverges to $-\infty$ as the size of the system gets larger, must be attributed to the work done by external fields that are required to exist outside our system to prevent the system from collapsing. 

Now let us look at the quasilocal charges associated with the rotational degrees of freedom and the tidal fields. 

From Fig.~\ref{fig:vanStockumEnergies} we observe that the $W_{\rm Tid}$ is everywhere negative, corresponding to tidal stretching of the surface on which the quasilocal observers are located. As the size of the system increases, so does the energy density of dust according to $\tilde{\rho} =\omega ^2e^{\omega ^2x^2}/(2\pi)$. This requires greater negative work done by the tidal field. The magnitude of $W_{\rm Tid}$ is exactly equal to the energy associated with the rotational degrees of freedom as shown in Fig.~\ref{fig:vanStockumEnergies}. We note that the observers who determine the quasilocal quantities are timelike geodesic observers, i.e., with acceleration $a^\mu=D_{\t {E}{_{\hat{0}}}}\t {E}{^\mu _{\hat{0}}}=0$ and furthermore they are comoving with the dust. In other words, the orbital angular velocity of the observers is zero with respect to the given coordinate system. In such a case one might expect to get zero energy associated with the rotational degrees of freedom of the system. However, for this set of observers, the vorticity of the timelike geodesics is nonzero. Indeed, the vorticity vector and vorticity scalar are given by
\begin{eqnarray}
\mathcal{w}^\mu &=&\frac{1}{2}\t {\eta} {^{\mu} _{\nu \alpha \beta}}g^{\nu \gamma}g^{\alpha \rho}\t {E}{^\beta _{\hat{0}}} D_\rho \t {E} {_{\gamma \hat{0}}}\nonumber \\
&=& \frac{2\omega \sin \theta \zeta ^{-2}}{rR}\partial _r+\frac{2\omega \zeta ^{-2}\cos \theta}{r^2R}\partial _\theta\\
\mathcal{w}&=&\sqrt{\mathcal{w}^\mu \mathcal{w}_\mu}=\frac{2\omega \zeta ^{-3/2}}{rR},
\end{eqnarray}
where $\t {\eta} {^{\mu} _{\nu \alpha \beta}}$ is the Levi-Civita tensor, $g_{\mu \nu}$ is the spacetime metric and we set the observer 4-velocity $u^\mu = \t {E}{^\mu _{\hat{0}}}=\partial _t$. This shows that every dust particle swirls around its own axis. Recall that vorticity is a measure of global rotation of a spacetime. Also previously it was shown by Chrobok \emph{et al.} \citep{Chroboketal:2001} that the rotation of the local matter elements, i.e. spin, can be directly linked to the global rotation of the spacetime, i.e. vorticity. Therefore even though the system we investigate here is defined by the set of observers with zero orbital angular velocity we can still calculate the energy associated with the rotational degrees of freedom of the system.

As the size of the system reaches $1/\omega$, the density of the dust reaches its maximum possible value. Accordingly, one might expect $E_{\rm Rot}$ and $W_{\rm Tid}$ to diverge to $+\infty$ and $-\infty$ respectively as the system size gets closer to the singularity point $1/\omega$. Note that, in Fig.~\ref{fig:vanStockumEnergies}, we observe that $E_{\rm Rot}$ and $W_{\rm Tid}$ tend to $+\infty$ and $-\infty$ respectively, as the size of the system gets larger.

\section{The challenge of stationary, axially symmetric spacetimes}\label{Delicate}
After considering those somewhat unrealistic scenarios one might wonder whether we can apply our formalism to more realistic cases. For example, can we calculate the quasilocal charges of a rotating black hole? The short answer is yes, we can. However it poses an immense technical challenge. 

Recall that we need to satisfy three null tetrad conditions, namely, $\{\rho =\c{\rho}, \mu=\c{\mu}, \pi + \c{\tau}=0 \}$. It is known that in general, the divergence of a null congruence around the vector $\b{l}$ can be written as the linear combination of the expansion and the twist of the congruence, i.e., $\rho =\Theta + i \omega$. This means that we need to have nontwisting null congruences for our formalism to hold. 

Let us consider the case of the Kerr spacetime \citep{Kerr:1963}. The circular orbits are the mostly studied world lines of Kerr because the trajectories follow the Killing vector fields and this simplifies the investigations considerably. Note that in this case, the Killing vectors $\partial _t$ and $\partial _\phi$ have nonzero twist. Moreover, the Kerr metric can be obtained by taking the $r$ coordinate of Schwarzschild to $r+ia\cos \theta$ \citep{Newman_Janis:1965}, where $a$ is the dimensionless angular momentum parameter. This automatically means that for a \emph{principal null tetrad} of a static black hole, by transforming the real divergence, $\rho =-1/r$ into a complex divergence $\rho =-1/(r+ia\cos \theta)$, we obtain a rotating black hole.\footnote{See \citep{Adamo_Newman:2014} for a recent review.} Our problem here is that investigations of a rotating black hole are done mostly using the principal null directions of the spacetime. We should also mention that there are other transverse tetrads such as the quasi-Kinnersely tetrad, which is a powerful tool for exploring Kerr \citep{Zhangetal:2012}. However, once we focus on such null \emph{geodesics}, that aid in the construction of a principal or transverse tetrad, then we have no hope of finding null congruences with a real divergence.

On the other hand, twist-free -- i.e., surface forming -- null congruences exist in \emph{all} Lorentzian spacetimes \citep{Adamoetal:2012}. It is just that we do not require them to be geodesic. Brink \emph{et al.} \citep{Brinketal:2013} have given a detailed investigation of axisymmetric spacetimes, focusing on the twist-free Killing vectors of the stationary axially symmetric spacetimes. We note that there are very few studies in the literature that investigate such a property. Bilge has found an exact twist-free solution whose principal null directions are not geodesic \citep{Bilge:1989}. It was also shown by Bilge and G{\"{u}}rses that those spacetimes are not asymptotically flat and include generalized Kerr-Schild metrics \citep{Bilge_Gurses:1986}. Gergely and Perj{\'{e}}s later concluded that those solutions are actually homogeneous and anisotropic Kasner solutions \citep{Gergely_Perjes:1994} and thus they are not physical. Therefore Brink \emph{et al.} conclude that ``Future studies which aim to extract physical information about isolated dynamical, axisymmetric spacetimes will have to focus on general spacetimes, where none of the principal null directions are geodesics, and which do not fall within Bilge's class of metrics."

In our case we are looking for \emph{a} null congruence, constructed from the timelike dyad that resides on $\mathbb{T}$, which does not even have to be aligned with the principal null directions. It is not necessarily composed of geodesics and it is not required to be composed solely of Killing vectors. All we want from our null tetrad is for it to satisfy the three integrability conditions. To the best of our knowledge, for the case of Kerr, none of the null tetrads introduced in the literature satisfies those conditions.

In order to find such a desired tetrad for the case of Kerr, one might consider the transformations of the quasi-Kinnersely tetrad, for example, by applying two successive Lorentz transformations to the null tetrad. First, apply a type-II Lorentz transformation around $\b{n}$  with parameter $A=a+ib$ and then a type-I Lorentz transformation around $\b{l}$ with parameter $B=c+id$ where $\{a,~b,~c,~d \}$ are all real. Then for the twice transformed spin coefficients we need to satisfy $\{\rho '' =\c{\rho} '', \mu ''=\c{\mu} '', \pi ''+ \c{\tau} ''=0, \c{\pi} ''+ \tau ''=0 \}$ where $''$ denotes the fact that the spin coefficients are transformed twice. After such a procedure we end up with four complex, highly coupled, nonlinear first order differential equations. The unknowns appear in the transformed tetrad condition equations with a polynomial order that goes up to order 5. This system of equations cannot be solved by any iterative method that we are aware of.

Therefore, we observe that our formalism should, in principle, be applicable for more realistic generic spacetimes than the ones we have presented here. However, the less symmetry the system possesses, the more mathematically challenging it becomes to find a null tetrad which satisfies our integrability conditions. Arbitrary nontwisting null congruences  of twisting spacetimes are the key to resolving this issue.

The discussion we presented in Sec.~\ref{Null tetrad gauge} should now be more clear for the reader. In the case of gravitational wave detection, one's ultimate aim is to extract information about the properties of the astrophysical objects that are the sources of radiation. Those properties, such as mass-energy and angular momentum are at best defined quasilocally in general relativity. Therefore the local tetrads of observers should be chosen in such a manner that the quasilocal properties of the system can be well defined throughout the evolution. In \citep{Zhangetal:2012}, Zhang \emph{et al.} showed that the wave fronts of passing gravitational radiation are aligned with the quasi-Kinnersley tetrad. This means that the observers can measure the gravitational radiation locally. However, since quasi-Kinnersley tetrad does not satisfy the integrability conditions of $\mathbb{S}$ and $\mathbb{T}$, the quasilocal charges corresponding to the quasi-Kinnersley tetrad are not well defined. Therefore we conclude that even though one can measure the gravitational radiation locally, there is not always a guarantee that one can extract the properties of its source consistently.
\section{Discussion and Summary}\label{Discussion}
According to many researchers, including the authors of Refs.~\citep{dInverno_Smallwood:1980, Smallwood:1983, Torre:1985, Yoon:2004}, the 2+2 picture of general relativity might be more fundamental than the 3+1 approach. Although one might debate this point, the existence of a nonvanishing boundary Hamiltonian leads to the necessity of modifying the symplectic structure of the Arnowitt--Deser--Misner formalism in phase space to obtain a covariant formalism which can directly be linked to the quasilocal charges \citep{Kijowski:1997, Nester:1991}. Energy definitions, which do not conflict with  the equivalence principle, generically involve the extrinsic or/and intrinsic geometry of a closed spacelike 2-surface. However, defining quasilocal charges that are measures of energy and angular momentum for a generic spacetime is often a challenge.

The energy and energy flux definitions that are made locally, globally or quasilocally, are sometimes compared and contrasted without questioning for which system those definitions are made. Actually, there exist well-defined quasilocal energy definitions that can be directly linked to the action principle of general relativity. What is ill-defined is the specification of the system that is enclosed by a boundary surface on which the quasilocal charges are to be integrated. 

Let us make an analogy with classical thermodynamics and consider two systems with same number of gas molecules: (i) a constant pressure system which is expanding and (ii) a constant volume system which has increasing pressure. If we use a barometer to measure the pressure values obtained within these two systems, the readings will of course be different. However, this is not because the barometer is not working properly, rather it is because the barometer is not sensitive to the defining properties (or symmetries) of the two systems in question. In other words, the measuring agent is indifferent to how the two systems are ``isolated.'' Moreover, even if we find a way to define the system consistently there exist many energies one can associate with a system. Going back to our analogy, let us say we keep track of the pressure value and make sure that we are actually investigating a system with constant pressure. Now we can define the internal energy of that system or define the average kinetic energy of the particles which is not necessarily related to internal energy unless there exists equilibrium. We can also define work done by the system on the surroundings throughout the expansion process etc. In that situation we would not expect all of those energies to give us the same value.

In this paper, we presented a quasilocal work-energy relation which can be applied to generic spacetimes in order to discuss quasilocal energy exchange. We identified the quasilocal charges associated with the rotational and  nonrotational degrees of freedom, in addition to a work term associated with the tidal fields. This construction was possible only after we defined  a quasilocal system by constraining the double dyad of the quasilocal observers, which is highly dependent on the symmetries of the spacetime in question.

Our present investigation emerged from three questions:
\begin{itemize}
\item[(i)] Is there something inherently fundamental about the 2+2 formalism in terms of quasilocal energy definitions?\\
\item[(ii)] quasilocal energy resides on the intrinsic and/or extrinsic curvature of a closed 2-dimensional spacelike surface, what do the other extrinsic properties of that surface correspond to?\\
\item[(iii)] Can the Raychaudhuri and the other geodesic deviation equations, which have proved their usefulness in terms of physically relevant observables in a 3+1 formalism, be investigated in a 2+2 formalism so that they can be linked to physically meaningful quasilocal charges?
\end{itemize}
To answer these questions, we considered Capovilla and Guven's generalized Raychaudhuri equation given in \citep{Capovilla_Guven:1994} for a 2-dimensional world sheet that is embedded in a 4-dimensional spacetime. Previously, for spherically symmetric systems, we investigated the Raychaudhuri equation of the world sheet at quasilocal thermodynamic equilibrium \cite{Uzun_Wiltshire:2015}, i.e., when the observers are located at the apparent horizon. In the present paper we considered more generic spacetimes that are in nonequilibrium with their surroundings. We also relaxed the spherically symmetric condition.

By transforming our equations from Capovilla and Guven's formalism, which is constructed on an orthogonal double dyad, to the Newman-Penrose formalism, which is based on a complex null tetrad, we were able to present the contracted Raychaudhuri equation in terms of the combinations of spin coefficients, their relevant directional derivatives and some of the curvature scalars. We also imposed three null tetrad gauge conditions which result from the integrability conditions of the 2-dimensional timelike surface $\mathbb{T}$ and the 2-dimensional spacelike surface $\mathbb{S}$. This spacelike 2-surface is defined instantaneously and is orthogonal to $\mathbb{T}$ at every point. Our null tetrad gauge conditions are shown to be invariant under type-III Lorentz transformations which basically correspond to boosting of the quasilocal observers in the spacelike direction orthogonal to $\mathbb{S}$. Ultimately we realized that, under such gauge conditions, the contracted Raychaudhuri equation is a linear combination of two of the spin field equations of the Newman-Penrose formalism.

Later, we defined certain quasilocal charges via the geometric variables that appear in the contracted Raychaudhuri equation. Our motivation is that there exists a direct link between the mean extrinsic curvature of $\mathbb{S}$ that encloses the system and the variables of the contracted Raychaudhuri equation. Note that mean extrinsic curvature of such a smooth, closed, spacelike 2-surface, $\mathbb{S}$, is the main object of most of the quasilocal energy definitions which are derived by a Hamiltonian approach. By choosing the quasilocal energy definitions made by Kijowski \citep{Kijowski:1997} as our anchor, we were able to define relevant quasilocal charges for which a physical interpretation would be found. We also showed in Appendix~\ref{App:SpinBoost} that all of those quasilocal charges are invariant under type-III Lorentz transformations. Note that this property is desired for a well-defined quasilocal construction, as boosted observers should agree on the fact that they are measuring the charges of the same system. 

We applied our formalism to a radiating Vaidya spacetime, a $C$-metric and an interior solution of the Lanczos-van Stockum dust cylinder. For the case of Vaidya we concluded that the usable energy of the system decreases purely due to radiation. For a $C$-metric we observed that the greater the acceleration of the black hole is, the more energy should be provided \emph{to} the system \emph{by} an external agent. We concluded that the decreasing trend in the total energy is due to the nonmonotonic, repulsive gravitational potential that can be observed at the exterior region of an extremely accelerated black hole. For the Lanczos-van Stockum dust we considered a nonasymptotically flat case. As the size of the system got larger, we obtained negative mass energy for the usable dilatational energy of the system. We concluded that this must be attributed to external fields doing work on the system in order to prevent it from collapse. We were also able to obtain the quasilocal energy associated with the rotational degrees of freedom whose magnitude is exactly equal to the one of work done by the tidal fields.

This paper can be seen as a first attempt to investigate the Raychaudhuri equation in 2+2 picture in terms of the quasilocal charges. There exist various open problems and delicate issues. To start with, at a given spacetime point one has six tetrad degrees of freedom and we imposed only three null tetrad gauge conditions to our system. That means we have additional freedom to specify a gauge, i.e, to define the quasilocal system. Although there exists no geometrically motivated reason we are aware of in our current approach, one can $choose$ additional conditions in order to compare the quasilocal charges of different spacetimes constructed with other well-known null tetrad gauges. 

Another delicate issue which may or may not be related to our null tetrad gauge freedom is shear. There is no $a\, priori$ reason for us to impose the shear-free condition to the null congruences, constructed from the timelike dyad that resides on $\mathbb{T}$. However, for generic spacetimes, one can find a gauge which satisfies our three gauge conditions more easily once the shear-free condition is imposed. This is primarily because our gauge conditions are trying to locate the set of quasilocal observers in such a configuration that the surface $\mathbb{S}$ is always orthogonal to $\mathbb{T}$. That is natural for radially moving observers of a spherically symmetric system but may hold even if the spacetime is not spherically symmetric. The shear-free condition locates the quasilocal observers as close to as they can get to such a configuration. Note that shear is the fundamental concept of Bondi's mass loss \cite{Bondietal:1962} without which gravitational radiation at null infinity cannot be defined. Thus, this automatically raises an issue for quasilocal observers at infinity who would like to measure the Bondi mass loss associated with gravitational radiation. Investigation of whether or not there exists a gauge which satisfies both the Bondi tetrad and our gauge conditions is left for future work.

Finally, we note that it is technically difficult to satisfy our null tetrad conditions for more realistic, axially symmetric, stationary spacetimes such as Kerr. This difficulty arises from the fact that our approach demands twist-free null congruences constructed by the tangent vectors of $\mathbb{T}$. However, finding twist-free null congruences for spacetimes whose principal null directions are twisting is a challenge. Although those nongeodesic null congruences that we are after are not physical, their existence will guarantee the fact that the quasilocal system, and the associated quasilocal charges, are all consistently defined.

Recently, a quasilocal energy for the Kerr spacetime has been calculated for stationary observers \citep{Liu_Tam:2016} by using the definition of \citep{Sunetal:2003} both for the quasilocal energy and the embedding method for the reference energy. Liu and Tam show that this energy is exactly equal to Brown and York's (BY) quasilocal energy \citep{Brown_York:1992}. One might wonder how our construction is compared to such an investigation. To start with, the null tetrad constructed from the orthonormal double dyad of the stationary observers in Boyer-Lindquist coordinates has imaginary divergence and hence does not satisfy our null tetrad gauge conditions. Recall that the tetrad conditions we introduced here guarantees the existence of well-defined, boost-invariant quasilocal charges. Also note that BY quasilocal energy is not invariant under boosts. Therefore, the fact that Liu and Tam end up with the BY quasilocal energy for their quasilocal system defined by stationary observers in Boyer-Lindquist coordinates is no surprise. Therefore, in our view, the calculations of Liu and Tam do not satisfy all the requirements of a genuine quasilocal construction. In fact, this is exactly the point that we tried to emphasize throughout the paper. Without a well-defined quasilocal system, there is no consistent definition of energy.
\section{Acknowledgements}
Many thanks to David L. Wiltshire for his critical suggestions and his careful reading of the manuscript.
\section*{APPENDIX}
\appendix
\setcounter{equation}{0}  
\section{Newman-Penrose Formalism}\label{Appendix:A}
For a complex null tetrad $\{ \b{l}, \b{n}, \b{m}, \c{\b{m}}\}$ , the Newman-Penrose spin coefficients are defined as \cite{Newman_Penrose:1961}\footnote{Note that we are using $\{-,+,+,+\}$ signature for the spacetime metric throughout the paper. Therefore our spin coefficients and the curvature scalars have an extra negative sign when compared to Newman-Penrose's original notation in \citep{Newman_Penrose:1961}.}
\begin{eqnarray}
\kappa = -\inner{D_{\b{l}}\b{l},\b{m}}, \qquad 
\nu = \inner{D_{\b{n}}\b{n},\c{\b{m}}} \label{kappanu},\\
\rho = -\inner{D_{\c{\b{m}}}\b{l},\b{m}}, \qquad
\mu = \inner{D_{\b{m}}\b{n},\c{\b{m}}}\label{rhomu},\\
\sigma = -\inner{D_{\b{m}}\b{l},\b{m}}, \qquad
\lambda = \inner{D_{\c{\b{m}}}\b{n},\c{\b{m}}}\label{sigmalambda},\\
\tau = -\inner{D_{\b{n}}\b{l},\b{m}},\qquad
\pi = \inner{D_{\b{l}}\b{n},\c{\b{m}}}\label{taupi},\\
\varepsilon =\frac{1}{2}\left[-\inner{D_{\b{l}}\b{l},\b{n}}+\inner{D_{\b{l}}\b{m},\c{\b{m}}}\right]\label{epsilon},\\
\gamma = \frac{1}{2}\left[\inner{D_{\b{n}}\b{n},\b{l}}-\inner{D_{\b{n}}\c{\b{m}},\b{m}}\right]\label{gamma},\\
\beta = \frac{1}{2}\left[-\inner{D_{\b{m}}\b{l},\b{n}}+\inner{D_{\b{m}}\b{m},\c{\b{m}}}\right]\label{beta},\\
\alpha = \frac{1}{2}\left[\inner{D_{\c{\b{m}}}\b{n},\b{l}}-\inner{D_{\c{\b{m}}}\c{\b{m}},\b{m}}\right]\label{alpha}.
\end{eqnarray}
The propagation equations are
\begin{eqnarray}
D_{\b{l}}\b{l}&=&\left(\varepsilon+\c{\varepsilon}\right)\b{l}-\c{\kappa}\b{m}-\kappa \c{\b{m}}\label{Dll},\\
D_{\b{n}}\b{l}&=&\left(\gamma+\c{\gamma}\right)\b{l}-\c{\tau}\b{m}-\tau \c{\b{m}}\label{Dnl},\\
D_{\b{m}}\b{l}&=&\left(\c{\alpha}+\beta\right)\b{l}-\c{\rho}\b{m}-\sigma \c{\b{m}}\label{Dml},\\
D_{\b{l}}\b{n}&=&-\left(\varepsilon+\c{\varepsilon}\right)\b{n}+\pi \b{m}+\c{\pi} \c{\b{m}}\label{Dln},\\
D_{\b{n}}\b{n}&=&-\left(\gamma+\c{\gamma}\right)\b{n}+\nu \b{m}+\c{\nu} \c{\b{m}}\label{Dnn},\\
D_{\b{m}}\b{n}&=&-\left(\c{\alpha}+\beta \right)\b{n}+\mu \b{m}+\c{\lambda} \c{\b{m}}\label{Dmn},\\
D_{\b{l}}\b{m}&=&\c{\pi}\b{l}-\kappa \b{n}+\left(\varepsilon - \c{\varepsilon}\right)\b{m}\label{Dlm},\\
D_{\b{n}}\b{m}&=&\c{\nu}\b{l}-\tau \b{n}+\left(\gamma - \c{\gamma}\right)\b{m}\label{Dnm},\\
D_{\b{m}}\b{m}&=&\c{\lambda}\b{l}-\sigma \b{n}+\left(-\c{\alpha}+\beta\right)\b{m}\label{Dmm},\\
D_{\b{m}}\c{\b{m}}&=&\mu \b{l}-\c{\rho} \b{n}+\left(\c{\alpha}-\beta\right)\c{\b{m}}\label{Dmm_}.
\end{eqnarray}
Commutation relations, $\left[\b{X},\b{Y}\right] = D_{\b{X}}\b{Y}-D_{\b{Y}}\b{X}$, for the null vectors are
\begin{align}
\left[\b{l},\b{n}\right]&=-\left(\gamma + \c{\gamma}\right)\b{l} -\left(\varepsilon + \c{\varepsilon}\right)\b{n} \nonumber \\
&\qquad 
+\left(\pi + \c{\tau}\right)\b{m}+\left(\c{\pi} + \tau \right)\c{\b{m}} 
\label{com_ln},\\
\left[\b{l},\b{m}\right]&=\left(\c{\pi} - \c{\alpha} -\beta \right)\b{l} -\kappa \b{n}+\left(\varepsilon - \c{\varepsilon} +\c{\rho}\right)\b{m}+\sigma \c{\b{m}} \label{com_lm},\\
\left[\b{n},\b{m}\right]&=\c{\nu}\b{l} + \left( \c{\alpha} +\beta -\tau \right)\b{n}+\left(\gamma - \c{\gamma} -\mu \right)\b{m}-\c{\lambda}\c{\b{m}} \label{com_nm},\\
\left[\b{m},\c{\b{m}}\right]&= \left(\mu - \c{\mu}\right)\b{l} +\left(\rho - \c{\rho}\right)\b{n} \nonumber \\
&\qquad 
+\left(\c{\beta}-\alpha \right)\b{m}+\left(\c{\alpha} - \beta \right)\c{\b{m}}
\label{com_mm_}.
\end{align}
Newman and Penrose introduce two sets of curvature scalars, Weyl scalars and Ricci scalars, which carry the same  information as in the Riemann curvature tensor. The Ricci scalars are defined as
\begin{align}
\Phi_{00}&:=\frac{1}{2}R_{\mu \nu}l^\mu l^\nu ,\,\,\, \qquad 
\Phi_{11}:=\frac{1}{4}R_{\mu \nu}(\,l^\mu n^\nu +m^\mu \bar{m}^\nu),\nonumber \\
\Phi_{01}&:=\frac{1}{2}R_{\mu \nu}l^a m^\nu \,, \qquad 
\Phi_{10}:=\frac{1}{2}R_{\mu \nu}l^\mu \bar{m}^\nu =\overline{\Phi_{01}}\,, \nonumber \\
\Phi_{02}&:=\frac{1}{2}R_{\mu \nu}m^\mu m^\nu  \qquad
\Phi_{20}:=\frac{1}{2}R_{\mu \nu}\bar{m}^\mu \bar{m}^\nu =\overline{\Phi_{02}}\, \nonumber \\
\Phi_{12}&:=\frac{1}{2}R_{\mu \nu} m^\mu n^\nu , \qquad
\Phi_{21}:=\frac{1}{2}R_{\mu \nu} \bar{m}^\mu n^\nu =\overline{\Phi_{12}}\,, \nonumber \\
\Phi_{22}&:=\frac{1}{2}R_{\mu \nu}n^\mu n^\nu \,, \qquad
\Lambda:=\frac{R}{24}, \nonumber\\
\end{align}
in which $R_{\mu \nu}$ is the Ricci tensor of the spacetime, $\Phi_{00},\, \Phi_{11},\, \Phi_{22},\, \Lambda$ are real scalars and $\Phi_{10},\, \Phi_{20},\, \Phi_{21}$ are complex scalars.
The Weyl scalars are defined as
\begin{eqnarray}
\psi _0&=&\t {C}{_\mu _\nu _\alpha _\beta}l^\mu m^\nu l^\alpha m^\beta ,\label{Psi0}\\
\psi _1&=&\t {C}{_\mu _\nu _\alpha _\beta}l^\mu n^\nu l^\alpha m^\beta ,\label{Psi1}\\
\psi _2&=&\t {C}{_\mu _\nu _\alpha _\beta}l^\mu m^\nu \c{m}^\alpha n^\beta ,\label{Psi2}\\
\psi _3&=&\t {C}{_\mu _\nu _\alpha _\beta}l^\mu n^\nu \c{m}^\alpha n^\beta ,\label{Psi3}\\
\psi _4&=&\t {C}{_\mu _\nu _\alpha _\beta}n^\mu \c{m}^\nu n^\alpha \c{m}^\beta ,\label{Psi4}
\end{eqnarray}
with $\t {C}{_\mu _\nu _\alpha _\beta}$ being the Weyl tensor.
\subsection{Type-III Lorentz transformations}
A type-III Lorentz transformation represents a boosting in the direction of $\b{l}$ and $\b{n}$ and a rotation in the $\b{m}$ and $\c{\b{m}}$ directions, i.e., the tetrad vectors transform as
\begin{eqnarray}
\b{l} &\rightarrow & a^2\b{l}\label{lIII}, \\
\b{n} &\rightarrow & \frac{1}{a^2}\b{n}\label{nIII},\\
\b{m} &\rightarrow & e^{2i\theta} \b{m}\label{mIII},\\
\c{\b{m}} &\rightarrow & e^{-2i\theta}\c{\b{m}}\label{m_III}.
\end{eqnarray}
Here both $a$ and $\theta$ are real functions.
Accordingly the spin coefficients transform as
\begin{eqnarray}
\nu &\rightarrow & a^{-4}e^{-2i\theta} \nu  \label{nuIII},\\
\tau &\rightarrow & e^{2i\theta} \tau \label{tauIII},\\
\gamma &\rightarrow & a^{-2}\left(\gamma +D_{\b{n}}\left[\ln a+i\theta \right] \right) \label{gammaIII},\\
\mu &\rightarrow & a^{-2} \mu \label{muIII},\\
\sigma &\rightarrow & a^2 e^{4i\theta}\sigma \label{sigmaIII},\\
\beta &\rightarrow & e^{2i\theta}\left(\beta +D_{\b{m}}\,\left[\ln a+i\theta \right]\right) \label{betaIII},\\
\lambda &\rightarrow & a^{-2} \,e^{-4i\theta}\lambda\label{lambdaIII},\\
\rho &\rightarrow & a^2\rho \label{rhoIII},\\
\alpha &\rightarrow & e^{-2i\theta}\left(\alpha +D_{\c{\b{m}}}\,\left[\ln a+i\theta \right]\right) \label{alphaIII},\\
\kappa &\rightarrow & a^4 e^{2i\theta}\kappa \label{kappaIII},\\
\varepsilon &\rightarrow & a^2\left(\varepsilon +D_{\b{l}}\,\left[\ln a+i\theta \right]\right)\label{varepsilonIII},\\
\pi &\rightarrow & e^{-2i\theta}\pi \label{piIII}.
\end{eqnarray}
The transformations of Ricci scalars are given by
\begin{eqnarray}
\Phi _{00} &\rightarrow & a^4\Phi _{00},\\
\Phi _{01} &\rightarrow & a^2 e^{2i\theta}\Phi _{01},\\
\Phi _{10} &\rightarrow & a^2 e^{-2i\theta}\Phi _{10},\\
\Phi _{02} &\rightarrow & e^{4i\theta}\Phi _{02},\\
\Phi _{20} &\rightarrow & e^{-4i\theta}\Phi _{20},\\
\Phi _{11} &\rightarrow & \Phi _{11},\\
\Phi _{12} &\rightarrow & a^{-2}e^{2i\theta}\Phi _{12} ,\\
\Phi _{21} &\rightarrow & a^{-2}e^{-2i\theta}\Phi _{21},\\
\Phi _{22} &\rightarrow & a^{-4}\Phi _{22},
\end{eqnarray}
and the transformations of Weyl scalars are given by
\begin{eqnarray}
\Psi _{0} &\rightarrow & a^4 e^{4i\theta}\Psi _{0}, \\
\Psi _{1} &\rightarrow & a^2 e^{2i\theta}\Psi _{1},\\
\Psi _{2} &\rightarrow & \Psi _{2} \label{Psi2III},\\
\Psi _{3} &\rightarrow & a^{-2}e^{-2i\theta}\Psi _{3},\\
\Psi _{4} &\rightarrow & a^{-4}e^{-4i\theta} \Psi _{4}.
\end{eqnarray}
\section{Raychaudhuri equation in Newman-Penrose formalism}\label{Appendix:B}
\subsection{Useful expressions}
The following expressions are used many times in our transformation to the NP formalism:
\begin{align}
\t {\eta}{^a^b}\t {E}{^\rho _b}\t {E}{^\gamma _a}&=-\t {E}{^\rho _{\hat{0}}}\t {E}{^\gamma _{\hat{0}}}+\t {E}{^\rho _{\hat{1}}}\t {E}{^\gamma _{\hat{1}}}\nonumber \\
&=-\left(\frac{1}{\sqrt{2}}\right)^2\left(l^\rho +n^\rho \right)\left(l^\gamma +n^\gamma \right)
\nonumber \\
&\qquad 
+\left(\frac{1}{\sqrt{2}}\right)^2\left(l^\rho -n^\rho \right)\left(l^\gamma -n^\gamma \right)\nonumber \\
&= -\left(l^\rho n^\gamma + l^\gamma n^\rho \right) \label{exp:eta_E_E}.\\
\t {\delta}{^i^j}\t {N}{^\nu _i}\t {N}{^\beta _j}&=\t {N}{^\nu _{\hat{2}}}\t {N}{^\beta _{\hat{2}}}+\t {N}{^\nu _{\hat{3}}}\t {N}{^\beta _{\hat{3}}}\nonumber \\
&= \left(\frac{1}{\sqrt{2}}\right)^2\left(m^\nu +\c{m}^\nu \right)\left(m^\beta +\c{m}^\beta \right)
\nonumber \\
&\qquad
+\left(\frac{-i}{\sqrt{2}}\right)^2\left(m^\nu -\c{m}^\nu \right)
\nonumber \\
&\qquad \qquad \qquad \times
\left(m^\beta -\c{m}^\beta \right)\nonumber \\
&= \left(m^\nu \c{m}^\beta + m^\beta \c{m}^\nu \right) \label{exp:delt_n_n}.\\
\t {\eta}{^a^b}\t {E}{^\beta _a}D_\alpha \t {E}{^\mu _b}&=-\t {E}{^\beta _{\hat{0}}}D_\mu \t {E}{^\beta _{\hat{0}}}+\t {E}{^\mu _{\hat{1}}}D_\alpha \t {E}{^\rho _{\hat{1}}}\nonumber \\
&=-\frac{1}{2}\left(l^\beta +n^\beta \right)D_\alpha \left(l^\mu +n^\mu\right)
\nonumber \\
&\qquad
+\frac{1}{2}\left(l^\beta -n^\beta \right)D_\alpha \left(l^\mu -n^\mu\right)\nonumber \\
&=-\left(l^\beta D_\alpha n^\mu +n^\beta D_{\alpha} l^\mu\right)\label{exp:eta_E_D_E}.\\
\t {\eta}{^a^b}\t {E}{^\alpha _a}D_\alpha \t {E}{^\mu _b}&=-\left( D_\b{l} n^\mu +D_\b{n} l^\mu\right) \label{exp:eta_E_D_E_cont}.\\
\t {\delta}{^i^j}\t {N}{^{\alpha} _i}D_{\beta}\t {N}{^{\nu} _j}
&=\t {N}{^{\alpha} _{\hat{2}}}D_{\beta} \t {N}{^{\nu} _{\hat{2}}}+\t {N}{^{\alpha} _{\hat{3}}}D_{\beta} \t {N}{^{\nu} _{\hat{3}}}\nonumber \\
&=\frac{1}{2}\left(m^\alpha +\c{m}^\alpha \right)D_\beta \left(m^\nu +\c{m}^\nu \right)
\nonumber \\
&\qquad
-\frac{1}{2}\left(m^\alpha -\c{m}^\alpha \right)
\nonumber \\
&\qquad \qquad \times
D_\beta \left(m^\nu -\c{m}^\nu \right)\nonumber \\
&= m^\alpha D_\beta \c{m}^\nu + \c{m}^\alpha D_\beta m^\nu \label{exp:delt_n_D_n}.\\
\t {\delta}{^i^j}\t {N}{^{\alpha} _i}D_{\alpha}\t {N}{^{\nu} _j}&= D_{\b{m}}\c{m}^\nu + D_{\c{\b{m}}}m^\nu \label{exp:delt_n_D_n_cont}.\\
\t {\eta}{^c^d}\left(D_\rho \t {E}{^\mu _c}\right)\left(D_\gamma \t {E}{^\alpha _d}\right)
&=-\left(D_\rho \t {E}{^\mu _{\hat{0}}}\right)\left(D_\gamma \t {E}{^\alpha _{\hat{0}}}\right)
\nonumber \\
&\qquad
+\left(D_\rho \t {E}{^\mu _{\hat{1}}}\right)\left(D_\gamma \t {E}{^\alpha _{\hat{1}}}\right)\nonumber \\
&=-\frac{1}{2}\left(D_\rho l^\mu + D_\rho n^\mu \right)
\nonumber \\
&\qquad \qquad \times
\left(D_\gamma l^\alpha + D_\gamma n^\alpha \right)
\nonumber \\
&\qquad
+\frac{1}{2}\left(D_\rho l^\mu - D_\rho n^\mu \right)
\nonumber \\
&\qquad \qquad \times
\left(D_\gamma l^\alpha - D_\gamma n^\alpha \right)\nonumber \\
&= -\left[\left(D_\rho l^\mu \right)\left(D_\gamma n^\alpha\right) 
\right.  \nonumber \\
&\qquad \qquad \left.
+\left(D_\rho n^\mu \right)\left(D_\gamma l^\alpha \right)\right]\label{exp:eta_D_E_D_E}.
\end{align}
\begin{align}
\t {\eta}{^a^b}\t {E}{^\beta _b}D_\beta D_\gamma \t {E}{^\mu _a}
&= -\t {E}{^\beta _{\hat{0}}}D_\beta D_\gamma \t {E}{^\mu _{\hat{0}}}+\t {E}{^\beta _{\hat{1}}}D_\beta D_\gamma \t {E}{^\mu _{\hat{1}}}\nonumber \\
&=-\frac{1}{2}\left(l^\beta +n^\beta \right)D_\beta D_\gamma \left(l^\mu + n^\mu \right)
\nonumber \\
&\qquad
+\frac{1}{2}\left(l^\beta - n^\beta \right)D_\beta D_\gamma \left(l^\mu - n^\mu \right)\nonumber \\
&=-\frac{1}{2}\left[D_{\b{l}}D_\gamma \left(l^\mu + n^\nu \right)
\right.  \nonumber \\
&\qquad  \qquad \left.
+D_{\b{n}}D_\gamma \left(l^\mu + n^\nu \right)\right]
\nonumber \\
&\qquad
+ \frac{1}{2}\left[D_{\b{l}}D_\gamma \left(l^\mu - n^\nu \right) 
\right. \nonumber  \\
&\qquad \qquad  \left.
- D_{\b{n}}D_\gamma \left(l^\mu - n^\nu \right)\right]\nonumber \\
&= -\left(D_{\b{l}}D_\gamma n^\mu + D_{\b{n}}D_\gamma l^\mu \right)\label{exp:eta_E_D_D_E}.
\end{align}
\subsection{Derivation of $\tilde{\nabla} _{\mathbb{T}}\mathcal{J}$}
Consider the left-hand side of the Raychaudhuri equation (\ref{Raych_simple}), and the world sheet covariant derivative of $\t {J}{_a_i_j}$ defined in relation (\ref{eq:CurlyCovJ}), i.e.,
\begin{align}\label{eq:tilD_J_a_i_j}
\tilde{\nabla} _{\mathbb{T}}\mathcal{J}
&:=\t {\eta}{^a^b}\t {\delta}{^i^j}\t {\tilde{\nabla}}{_b}\t {J}{_a_i_j} \nonumber \\
&=\t {\eta}{^a^b}\t {\delta}{^i^j}\left(\underbrace{\t {\nabla}{_b}\t {J}{_a_i_j}}_\text{$\t {D}{_b}\t {J}{_a_i_j}-\t {\gamma}{_b_a^c}\t {J}{_c_i_j}$}-\t {w}{_b_i^k}\t {J}{_a_k_j}-\t {w}{_b_j^k}\t {J}{_a_i_k}\right).
\end{align}
By using the definition of $\t {J}{_a_i_j}$, Eq.~(\ref{eq:J_a^i^j}), the first term of the equation (\ref{eq:tilD_J_a_i_j}) becomes
\begin{align}
\t {\eta}{^a^b}\t {\delta}{^i^j}D{_b}\t {J}{_a_i_j}
&= \t {\eta}{^a^b}\t {\delta}{^i^j}D_b \left[\t {g}{_\mu _\nu}D_i\left(\t {E}{^\mu _a}\right)\t {N}{^\nu _j}\right]\nonumber \\
&= \t {g}{_\mu _\nu} \t {\eta}{^a^b}\t {\delta}{^i^j}\left(D_b \t {N} {^\gamma _i}\right)\left(D_\gamma \t {E}{^\mu _a}\right)\t {N}{^\nu _j}
\nonumber \\
&\qquad
+\t {g}{_\mu _\nu} \t {\eta}{^a^b}\t {\delta}{^i^j} \t {N}{^\gamma _i}\left(D_b D_\gamma \t {E}{^\mu _a}\right)\t {N}{^\nu _j} 
\nonumber \\
&\qquad
+\t {g}{_\mu _\nu} \t {\eta}{^a ^b}\t {\delta}{^i ^j}\t {N}{^\gamma _i}\left(D_\gamma \t {E}{^\mu _a} \right) \t {E}{^\beta _b} \left(D_\beta \t {N}{^\nu _j}\right)
\nonumber \\
&= \t {g}{_\mu _\nu} \left(\t {\delta}{^i ^j}\t {N}{^\nu _j} D_\beta \t {N}{^\gamma _i}\right)\left(\t {\eta}{^a ^b} \t {E}{^\beta _b} D_\gamma \t {E}{^\mu _a}\right)
\nonumber \\
&\qquad
+\t {g}{_\mu _\nu}\left(\t {\delta}{^i ^j} \t {N}{^\gamma _i} \t {N}{^\nu _j}\right)\left(\t {\eta}{^a ^b}\t {E}{^\beta _b}D_\beta D_\gamma \t {E}{^\mu _a} \right)
\nonumber \\
&\qquad
+ \t {g}{_\mu _\nu} \left(\t {\delta}{^i ^j}\t {N}{^\gamma _i} D_\beta \t {N}{^\nu _j}\right)\left(\t {\eta}{^a ^b}\t {E}{^\beta _b}D_\gamma \t {E}{^\mu _a} \right),\nonumber 
\end{align}
and by making use of Eqs.~(\ref{exp:delt_n_n}), (\ref{exp:eta_E_D_E}), (\ref{exp:delt_n_D_n}) and (\ref{exp:eta_E_D_D_E}) we obtain
\begin{align}
\t {\eta}{^a^b}\t {\delta}{^i^j}D{_b}\t {J}{_a_i_j}
&= -\t {g}{_\mu _\nu}\left(m^\nu D_\beta \c{m}^\gamma + \c{m}^\nu D_\beta m^\gamma \right)
\nonumber \\
&\qquad \qquad \times
\left(l^\beta D_\gamma n^\mu +n^\beta D_{\gamma} l^\mu\right)
\nonumber \\
& \qquad
- \left(m^\gamma \c{m}^\nu + m^\nu \c{m}^\gamma \right)
\nonumber \\
&\qquad \qquad \times
\left(D_{\b{l}}D_\gamma n^\mu + D_{\b{n}}D_\gamma l^\mu \right)
\nonumber \\
& \qquad
- \t {g}{_\mu _\nu}\left(m^\gamma D_\beta \c{m}^\nu + \c{m}^\gamma D_\beta m^\nu \right)
\nonumber \\
&\qquad \qquad \times
\left(l^\beta D_\gamma n^\mu +n^\beta D_{\gamma} l^\mu\right)\nonumber.
\end{align}
Then
\begin{align}
\t {\eta}{^a^b}\t {\delta}{^i^j}D{_b}\t {J}{_a_i_j}
&=-\t {g}{_\mu _\nu}\left[\c{m}^\nu \left(D_\beta m^\gamma \right)l^\beta\left(D_\gamma n^\mu \right)
\right. \nonumber  \\
&\qquad \qquad  \left.
+\c{m}^\nu m^\gamma D_{\b{l}}D_\gamma n^\mu \right] 
\nonumber \\
& \qquad
- \t {g}{_\mu _\nu}\left[m^\nu \left(D_\beta \c{m}^\gamma \right) l^\beta \left(D_\gamma n^\mu \right) 
\right. \nonumber  \\
&\qquad \qquad  \left.
+ m^\nu \c{m}^\gamma D_{\b{l}}D_\gamma n^\mu\right] \nonumber \\
& \qquad
-\t {g}{_\mu _\nu}\left[ \c{m}^\nu \left(D_\beta m^\gamma \right)n^\beta \left(D_\gamma l^\mu \right)
\right. \nonumber  \\
&\qquad \qquad  \left.
+ \c{m}^\nu m^\gamma D_{\b{n}}D_\gamma l^\mu \right]
\nonumber \\
& \qquad
- \t {g}{_\mu _\nu}\left[m^\nu \left( D_\beta \c{m}^\gamma \right) n^\beta \left(D_\gamma l^\mu \right)
\right. \nonumber  \\
&\qquad \qquad  \left.
+ m^\nu \c{m}^\gamma D_{\b{n}}D_\gamma l^\mu \right]
\nonumber \\
& \qquad
-\t {g}{_\mu _\nu} \left[ \left( D_{\b{l}}\c{m}^\nu \right)\left( D_{\b{m}}n^\mu \right) 
\right. \nonumber  \\
&\qquad \qquad  \left.
+ \left( D_{\b{n}}\c{m}^\nu \right)\left(D_{\b{m}}l^\mu \right)\right] \nonumber \\
& \qquad
-\t {g}{_\mu _\nu} \left[\left(D_{\b{l}}{m}^\nu \right)\left(D_{\c{\b{m}}}n^\mu \right)
\right. \nonumber  \\
&\qquad \qquad  \left.
+\left(D_{\b{n}}m^\nu \right) \left(D_{\c{\b{m}}}l^\mu \right)\right]
\nonumber \\
&= -\left[\inner{\c{\b{m}}, D_{\b{l}}D_{\b{m}}\b{n}}+
\inner{\b{m}, D_{\b{l}}D_{\c{\b{m}}}\b{n}}\right]
\nonumber  \\
&\qquad 
-\left[\inner{\c{\b{m}}, D_{\b{n}}D_{\b{m}}\b{l}}+
\inner{\b{m}, D_{\b{n}}D_{\c{\b{m}}}\b{l}}\right]
\nonumber \\
& \qquad
-\left[\inner{D_{\b{l}}\c{\b{m}},D_{\b{m}}\b{n}} + \inner{D_{\b{n}}\c{\b{m}}, D_{\b{m}}\b{l}}\right]
\nonumber  \\
&\qquad 
-\left[\inner{D_{\b{l}}\b{m},D_{\c{\b{m}}}\b{n}} + \inner{D_{\b{n}}\b{m}, D_{\c{\b{m}}}\b{l}}\right].\nonumber 
\end{align}
Now we can use Eqs.~(\ref{Dml}), (\ref{Dln}), (\ref{Dmn}), (\ref{Dlm}) and (\ref{Dnm}) to obtain
\begin{align}
\t {\eta}{^a^b}\t {\delta}{^i^j}D{_b}\t {J}{_a_i_j}
&= \left[D_{\b{n}}\left(\rho +\c{\rho }\right)-D_{\b{l}}\left(\mu +\c{\mu }\right)\right]
\nonumber  \\
&\qquad
+\left[\left(\c{\alpha}+\beta \right)\left( \pi +\c{\tau} \right)+ \left(\alpha +\c{\beta} \right)\left( \c{\pi} +\tau \right)\right]
\nonumber  \\
&\qquad
-\left[\left(\varepsilon - \c{\varepsilon}\right)\left(\mu -\c{\mu} \right)+\left(\gamma - \c{\gamma}\right)\left(\rho -\c{\rho}\right)\right]
\nonumber  \\
&\qquad
-\left[\left(\c{\alpha}+\beta \right)\left( \pi +\c{\tau} \right)+ \left(\alpha +\c{\beta} \right)\left( \c{\pi} +\tau \right)\right]
\nonumber  \\
&\qquad
+\left[\left(\varepsilon - \c{\varepsilon}\right)\left(\mu -\c{\mu} \right)+\left(\gamma - \c{\gamma}\right)\left(\rho -\c{\rho}\right)\right]\nonumber \\
&=\left[D_{\b{n}}\left(\rho +\c{\rho }\right)-D_{\b{l}}\left(\mu +\c{\mu }\right)\right].\label{DeltaJ1}
\end{align}
In order to derive the second term of Eq.~(\ref{eq:tilD_J_a_i_j}), we will use the definitions in Eq.~(\ref{eq:gamma_a_b_c}) and Eq.~(\ref{eq:J_a^i^j}). Then we get
\begin{align}
\t {\eta}{^a^b}\t {\delta}{^i^j}\t {\gamma}{_b_a^c}\t {J}{_c_i_j}&= \t {\eta}{^a^b}\t {\eta}{^c^d}\t {\delta}{^i^j}\left(\t {g}{_\mu _\nu}\left[\t {D}{_b}\t {E}{^\mu _a}\right]  \t {E}{^\nu _d}\right) 
\nonumber \\
&\qquad \qquad \times
\left( \t {g}{_\alpha _\beta}\left[\t {D}{_i}\t {E}{^\alpha _c}\right]  \t {N}{^\beta _j}\right)\nonumber \\
&=\t {g}{_\mu _\nu}\t {g}{_\alpha _\beta}\left( \t {N}{^\gamma _i} \t {N}{^\beta _j}\t {\delta}{^i^j}\right)
\nonumber \\
&\qquad \qquad \times
\left( \t {\eta}{^a^b} \t {E}{^\rho _b}D_\rho \t {E}{^\mu _a}\right)
\nonumber \\
&\qquad \qquad \times
\left(\t {\eta}{^c^d} \t {E}{^\nu _d} D_\gamma \t {E}{^\alpha _c}\right).\nonumber
\end{align}
Then by using relations (\ref{exp:delt_n_n}), (\ref{exp:eta_E_D_E}) and (\ref{exp:eta_E_D_E_cont}) we obtain
\begin{align}
\t {\eta}{^a^b}\t {\delta}{^i^j}\t {\gamma}{_b_a^c}\t {J}{_c_i_j}
&= 
\t {g}{_\mu _\nu}\t {g}{_\alpha _\beta}\left(m^\gamma \c{m}^\beta + m^\beta \c{m}^\gamma \right) 
\nonumber \\
&\qquad \qquad \times
\left( D_\b{l} n^\mu +D_\b{n} l^\mu\right)
\nonumber \\
&\qquad \qquad \times
\left(l^\nu D_\gamma n^\alpha +n^\nu D_{\gamma} l^\alpha \right)\nonumber \\
&= \t {g}{_\mu _\nu}\t {g}{_\alpha _\beta}\left(D_{\b{l}}n^\mu + D_{\b{n}} l^\mu \right)
\nonumber \\
&\qquad \qquad \times
\left(\c{m}^\beta l^\nu D_{\b{m}}n^\alpha + \c{m}^\beta n^\nu D_{\b{m}}l^\alpha 
\right. \nonumber \\
&\qquad \qquad \left.
+ m^\beta l^\nu D_{\c{\b{m}}}n^\alpha + m^\beta n^\nu D_{\c{\b{m}}}l^\alpha \right) \nonumber.
\end{align}
Hence,
\begin{align}
\t {\eta}{^a^b}\t {\delta}{^i^j}\t {\gamma}{_b_a^c}\t {J}{_c_i_j}
&= \inner{D_{\b{m}}\b{n},\c{\b{m}}}\left( \inner{D_{\b{l}}\b{n},\b{l}}
+\inner{D_{\b{n}}\b{l},\b{l}}\right)
\nonumber \\
&\qquad 
+\inner{D_{\b{m}}\b{l},\c{\b{m}}}\left( \inner{D_{\b{l}}\b{n},\b{n}}+\inner{D_{\b{n}}\b{l},\b{n}}\right)
\nonumber \\
&\qquad
+ \inner{D_{\c{\b{m}}}\b{n},\b{m}}\left( \inner{D_{\b{l}}\b{n},\b{l}}+\inner{D_{\b{n}}\b{l},\b{l}}\right)
\nonumber \\
&\qquad
+\inner{D_{\c{\b{m}}}\b{l},\b{m}}\left( \inner{D_{\b{l}}\b{n},\b{n}}+\inner{D_{\b{n}}\b{l},\b{n}}\right),\nonumber 
\end{align}
and by using Eqs.~(\ref{Dnl}), (\ref{Dml}), (\ref{Dln}) and (\ref{Dmn}) we have
\begin{align}
\t {\eta}{^a^b}\t {\delta}{^i^j}\t {\gamma}{_b_a^c}\t {J}{_c_i_j}
&= \left(\varepsilon +\c{\varepsilon}\right)\left(\mu +\c{\mu}\right)+\left(\gamma +\c{\gamma}\right)\left(\rho +\c{\rho}\right).\label{DeltaJ2}
\end{align}
In order to derive the third term of Eq.~(\ref{eq:tilD_J_a_i_j}) one uses the definitions in Eq.~(\ref{eq:w_a^i^j}) and Eq.~(\ref{eq:J_a^i^j}).
Then we write
\begin{align}
\t {\eta}{^a^b}\t {\delta}{^i^j}\t {w}{_b_i^k}\t {J}{_a_k_j} &= \t {\eta}{^a^b}\t {\delta}{^i^j}\t {\delta}{^k^l}\left[\t {g}{_\mu _\nu}\left(\t {D}{_b}\t {N}{^\mu _i}\right)  \t {N}{^\nu _k}\right]
\nonumber \\
&\qquad \qquad \times
\left[ \t {g}{_\alpha _\beta}\left(\t {D}{_l}\t {E}{^\alpha _a}\right)  \t {N}{^\beta _j}\right]\nonumber\\ 
&=\t {g}{_\mu _\nu}\t {g}{_\alpha _\beta}\left(\t {\delta}{^k ^l}\t {N}{^\gamma _l}\t {N}{^\nu _k}\right)\left(\t {\delta}{^i ^j}\t {N}{^\beta _j}D_\rho \t {N}{^\mu _i}\right)
\nonumber \\
&\qquad \qquad \times
\left( \t {\eta}{^a ^b}\t {E}{^\rho _b}D_\gamma \t {E}{^\alpha _a}\right).\nonumber
\end{align}
Now using Eqs.~(\ref{exp:delt_n_n}), (\ref{exp:eta_E_D_E}) and (\ref{exp:delt_n_D_n}) results in
\begin{align}
\t {\eta}{^a^b}\t {\delta}{^i^j}\t {w}{_b_j^k}\t {J}{_a_k_i}
&=-\t {g}{_\mu _\nu}\t {g}{_\alpha _\beta}\left(m^\gamma \c{m}^\nu + m^\nu \c{m}^\gamma \right)
\nonumber \\
&\qquad \qquad \times
\left(m^\beta D_\rho \c{m}^\mu + \c{m}^\beta D_\rho m^\mu \right)
\nonumber \\
&\qquad \qquad \times
\left(l^\rho D_\gamma n^\alpha +n^\rho D_{\gamma} l^\alpha \right)\nonumber\\
&=-\bigl[\inner{D_{\b{m}}\b{n},\b{m}}\inner{D_{\b{l}}\c{\b{m}},\c{\b{m}}}
 \nonumber  \\
&\qquad 
+\inner{D_{\b{l}}\b{m},\c{\b{m}}}\inner{D_{\b{m}}\b{n},\c{\b{m}}}
\nonumber  \\
&\qquad 
+\inner{D_{\b{n}}\c{\b{m}},\c{\b{m}}}\inner{D_{\b{m}}\b{l},\b{m}}
\nonumber  \\
&\qquad 
+\inner{D_{\b{n}}\b{m},\c{\b{m}}}\inner{D_{\b{m}}\b{l},\c{\b{m}}}
\nonumber  \\
&\qquad 
+ \inner{D_{\b{l}}\b{m},\b{m}}\inner{D_{\c{\b{m}}}\b{n},\b{m}}
\nonumber  \\
&\qquad 
+ \inner{D_{\b{l}}\b{m},\b{m}}\inner{D_{\c{\b{m}}}\b{n},\c{\b{m}}}
 \nonumber  \\
&\qquad 
+ \inner{D_{\b{n}}\c{\b{m}},\b{m}}\inner{D_{\c{\b{m}}}\b{l},\b{m}}
\nonumber  \\
&\qquad 
+ \inner{D_{\b{n}}\b{m},\b{m}}\inner{D_{\c{\b{m}}}\b{l},\c{\b{m}}}\bigr],\nonumber 
\end{align}
and by Eqs.~(\ref{Dml}), (\ref{Dmn}), (\ref{Dlm}) and (\ref{Dnm}) we obtain
\begin{align}
\t {\eta}{^a^b}\t {\delta}{^i^j}\t {w}{_b_j^k}\t {J}{_a_k_i}
&=-\left[\left(\varepsilon -\c{\varepsilon}\right)\left(\mu -\c{\mu}\right)
+\left(\gamma -\c{\gamma}\right)\left(\rho -\c{\rho}\right)\right].\label{DeltaJ3}
\end{align}
Similarly, the fourth term in Eq.~(\ref{eq:tilD_J_a_i_j}) follows from
\begin{align}
\t {\eta}{^a^b}\t {\delta}{^i^j}\t {w}{_b_j^k}\t {J}{_a_i_k} &= \t {\eta}{^a^b}\t {\delta}{^i^j}\t {\delta}{^k^l}\left[\t {g}{_\mu _\nu}\left(\t {D}{_b}\t {N}{^\mu _j}\right)  \t {N}{^\nu _k}\right] 
\nonumber \\
&\qquad \times
\left[ \t {g}{_\alpha _\beta}\left(\t {D}{_i}\t {E}{^\alpha _a}\right)  \t {N}{^\beta _l}\right]\nonumber\\ 
&=\t {g}{_\mu _\nu}\t {g}{_\alpha _\beta}\left(\t {\delta}{^k ^l}\t {N}{^\nu _k}\t {N}{^\beta _l}\right)
\nonumber \\
&\qquad \times
\left(\t {\delta}{^i ^j}\t {N}{^\gamma _i}D_\rho \t {N}{^\mu _j}\right)
\nonumber \\
&\qquad \times
\left( \t {\eta}{^a ^b}\t {E}{^\rho _b}D_\gamma \t {E}{^\alpha _a}\right).\nonumber
\end{align}
Then by using relations (\ref{exp:delt_n_n}), (\ref{exp:eta_E_D_E}) and (\ref{exp:delt_n_D_n}),
\begin{align}
\t {\eta}{^a^b}\t {\delta}{^i^j}\t {w}{_b_j^k}\t {J}{_a_i_k}
&=-\t {g}{_\mu _\nu}\t {g}{_\alpha _\beta}\left(m^\nu \c{m}^\beta + m^\beta \c{m}^\nu \right)
\nonumber \\
&\qquad \times
\left(m^\gamma D_\rho \c{m}^\mu + \c{m}^\gamma D_\rho m^\mu \right)
\nonumber \\
&\qquad \times
\left(l^\rho D_\gamma n^\alpha +n^\rho D_{\gamma} l^\alpha \right)\nonumber\\
&=-\bigl[\inner{D_{\b{m}}\b{n},\b{m}}\inner{D_{\b{l}}\c{\b{m}},\c{\b{m}}}
\nonumber  \\
&\qquad 
+\inner{D_{\b{l}}\c{\b{m}},\b{m}}\inner{D_{\b{m}}\b{n},\c{\b{m}}}
\nonumber  \\
&\qquad 
+\inner{D_{\b{n}}\c{\b{m}},\b{m}}\inner{D_{\b{m}}\b{l},\c{\b{m}}}
\nonumber  \\
&\qquad
+\inner{D_{\b{n}}\b{m},\b{m}}\inner{D_{\c{\b{m}}}\b{l},\c{\b{m}}}
\nonumber  \\
&\qquad
+ \inner{D_{\b{l}}\b{m},\c{\b{m}}}\inner{D_{\c{\b{m}}}\b{n},\b{m}}
\nonumber  \\
&\qquad
+ \inner{D_{\b{l}}\b{m},\b{m}}\inner{D_{\c{\b{m}}}\b{n},\c{\b{m}}}
\nonumber  \\
&\qquad
+ \inner{D_{\b{n}}\c{\b{m}},\c{\b{m}}}\inner{D_{\b{m}}\b{l},\b{m}}
\nonumber  \\
&\qquad
+ \inner{D_{\b{n}}\b{m},\c{\b{m}}}\inner{D_{\c{\b{m}}}\b{l},\b{m}}\bigr],\nonumber 
\end{align}
and by further using Eqs.~(\ref{Dml}), (\ref{Dmn}), (\ref{Dlm}) and (\ref{Dnm}) we obtain the same result as in (\ref{DeltaJ3}), i.e.,
\begin{align}
\t {\eta}{^a^b}\t {\delta}{^i^j}\t {w}{_b_j^k}\t {J}{_a_i_k}
&=-\left[\left(\varepsilon -\c{\varepsilon}\right)\left(\mu -\c{\mu}\right)+\left(\gamma -\c{\gamma}\right)\left(\rho -\c{\rho}\right)\right].\label{DeltaJ4}
\end{align}
Hence, substitution of the relations (\ref{DeltaJ1}), (\ref{DeltaJ2}), (\ref{DeltaJ3}) and (\ref{DeltaJ4}) into Eq.~(\ref{eq:tilD_J_a_i_j}) results in
\begin{align}
\tilde{\nabla} _{\mathbb{T}}\mathcal{J}&= \left[D_{\b{n}}\left(\rho +\c{\rho }\right)-D_{\b{l}}\left(\mu +\c{\mu }\right)\right]
\nonumber  \\
&\qquad
-\left[\left(\varepsilon +\c{\varepsilon}\right)\left(\mu +\c{\mu}\right)+\left(\gamma +\c{\gamma}\right)\left(\rho +\c{\rho}\right)\right]
\nonumber  \\
&\qquad
+2\left[\left(\varepsilon - \c{\varepsilon}\right)\left(\mu -\c{\mu} \right)+\left(\gamma - \c{\gamma}\right)\left(\rho -\c{\rho}\right)\right].
\end{align}
\subsection{Derivation of $\tilde{\nabla} _{\mathbb{S}}\mathcal{K}$}
Consider the first term on the right-hand side of the Raychaudhuri equation (\ref{Raych_simple}), and the covariant derivative of $\t {K}{_a_b_j}$ on the spacelike 2-surface defined in relation (\ref{eq:CurlyCovK}), i.e.,
\begin{align}\label{eq:tilD_K_a_b_j}
\tilde{\nabla} _{\mathbb{S}}\mathcal{K}
&:=\t {\eta}{^a^b}\t {\delta}{^i^j}\t {\tilde{\nabla}}{_i}\t {K}{_a_b_j}
\nonumber \\
&=\t {\eta}{^a^b}\t {\delta}{^i^j} \left(\underbrace{\t {\nabla}{_i}\t {K}{_a_b_j}}_\text{$\t {D}{_i}\t {K}{_a_b_j}-\t {\gamma}{_i_j_k}\t {K}{_a_b^k}$}-\t {S}{_a_c_i}\t {K}{_b^c_j}-\t {S}{_b_c_i}\t {K}{_a^c_j}\right).
\end{align}
Then, by making use of the definition (\ref{eq:K_a_b^i}), the first term of Eq. (\ref{eq:tilD_K_a_b_j}) is as follows 
\begin{align}
D_i \t {K}{_a_b_j}\t {\eta}{^a ^b}\t {\delta} {^i ^j}&=\t {\eta}{^a ^b}\t {\delta} {^i ^j}D_i\left[-\t {g}{_\mu _\nu}\left(D_a \t {E}{^\mu _b}\right)\t {N}{^\nu _j}\right]\nonumber \\
&= -\t {\eta}{^a ^b}\t {\delta} {^i ^j}\left[\t {N}{^\nu _j} \t {N}{^\gamma _i}D_\gamma \left(\t {g}{_\mu _\nu}\left(D_a \t {E}{^\mu _b}\right)\right)\right]
\nonumber  \\
&\qquad
-\t {\eta}{^a ^b}\t {\delta} {^i ^j}\left[ \t {g}{_\mu _\nu}\left(D_a \t {E}{^\mu _b}\right)\t {N}{^\gamma _i}D_\gamma \t {N}{^\nu _j} \right]
\nonumber \\
&=-\t {g}{_\mu _\nu}\left[\left(\t {\delta}{^i ^j}\t {N}{^\nu _j}\t {N}{^\gamma _j} \right)\t {\eta}{^a^b}D_\gamma \left( \t {E}{^\beta _a}D_\beta \t {E}{^\mu _b}\right)\right]
\nonumber  \\
&\qquad
-\t {g}{_\mu _\nu}\left[\left(\t {\eta}{^a^b}\t {E}{^\beta _a}D_\beta \t {E}{^\mu _b}\right)
\right. \nonumber  \\
&\qquad \qquad  \left.
\times \left(\t {\delta}{^i^j}\t {N}{^{\gamma} _i}D_{\gamma}\t {N}{^{\nu} _j}\right)\right].\nonumber
\end{align}
By using Eqs.~(\ref{exp:delt_n_n}), (\ref{exp:eta_E_D_E_cont}) and (\ref{exp:delt_n_D_n_cont}) we write 
\begin{align}
D_i \t {K}{_a_b_j}\t {\eta}{^a ^b}\t {\delta} {^i ^j}
&= \t {g}{_\mu _\nu} \left[\left(m^\gamma \c{m}^\nu + m^\nu \c{m}^\gamma \right)
\right. \nonumber  \\
&\qquad \qquad \left. \times
D_\gamma \left( D_\b{l} n^\mu +D_\b{n} l^\mu \right)\right]
\nonumber  \\
&\qquad
+ \t {g}{_\mu _\nu} \left[\left( D_\b{l} n^\mu +D_\b{n} l^\mu\right) 
\right. \nonumber  \\
&\qquad \qquad \qquad \left. \times
\left(D_{\b{m}}\c{m}^\nu + D_{\c{\b{m}}}m^\nu \right)\right]\nonumber \\
&= \inner{\c{\b{m}}, D_{\b{m}}D_{\b{l}}\b{n}}+ \inner{\c{\b{m}}, D_{\b{m}}D_{\b{n}}\b{l}}
\nonumber  \\
&\qquad
+\inner{\b{m}, D_{\c{\b{m}}}D_{\b{l}}\b{n}}+
\inner{\b{m}, D_{\c{\b{m}}}D_{\b{n}}\b{l}}
\nonumber  \\
&\qquad
+\inner{D_{\b{l}}\b{n},D_{\b{m}}\c{\b{m}}} + \inner{D_{\b{l}}\b{n}, D_{\c{\b{m}}}\b{m}}
\nonumber  \\
&\qquad
+\inner{D_{\b{n}}\b{l},D_{\b{m}}\c{\b{m}}} + \inner{D_{\b{n}}\b{l}, D_{\c{\b{m}}}\b{m}}, \nonumber
\end{align}
and by further using Eqs.~(\ref{Dnl}), (\ref{Dln}) and (\ref{Dmm_}) we obtain
\begin{align}
D_i \t {K}{_a_b_j}\t {\eta}{^a ^b}\t {\delta} {^i ^j}
&= D_{\b{m}}\left(\pi -\c{\tau} \right)+D_{\c{\b{m}}}\left(\c{\pi} -\tau \right)
\nonumber  \\
&\qquad
-\left[\left(\alpha - \c{\beta} \right)\left( \c{\pi} - \tau \right)+\left(\c{\alpha}-\beta \right)\left( \pi - \c{\tau} \right)\right]
\nonumber  \\
&\qquad
-\left[\left(\varepsilon +\c{\varepsilon}\right)\left(\mu +\c{\mu}\right)+\left(\gamma +\c{\gamma}\right)\left(\rho +\c{\rho}\right)\right]
\nonumber  \\
&\qquad
+\left[\left(\varepsilon +\c{\varepsilon}\right)\left(\mu +\c{\mu}\right)+\left(\gamma +\c{\gamma}\right)\left(\rho +\c{\rho}\right)\right]
\nonumber  \\
&\qquad
+\left[\left(\alpha - \c{\beta} \right)\left( \c{\pi} - \tau \right)+\left(\c{\alpha}-\beta \right)\left( \pi - \c{\tau} \right)\right]\nonumber \\
&= D_{\b{m}}\left(\pi -\c{\tau} \right)+D_{\c{\b{m}}}\left(\c{\pi} -\tau \right).\label{DeltaK1}
\end{align}
The second term in Eq.~(\ref{eq:tilD_K_a_b_j}) is obtained by using the definitions (\ref{eq:K_a_b^i}) and (\ref{gamma_i_j_k}). The derivation follows as
\begin{align}
\t {\gamma}{_i_j_k}\t {K}{_a_b_l}\t {\delta}{^i ^j}\t {\delta}{^k ^l}\t {\eta}{^a ^b}
&= \left[\t {g}{_\alpha _\beta}\left(\t {D}{_i}\t {N}{^\alpha _j}\right)  \t {N}{^\beta _k}\right]
\nonumber  \\
&\qquad \times
\left[-\t {g}{_\mu _\nu}\left(\t {D}{_a}\t {E}{^\mu _b}\right)  \t {N}{^\nu _l}\right]\t {\delta}{^i ^j}\t {\delta}{^k ^l}\t {\eta}{^a ^b}\nonumber \\
&= -\t {g}{_\alpha _\beta}\t {g}{_\mu _\nu}\left(\t {\delta}{^k ^l}\t {N}{^\beta _k}\t {N}{^\nu _l}\right)
\nonumber  \\
&\qquad \times
\left(\t {\delta}{^i ^j}\t {N}{^{\rho} _i}D_{\rho}\t {N}{^{\alpha} _j}\right)\left(\t {\eta}{^a ^b}\t {E}{^\gamma _a}D_\gamma \t {E}{^\mu _b} \right).\nonumber
\end{align}
Now let us use Eqs.~(\ref{exp:delt_n_n}), (\ref{exp:eta_E_D_E_cont}) and (\ref{exp:delt_n_D_n}) to write
\begin{align}
\t {\gamma}{_i_j_k}\t {K}{_a_b_l}\t {\delta}{^i ^j}\t {\delta}{^k ^l}\t {\eta}{^a ^b}
&= \t {g}{_\alpha _\beta}\t {g}{_\mu _\nu}
\left(m^\beta \c{m}^\nu + m^\nu \c{m}^\beta \right)
\nonumber  \\
&\qquad \times
\left(m^\rho D_\rho \c{m}^\alpha + \c{m}^\rho D_\rho m^\alpha \right)
\nonumber  \\
&\qquad \times
\left( D_\b{l} n^\mu +D_\b{n} l^\mu\right) \nonumber \\
&= \inner{D_{\b{m}}\c{\b{m}},\b{m}} \inner{D_{\b{l}}\b{n},\c{\b{m}}}
\nonumber  \\
&\qquad 
+\inner{D_{\c{\b{m}}}\b{m},\b{m}}\inner{D_{\b{l}}\b{n},\c{\b{m}}}
\nonumber  \\
&\qquad 
+\inner{D_{\b{m}}\c{\b{m}},\b{m}} \inner{D_{\b{n}}\b{l},\c{\b{m}}}
\nonumber  \\
&\qquad 
+\inner{D_{\c{\b{m}}}\b{m},\b{m}}\inner{D_{\b{n}}\b{l},\c{\b{m}}}
\nonumber  \\
&\qquad 
+\inner{D_{\b{m}}\c{\b{m}},\c{\b{m}}} \inner{D_{\b{l}}\b{n},\b{m}}
\nonumber  \\
&\qquad 
+\inner{D_{\c{\b{m}}}\b{m},\c{\b{m}}}\inner{D_{\b{l}}\b{n},\b{m}}
\nonumber  \\
&\qquad 
+\inner{D_{\b{m}}\c{\b{m}},\c{\b{m}}} \inner{D_{\b{n}}\b{l},\b{m}}
\nonumber  \\
&\qquad 
+\inner{D_{\c{\b{m}}}\b{m},\c{\b{m}}}\inner{D_{\b{n}}\b{l},\b{m}}.\nonumber
\end{align}
By using Eqs.~(\ref{Dnl}), (\ref{Dln}) and (\ref{Dmm_}) we obtain
\begin{align}
\t {\gamma}{_i_j_k}\t {K}{_a_b_l}\t {\delta}{^i ^j}\t {\delta}{^k ^l}\t {\eta}{^a ^b}
&=\left(\c{\alpha}-\beta \right)\left( \pi -\c{\tau} \right)+ \left(\alpha -\c{\beta} \right)\left( \c{\pi} -\tau \right).\label{DeltaK2}
\end{align}
Finally we derive the third term that appears in Eq.~(\ref{eq:tilD_K_a_b_j}). Note that the third term is equal to the fourth term since our $\t {\eta}{_a_b}$ is diagonal. Here we make use of the definitions (\ref{eq:S_a_b^i}) and (\ref{eq:K_a_b^i}) and get
\begin{align}
\t {S}{_a_c_i}\t {K}{_b_d_j}\t {\delta}{^i ^j}\t {\eta}{^a^b}\t {\eta}{^c^d}
&=\left[ \t {g}{_\mu _\nu}\left(\t {D}{_i}\t {E}{^\mu _a}\right)  \t {E}{^\nu _c}\right]
\nonumber  \\
&\qquad \times
\left[-\t {g}{_\alpha _\beta}\left(\t {D}{_b}\t {E}{^\alpha _d}\right)  \t {N}{^\beta _j}\right]\t {\delta}{^i ^j}\t {\eta}{^a^b}\t {\eta}{^c^d}\nonumber \\
&=-\t {g}{_\mu _\nu} \t {g}{_\alpha _\beta}\left(\t {\delta}{^i ^j}\t {N}{^\gamma _i}\t {N}{^\beta _j}\right)
\nonumber  \\
&\qquad \times
\left(\t {\eta}{^a ^b}\t {E}{^\rho _b}D_\gamma \t {E}{^\mu _a} \right)
\nonumber  \\
&\qquad \times
\left(\t {\eta}{^c ^d}\t {E}{^\nu _c}D_\rho \t {E}{^\alpha _d} \right).\nonumber
\end{align}
Also by using Eqs.~(\ref{exp:delt_n_n}) and (\ref{exp:eta_E_D_E}) we obtain
\begin{align}
\t {S}{_a_c_i}\t {K}{_b_d_j}\t {\delta}{^i ^j}\t {\eta}{^a^b}\t {\eta}{^c^d}
&=-\t {g}{_\mu _\nu} \t {g}{_\alpha _\beta} \left(m^\gamma \c{m}^\beta + m^\beta \c{m}^\gamma \right)
\nonumber  \\
&\qquad \times
\left(l^\rho D_\gamma n^\mu +n^\rho D_{\gamma} l^\mu \right)
\nonumber  \\
&\qquad \times
\left(l^\nu D_\rho n^\alpha +n^\nu D_{\rho} l^\alpha \right)\nonumber \\
&=-\left[\inner{D_{\b{m}}\b{n},\b{l}}\inner{D_{\b{l}}\b{n},\c{\b{m}}}
\right. \nonumber  \\
&\qquad \left.
+\inner{D_{\b{m}}\b{n},\b{n}}\inner{D_{\b{l}}\b{l},\c{\b{m}}}\right]\nonumber \\
&=-\left[\inner{D_{\b{m}}\b{l},\b{l}}\inner{D_{\b{n}}\b{n},\c{\b{m}}}
\right. \nonumber  \\
&\qquad  \left.
+\inner{D_{\b{m}}\b{l},\b{n}}\inner{D_{\b{n}}\b{l},\c{\b{m}}}\right]\nonumber \\
&=-\left[\inner{D_{\c{\b{m}}}\b{n},\b{l}}\inner{D_{\b{l}}\b{n},\b{m}}
\right. \nonumber  \\
&\qquad  \left.
+\inner{D_{\c{\b{m}}}\b{n},\b{n}}\inner{D_{\b{l}}\b{l},\b{m}}\right]\nonumber \\
&=-\left[\inner{D_{\c{\b{m}}}\b{l},\b{n}}\inner{D_{\b{n}}\b{l},\b{m}}
\right. \nonumber  \\
&\qquad  \left.
+\inner{D_{\c{\b{m}}}\b{l},\b{l}}\inner{D_{\b{n}}\b{n},\b{m}}\right].\nonumber 
\end{align}
Then by further using Eqs.~(\ref{Dll}), (\ref{Dnl}), (\ref{Dml}), \,(\ref{Dln}) and (\ref{Dnn}) we write
\begin{align}
\t {S}{_a_c_i}\t {K}{_b_d_j}\t {\delta}{^i ^j}\t {\eta}{^a^b}\t {\eta}{^c^d}
&=
-\left[ \left(\c{\alpha}+\beta \right)\left( \pi + \c{\tau} \right)
\right. \nonumber  \\
&\qquad  \qquad \left.
+\left(\alpha + \c{\beta} \right)\left( \c{\pi} + \tau \right) \right].
\label{DeltaK3-4}
\end{align}
Therefore substitution of relations (\ref{DeltaK1}), (\ref{DeltaK2}) and (\ref{DeltaK3-4}) into Eq.~(\ref{eq:tilD_K_a_b_j}) results in
\begin{align}
\tilde{\nabla} _{\mathbb{S}}\mathcal{K}
&=D_{\b{m}}\left(\pi -\c{\tau} \right)+D_{\c{\b{m}}}\left(\c{\pi} -\tau \right)
\nonumber  \\
&\qquad 
-\left[\left(\c{\alpha}-\beta \right)\left( \pi -\c{\tau} \right)
+ \left(\alpha -\c{\beta} \right)\left( \c{\pi} -\tau \right)\right]
\nonumber  \\
&\qquad
+ 2\left[ \left(\c{\alpha}+\beta \right)\left( \pi + \c{\tau} \right) 
+\left(\alpha + \c{\beta} \right)\left( \c{\pi} + \tau \right)\right].
\end{align}
\subsection{Derivation of $\mathcal{J}^2$}
In order to derive the second term that appears on the right-hand side of the Raychaudhuri equation (\ref{Raych_simple}), we start with the definition (\ref{eq:J_a^i^j}) and write
\begin{align}
\mathcal{J}^2
&:=\t {J}{_b_i_k}\t {J}{_a_l_j}{\t {\eta}{^a^b}}\t {\delta}{^i^j}\t {\delta}{^l^k}
\nonumber \\
&=\left[\t {g}{_\mu _\nu}\left(\t {D}{_i}\t {E}{^\mu _b}\right)  \t {N}{^\nu _k}\right] \left[\t {g}{_\alpha _\beta}\left(\t {D}{_l}\t {E}{^\alpha _a}\right)  \t {N}{^\beta _j}\right]
\nonumber  \\
&\qquad \times
\t {\eta}{^a^b}\t {\delta}{^i^j}\t {\delta}{^ l ^k}\nonumber \\ 
&= \t {g}{_\mu _\nu}\t {g}{_\alpha _\beta} \left(\t {\delta}{^ i ^j} \t {N} {^\rho _i}\t {N}{^\beta _j}\right) \left(\t {\delta}{^ k ^l} \t {N} {^\gamma _l}\t {N}{^\nu _k}\right)
\nonumber  \\
&\qquad \times
\left[\t {\eta}{^a^b}\left(D_\gamma \t {E}{^\alpha _a}\right)\left(D_\rho \t {E}{^\mu _b}\right)\right], \nonumber
\end{align}
then by 
Eqs.~(\ref{exp:delt_n_n}) and (\ref{exp:eta_D_E_D_E}),
\begin{align}
\mathcal{J}^2 
&=-\t {g}{_\mu _\nu}\t {g}{_\alpha _\beta} \left(m^\rho \c{m}^\beta + m^\beta \c{m}^\rho \right)\left(m^\gamma \c{m}^\nu + m^\nu \c{m}^\gamma \right)
\nonumber \\
&\qquad
\times \left[\left(D_\gamma l^\alpha \right)\left(D_\rho n^\mu \right) +\left(D_\gamma n^\alpha \right)\left(D_\rho l^\mu \right)\right]\nonumber \\
&= -\left[\inner{D_{\b{m}}\b{n},\c{\b{m}}} \inner{D_{\b{m}}\b{l},\c{\b{m}}} + \inner{D_{\b{m}}\b{n},\c{\b{m}}} \inner{D_{\b{m}}\b{l},\c{\b{m}}} \right]
\nonumber \\
&\qquad
-\left[\inner{D_{\b{m}}\b{n},\b{m}} \inner{D_{\c{\b{m}}}\b{l},\c{\b{m}}} + \inner{D_{\b{m}}\b{l},\b{m}} \inner{D_{\c{\b{m}}}\b{n},\c{\b{m}}} \right]
\nonumber \\
&\qquad
-\left[\inner{D_{\c{\b{m}}}\b{n},\c{\b{m}}} \inner{D_{\b{m}}\b{l},\b{m}} + \inner{D_{\b{m}}\b{n},\b{m}} \inner{D_{\c{\b{m}}}\b{l},\c{\b{m}}} \right]
\nonumber \\
&\qquad
-\left[\inner{D_{\c{\b{m}}}\b{l},\b{m}} \inner{D_{\c{\b{m}}}\b{n},\b{m}} + \inner{D_{\c{\b{m}}}\b{l},\b{m}} \inner{D_{\c{\b{m}}}\b{n},\b{m}} \right].\nonumber
\end{align}
Finally, by using Eqs.~(\ref{Dml}) and (\ref{Dmn}) we obtain
\begin{align}
\mathcal{J}^2
&= 2\left(\mu \c{\rho} + \c{\mu} \rho + \sigma \lambda + \c{\sigma} \c{\lambda} \right).
\end{align}
\subsection{Derivation of $\mathcal{K}^2$}
The third term that appears on the right-hand side of the Raychaudhuri equation (\ref{Raych_simple}), is obtained as the following once the definition (\ref{eq:K_a_b^i}) is considered:
\begin{align}
\mathcal{K}^2&:=\t {K}{_b_c_i}\t {K}{_a_d_j}\t {\eta}{^a^b}\t {\eta}{^c^d} \t {\delta}{^i^j} \nonumber \\
&=  \left[-\t {g}{_\mu _\nu}\left(\t {D}{_b}\t {E}{^\mu _c}\right)  \t {N}{^\nu _i}\right]\left[-\t {g}{_\alpha _\beta}\left(\t {D}{_a}\t {E}{^\alpha _d}\right)  \t {N}{^\beta _j}\right]
\nonumber  \\
&\qquad \times
\t {\eta}{^a^b}\t {\eta}{^c^d} \t {\delta}{^i^j}\nonumber \\
&=\t {g}{_\mu _\nu}\t {g}{_\alpha _\beta} \left(\t {\delta}{^ i ^j} \t {N} {^\nu _i}\t {N}{^\beta _j}\right)\left(\t {\eta}{^a^b}\t {E}{^\rho _b}\t {E}{^\gamma _a}\right)
\nonumber  \\
&\qquad \times
\left[\t {\eta}{^c^d}\left(D_\rho \t {E}{^\mu _c}\right)\left(D_\gamma \t {E}{^\alpha _d}\right)\right].\nonumber
\end{align}
Also by making use of Eqs.~(\ref{exp:eta_E_E}), (\ref{exp:delt_n_n}) and (\ref{exp:eta_D_E_D_E}) we write
\begin{align}
\mathcal{K}^2
&= \left(m^\nu \c{m}^\beta + m^\beta \c{m}^\nu \right)\left(l^\rho n^\gamma + l^\gamma n^\rho \right)
\nonumber  \\
&\qquad \times
\left[\left(D_\rho l^\mu \right)\left(D_\gamma n^\alpha \right) +\left(D_\rho n^\mu \right)\left(D_\gamma l^\alpha \right)\right]\nonumber \\
&=\left[\inner{D_{\b{l}}\b{l},\b{m}} \inner{D_{\b{n}}\b{n},\c{\b{m}}} + \inner{D_{\b{l}}\b{n},\b{m}} \inner{D_{\b{n}}\b{l},\c{\b{m}}} \right]
\nonumber  \\
&\qquad 
+\left[\inner{D_{\b{n}}\b{l},\b{m}} \inner{D_{\b{l}}\b{n},\c{\b{m}}} + \inner{D_{\b{n}}\b{n},\b{m}} \inner{D_{\b{l}}\b{l},\c{\b{m}}} \right]
\nonumber  \\
&\qquad 
+\left[\inner{D_{\b{l}}\b{l},\c{\b{m}}} \inner{D_{\b{n}}\b{n},\b{m}} + \inner{D_{\b{l}}\b{n},\c{\b{m}}} \inner{D_{\b{n}}\b{l},\b{m}} \right]
\nonumber  \\
&\qquad 
+\left[\inner{D_{\b{n}}\b{l},\c{\b{m}}} \inner{D_{\b{l}}\b{n},\b{m}} + \inner{D_{\b{n}}\b{n},\c{\b{m}}} \inner{D_{\b{l}}\b{l},\b{m}} \right].\nonumber
\end{align}
Then by Eqs.~(\ref{Dll}), (\ref{Dnl}), (\ref{Dln}) and (\ref{Dnn}) we obtain the final form as
\begin{align}
\mathcal{K}^2
&= -2\left(\kappa \nu + \c{\kappa} \c{\nu} + \pi \tau + \c{\pi} \c{\tau} \right).
\end{align}
\subsection{Derivation of $\mathcal{R_{\,W}}$} 
Now we derive the last term on the right-hand side of the Raychaudhuri equation (\ref{Raych_simple}), in terms of the variables of the Newman-Penrose formalism, i.e.,
\begin{align}
\mathcal{R_{\,W}}
&:=g(R(\t {E}{_b},\t {N}{_i})\t {E}{_a},\t {N}{_j}){\t {\eta}{^a^b}}\t {\delta}{^i^j}
\nonumber \\
=&\t {R}{_\alpha_\beta_\mu_\nu}\t {E}{^\mu _b}\t {N}{^\nu _i}\t {E}{^\beta _a}\t {N}{^\alpha_j}{\t {\eta}{^a^b}}\t {\delta}{^i^j}\nonumber \\
&=\t {R}{_\alpha_\beta_\mu_\nu}\left(\t {\eta}{^a^b}\t {E}{^\mu _b}\t {E}{^\beta _a}\right)\left(\t {\delta}{^i^j} \t {N}{^\nu _i} \t {N}{^\alpha_j} \right).\nonumber 
\end{align}
Then by using Eqs.~(\ref{exp:eta_E_E}) and (\ref{exp:delt_n_n}) we obtain
\begin{align}
\mathcal{R_{\,W}}
&=-\t {R}{_\alpha_\beta_\mu_\nu}\left(l^\mu n^\beta + l^\beta n^\mu \right)\left(m^\nu \c{m}^\alpha + m^\alpha \c{m}^\nu \right)\nonumber \\
&=-\left[\t {R}{_{\c{\b{m}}}_{\b{n}}_{\b{l}}_{\b{m}}} + \t {R}{_{\b{m}}_{\b{n}}_{\b{l}}_{\c{\b{m}}}}+ \t {R}{_{\c{\b{m}}}_{\b{l}}_{\b{n}}_{\b{m}}}+\t {R}{_{\b{m}}_{\b{l}}_{\b{n}}_{\c{\b{m}}}}\right].
\end{align}
Since, the Riemann tensor is defined as
\begin{align}
\t {R}{_{\b{x}}_{\b{y}}_{\b{v}}_{\b{w}}}
&= -\inner{D_{\b{x}}D_{\b{y}}\b{v},\b{w}}+\inner{D_{\b{y}}D_{\b{x}}\b{v},\b{w}}+\inner{D_{[\b{x},\b{y}]}\b{v},\b{w}},\nonumber
\end{align}
we write
\begin{align}
\mathcal{R_{\,W}}
&=-\left[\t {R}{_{\c{\b{m}}}_{\b{n}}_{\b{l}}_{\b{m}}} + \t {R}{_{\b{m}}_{\b{n}}_{\b{l}}_{\c{\b{m}}}}+ \t {R}{_{\c{\b{m}}}_{\b{l}}_{\b{n}}_{\b{m}}}+\t {R}{_{\b{m}}_{\b{l}}_{\b{n}}_{\c{\b{m}}}}\right]\nonumber \\
&=-\left[-\inner{D_{\c{\b{m}}}D_{\b{n}}\b{l},\b{m}}+\inner{D_{\b{n}}D_{\c{\b{m}}}\b{l},\b{m}}+\inner{D_{[\c{\b{m}},\b{n}]}\b{l},\b{m}}\right]
\nonumber  \\
&\qquad 
- \left[-\inner{D_{\b{m}}D_{\b{n}}\b{l},\c{\b{m}}}+\inner{D_{\b{n}}D_{\b{m}}\b{l},\c{\b{m}}}
\right. \nonumber  \\
&\qquad  \qquad \qquad \qquad \qquad \qquad \left.
+\inner{D_{[\b{m},\b{n}]}\b{l},\c{\b{m}}}\right]
\nonumber  \\
&\qquad 
-\left[-\inner{D_{\c{\b{m}}}D_{\b{l}}\b{n},\b{m}}+\inner{D_{\b{l}}D_{\c{\b{m}}}\b{n},\b{m}}
\right. \nonumber  \\
&\qquad  \qquad \qquad \qquad  \qquad \qquad \left.
+\inner{D_{[\c{\b{m}},\b{l}]}\b{n},\b{m}}\right]
\nonumber  \\
&\qquad 
- \left[-\inner{D_{\b{m}}D_{\b{l}}\b{n},\c{\b{m}}}+\inner{D_{\b{l}}D_{\b{m}}\b{n},\c{\b{m}}}
\right. \nonumber  \\
&\qquad  \qquad \qquad \qquad \qquad \qquad \left.
+\inner{D_{[\b{m},\b{l}]}\b{n},\c{\b{m}}}\right]\label{eq:Riemm_inner}.
\end{align}
Now we will make use of the commutation relations, (\ref{com_lm}) and (\ref{com_nm}), in order to write the inner products that involve the brackets in terms of the Newman-Penrose variables. In particular,
\begin{align}
D_{[\c{\b{m}},\b{n}]}\b{l}
&=-\nu D_{\b{l}}\b{l}-\left(\alpha + \c{\beta} - \c{\tau}\right)D_{\b{n}}\b{l}
\nonumber  \\
&\qquad \qquad 
-\left(\c{\gamma} -\gamma -\c{\mu}\right) D_{\c{\b{m}}}\b{l}+\lambda D_{\b{m}}\b{l},\nonumber \\
D_{[\b{m},\b{n}]}\b{l}
&= -\c{\nu} D_{\b{l}}\b{l}-\left(\c{\alpha} + \beta - \tau\right)D_{\b{n}}\b{l}
\nonumber  \\
&\qquad \qquad 
-\left(\gamma -\c{\gamma} -\mu \right) D_{\b{m}}\b{l}+\c{\lambda} D_{\c{\b{m}}}\b{l},\nonumber \\
D_{[\c{\b{m}},\b{l}]}\b{n}
&= -\left(\pi -\alpha - \c{\beta}\right) D_{\b{l}}\b{n} +\c{\kappa}D_{\b{n}}\b{n}
\nonumber  \\
&\qquad \qquad 
-\left(\c{\varepsilon}-\varepsilon +\rho \right)D_{\c{\b{m}}}\b{n}-\c{\sigma}D_{\b{m}}\b{n},\nonumber \\
D_{[\b{m},\b{l}]}\b{n}
&= -\left(\c{\pi} -\c{\alpha} - \beta\right) D_{\b{l}}\b{n} +\kappa D_{\b{n}}\b{n}
\nonumber  \\
&\qquad \qquad 
-\left(\varepsilon - \c{\varepsilon} +\c{\rho} \right)D_{\b{m}}\b{n}-\sigma D_{\c{\b{m}}}\b{n}.
\end{align}
At the next step of our derivation we make use of the propagation equations (\ref{Dll}), (\ref{Dnl}), (\ref{Dml}), (\ref{Dln}), (\ref{Dnn}) and (\ref{Dmn}). Then we obtain
\begin{align}
&\inner{D_{[\c{\b{m}},\b{n}]}\b{l},\b{m}}+\inner{D_{[\b{m},\b{n}]}\b{l},\c{\b{m}}} 
+\inner{D_{[\c{\b{m}},\b{l}]}\b{n},\b{m}}\nonumber  \\
&+\inner{D_{[\b{m},\b{l}]}\b{n},\c{\b{m}}}
= 2\left(\kappa \nu + \c{\kappa} \c{\nu} \right)-2\left(\tau \c{\tau}+\pi \c{\pi}\right) 
\nonumber  \\
& \qquad \qquad \qquad \qquad 
-2\left(\rho \c{\mu} +\c{\rho} \mu +\lambda \sigma + \c{\lambda} \c{\sigma} \right)
\nonumber  \\
& \qquad \qquad \qquad \qquad 
+\left[ \left(\c{\alpha}+\beta \right)\left( \pi + \c{\tau} \right) +\left(\alpha + \c{\beta} \right)\left( \c{\pi} + \tau \right)\right]
\nonumber \\
& \qquad \qquad \qquad \qquad 
-\left[ \left(\varepsilon -\c{\varepsilon}\right)\left(\mu -\c{\mu}\right)+\left(\gamma -\c{\gamma}\right)\left(\rho -\c{\rho}\right)\right],\nonumber
\end{align}
so that 
\begin{align}
\mathcal{R_{\,W}}
&=\left[\inner{D_{\c{\b{m}}}D_{\b{n}}\b{l},\b{m}}-\inner{D_{\b{n}}D_{\c{\b{m}}}\b{l},\b{m}}\right]
\nonumber \\
& \qquad
+ \left[\inner{D_{\b{m}}D_{\b{n}}\b{l},\c{\b{m}}}-\inner{D_{\b{n}}D_{\b{m}}\b{l},\c{\b{m}}}\right]
\nonumber \\
& \qquad
+\left[\inner{D_{\c{\b{m}}}D_{\b{l}}\b{n},\b{m}}-\inner{D_{\b{l}}D_{\c{\b{m}}}\b{n},\b{m}}\right]
\nonumber \\
& \qquad
+ \left[\inner{D_{\b{m}}D_{\b{l}}\b{n},\c{\b{m}}}-\inner{D_{\b{l}}D_{\b{m}}\b{n},\c{\b{m}}}\right]\
\nonumber \\
& \qquad
- 2\left(\kappa \nu + \c{\kappa} \c{\nu} \right)+2\left(\tau \c{\tau}+\pi \c{\pi}\right)
\nonumber \\
& \qquad
+2\left(\rho \c{\mu} +\c{\rho} \mu +\lambda \sigma + \c{\lambda} \c{\sigma} \right)
\nonumber \\
& \qquad
-\left[ \left(\c{\alpha}+\beta \right)\left( \pi + \c{\tau} \right) +\left(\alpha + \c{\beta} \right)\left( \c{\pi} + \tau \right)\right]
\nonumber \\
& \qquad
+\left[ \left(\varepsilon -\c{\varepsilon}\right)\left(\mu -\c{\mu}\right)+\left(\gamma -\c{\gamma}\right)\left(\rho -\c{\rho}\right)\right].\nonumber
\end{align}
Now we further use Eqs.~(\ref{Dnl}), (\ref{Dml}), (\ref{Dln}),\,(\ref{Dmn}), (\ref{Dlm}), (\ref{Dnm}) and (\ref{Dmm_}) and write
\begin{align}
\mathcal{R_{\,W}}
&=D_{\b{m}}\left(\pi -\c{\tau} \right)+D_{\c{\b{m}}}\left(\c{\pi} -\tau \right)
\nonumber \\
& \qquad
-\left[\left(\alpha - \c{\beta} \right)\left( \c{\pi} - \tau \right)+\left(\c{\alpha}-\beta \right)\left( \pi - \c{\tau} \right)\right]
\nonumber \\
& \qquad
-\left[\left(\varepsilon +\c{\varepsilon}\right)\left(\mu +\c{\mu}\right)+\left(\gamma +\c{\gamma}\right)\left(\rho +\c{\rho}\right)\right]
\nonumber \\
& \qquad
+\left[D_{\b{n}}\left(\rho +\c{\rho }\right)-D_{\b{l}}\left(\mu +\c{\mu }\right)\right]
\nonumber \\
& \qquad
+\left[\left(\c{\alpha}+\beta \right)\left( \pi +\c{\tau} \right)+ \left(\alpha +\c{\beta} \right)\left( \c{\pi} +\tau \right)\right]
\nonumber \\
& \qquad
-\left[\left(\varepsilon - \c{\varepsilon}\right)\left(\mu -\c{\mu} \right)+\left(\gamma - \c{\gamma}\right)\left(\rho -\c{\rho}\right)\right]
\nonumber \\
& \qquad
- 2\left(\kappa \nu + \c{\kappa} \c{\nu} \right)+2\left(\tau \c{\tau}+\pi \c{\pi}\right)
\nonumber \\
& \qquad
+ 2\left(\rho \c{\mu} +\c{\rho} \mu +\lambda \sigma + \c{\lambda} \c{\sigma} \right)
\nonumber \\
& \qquad
-\left[ \left(\c{\alpha}+\beta \right)\left( \pi + \c{\tau} \right) +\left(\alpha + \c{\beta} \right)\left( \c{\pi} + \tau \right)\right]
\nonumber \\
& \qquad
+\left[ \left(\varepsilon -\c{\varepsilon}\right)\left(\mu -\c{\mu}\right)+\left(\gamma -\c{\gamma}\right)\left(\rho -\c{\rho}\right)\right].\nonumber
\end{align}
Hence,
\begin{align}
\mathcal{R_{\,W}}
&=D_{\b{n}}\left(\rho +\c{\rho }\right)-D_{\b{l}}\left(\mu +\c{\mu }\right)
+ D_{\b{m}}\left(\pi -\c{\tau} \right)+D_{\c{\b{m}}}\left(\c{\pi} -\tau \right)
\nonumber \\
& \qquad
-\left[\left(\alpha - \c{\beta} \right)\left( \c{\pi} - \tau \right)+\left(\c{\alpha}-\beta \right)\left( \pi - \c{\tau} \right)\right]
\nonumber \\
& \qquad
-\left[\left(\varepsilon +\c{\varepsilon}\right)\left(\mu +\c{\mu}\right)+\left(\gamma +\c{\gamma}\right)\left(\rho +\c{\rho}\right)\right]
\nonumber \\
& \qquad
- 2\left(\kappa \nu + \c{\kappa} \c{\nu} \right)+2\left(\tau \c{\tau}+\pi \c{\pi}\right) 
\nonumber \\
& \qquad
+2\left(\rho \c{\mu} +\c{\rho} \mu +\lambda \sigma + \c{\lambda} \c{\sigma} \right).\nonumber
\end{align}
\subsection{Alternative derivation of $\mathcal{R_{\,W}}$} 
Here we will present a derivation of $\mathcal{R_{\,W}}$ by using the decomposition of the Riemann tensor into its fully traceless, $\t {C} {_\mu _\nu _\alpha _\beta}$, semitraceless, $\t{Y}{_\mu _\nu _\alpha _\beta}$, and the trace parts, $\t{S}{_\mu _\nu _\alpha _\beta}$. For a 4-dimensional spacetime, the decomposition is as follows \cite{Wald:1984}:
\begin{align}\label{Riccidecomp}
\t {R}{_\mu _\nu _\alpha _\beta}&=\t {C}{_\mu _\nu _\alpha _\beta}+\t {Y}{_\mu _\nu _\alpha _\beta}-\t {S}{_\mu _\nu _\alpha _\beta},
\end{align}
where $\t {C} {_\mu _\nu _\alpha _\beta}$ is the Weyl tensor, $R$ is the Ricci scalar of the spacetime and
\begin{align}
\t {Y}{_{\mu} _{\nu} _{\alpha} _{\beta}}
&=\frac{1}{2}\left(\t {g}{_{\mu} _{\alpha}}\t {R}{_\beta _\nu}-\t {g}{_\mu _\beta}\t {R}{_\alpha _\nu}-\t {g}{_\nu _\alpha}\t {R}{_\beta _\mu}+\t {g}{_\nu _\beta}\t {R}{_\alpha _\mu} \right)\label{semitraceRiem},\\
\t {S}{_\mu _\nu _\alpha _\beta}&=\frac{R}{6}\left(\t {g}{_\mu _\alpha}\t {g}{_\beta _\nu}-\t {g}{_\mu _\beta}\t {g}{_\alpha _\nu}\right)\label{traceRiem}.
\end{align}
The term we are after follows as
\begin{align}
\mathcal{R_{\,W}}
&:=g(R(\t {E}{_b},\t {N}{_i})\t {E}{_a},\t {N}{_j}){\t {\eta}{^a^b}}\t {\delta}{^i^j}
\nonumber \\
&=\t {R}{_\alpha_\beta_\mu_\nu}\t {E}{^\mu _b}\t {N}{^\nu _i}\t {E}{^\beta _a}\t {N}{^\alpha_j}{\t {\eta}{^a^b}}\t {\delta}{^i^j}\nonumber \\
&= \t {R}{_\alpha_\beta_\mu_\nu}\left(\t {\eta}{^a^b}\t {E}{^\mu _b}\t {E}{^\beta _a}\right)\left(\t {\delta}{^i^j} \t {N}{^\nu _i} \t {N}{^\alpha_j} \right).\nonumber 
\end{align}
Now by using Eqs.~(\ref{exp:eta_E_E}) and (\ref{exp:delt_n_n}) we obtain
\begin{align}
\mathcal{R_{\,W}}
&=-\t {R}{_\alpha_\beta_\mu_\nu}\left(l^\mu n^\beta + l^\beta n^\nu \right)\left(m^\nu \c{m}^\alpha + m^\alpha \c{m}^\nu \right)\nonumber \\
&=-\left(\t {R}{_{\c{\b{m}}}_{\b{n}}_{\b{l}}_{\b{m}}} + \t {R}{_{\b{m}}_{\b{n}}_{\b{l}}_{\c{\b{m}}}}+ \t {R}{_{\c{\b{m}}}_{\b{l}}_{\b{n}}_{\b{m}}}+\t {R}{_{\b{m}}_{\b{l}}_{\b{n}}_{\c{\b{m}}}}\right).\nonumber 
\end{align}
Symmetries of $\t {R}{_\mu _\nu _\alpha _\beta}$ allows us to write
\begin{align}
\mathcal{R_{\,W}}
&=-2\left(\t {R}{_{\c{\b{m}}}_{\b{n}}_{\b{l}}_{\b{m}}} + \t {R}{_{\c{\b{m}}} _{\b{l}}_{\b{n}}_{\b{m}}} \right),\nonumber 
\end{align}
and by using the decomposition (\ref{Riccidecomp}),
\begin{align}
\mathcal{R_{\,W}}
&= -2\left(\t {C}{_{\c{\b{m}}}_{\b{n}}_{\b{l}}_{\b{m}}} + \t {C}{_{\c{\b{m}}} _{\b{l}}_{\b{n}}_{\b{m}}}\right)
\nonumber \\
& \qquad
-2\left(\t {Y}{_{\c{\b{m}}}_{\b{n}}_{\b{l}}_{\b{m}}} + \t {Y}{_{\c{\b{m}}} _{\b{l}}_{\b{n}}_{\b{m}}}-\t {S}{_{\c{\b{m}}}_{\b{n}}_{\b{l}}_{\b{m}}} - \t {S}{_{\c{\b{m}}} _{\b{l}}_{\b{n}}_{\b{m}}} \right).\nonumber
\end{align}
Here we make use of the symmetries of $\t {C}{_\mu _\nu _\alpha _\beta}$ and the definition (\ref{Psi2}) to get
\begin{align}
\mathcal{R_{\,W}}
&= -2\left(\Psi _2 +\c{\Psi _2}\right)
\nonumber \\
& \qquad
-2\left(\t {Y}{_{\c{\b{m}}}_{\b{n}}_{\b{l}}_{\b{m}}} + \t {Y}{_{\c{\b{m}}} _{\b{l}}_{\b{n}}_{\b{m}}}-\t {S}{_{\c{\b{m}}}_{\b{n}}_{\b{l}}_{\b{m}}} - \t {S}{_{\c{\b{m}}} _{\b{l}}_{\b{n}}_{\b{m}}} \right).\nonumber \\
\end{align}
By using the definitions of $\t {Y}{_\mu _\nu _\alpha _\beta}$ and $\t {S}{_\mu _\nu _\alpha _\beta}$ given in (\ref{semitraceRiem}) and (\ref{traceRiem}) we write
\begin{align}
\t {Y}{_{\c{\b{m}}} _{\b{n}}_{\b{l}}_{\b{m}}}
&=\frac{1}{2}\left(\inner{\c{\b{m}},\b{l}}\t {R}{_{\b{m}}_{\b{n}}}-\inner{\c{\b{m}},\b{m}} \t {R}{_{\b{l}}_{\b{n}}}
\right. \nonumber  \\
&\qquad  \qquad \qquad \left.
-\inner{\b{n},\b{l}}\t {R}{_{\b{m}}_{\c{\b{m}}}}+\inner{\b{n},\b{m}} \t {R}{_{\b{l}}_{\c{\b{m}}}}\right),\\
\t {Y}{_{\c{\b{m}}} _{\b{l}}_{\b{n}}_{\b{m}}}
&=\frac{1}{2}\left(\inner{\c{\b{m}},\b{n}}\t {R}{_{\b{m}}_{\b{l}}}-\inner{\c{\b{m}},\b{m}} \t {R}{_{\b{n}}_{\b{l}}}
\right. \nonumber  \\
&\qquad  \qquad \qquad \left.
-\inner{\b{l},\b{n}}\t {R}{_{\b{m}}_{\c{\b{m}}}}+\inner{\b{l},\b{m}} \t {R}{_{\b{n}}_{\c{\b{m}}}}\right),\\
\t {S}{_{\c{\b{m}}} _{\b{n}}_{\b{l}}_{\b{m}}}
&=\frac{R}{6}\left(\inner{\c{\b{m}},\b{l}} \inner{\b{m},\b{n}}-\inner{\c{\b{m}},\b{m}}\inner{\b{l},\b{n}}\right),\\
\t {S}{_{\c{\b{m}}} _{\b{l}}_{\b{n}}_{\b{m}}}
&=\frac{R}{6}\left(\inner{\c{\b{m}},\b{n}} \inner{\b{m},\b{l}}-\inner{\c{\b{m}},\b{m}}\inner{\b{n},\b{l}}\right). 
\end{align}
Also, since the Ricci scalar is $R=\t {g}{^\mu ^\nu}\t {R}{_\mu _\nu}=2\left(-\t {R}{_{\b{l}}_{\b{n}}}+\t {R}{_{\b{m}}_{\c{\b{m}}}}\right)$ and the Ricci tensor is symmetric, we have
\begin{align}
\mathcal{R_{\,W}}&=-2\left(\Psi _2 +\c{\Psi _2}\right)-2\left(\frac{R}{2}-\frac{R}{3}\right).
\end{align}
In the NP formalism one defines a variable $\Lambda =R/24$, thus we conclude that
\begin{align}
\mathcal{R_{\,W}}=-2\left(\Psi _2 +\c{\Psi _2} +4\Lambda \right).
\end{align}
\section{Other derivations}\label{Otherderivations}
\subsection{Gauss equation of $\mathbb{S}$}
For a 2-dimensional spacelike surface embedded in a 4-dimensional spacetime, the Gauss equation reads as \citep{Spivak:1979}
\begin{align}\label{Gausseq}
g(R(N_k, N_l)N_j,N_i)&=\mathcal{R} _{ijkl} - \t {J}{_a_i_k}\t {J}{_b_j_l}{\t {\eta}{^a^b}} + \t {J}{_a_j_k}\t {J}{_b_i_l}{\t {\eta}{^a^b}}.
\end{align} 
When we contract Eq.~(\ref{Gausseq}) with $\t {\delta} {^i^k} \t {\delta} {^j^l}$ we get
\begin{align}\label{Gausssimple}
g(R(N^k, N^l)N_k,N_l)&=\mathcal{R}_{\,\mathbb{S}}-H^2+\mathcal{J}^2,
\end{align}
where $\mathcal{R}_{\,\mathbb{S}}$ is the intrinsic curvature scalar of $\mathbb{S}$, $H^2=\t {J}{_a_i_k}\t {J}{_b_j_l}{\t {\eta}{^a^b}}\t {\delta} {^i^k} \t {\delta} {^j^l}$ is the square of the mean extrinsic curvature scalar of $\mathbb{S}$ and $\mathcal{J}^2=\t {J}{_a_j_k}\t {J}{_b_i_l}{\t {\eta}{^a^b}} \t {\delta} {^i^k} \t {\delta} {^j^l}$ is one of the variables that appear in the contracted Raychaudhuri equation.
Then derivation of $g(R(N^k, N^l)N_k,N_l)$ in terms of the NP variables proceeds as follows:
\begin{align}
g(R(N_k, &N_l)N_j,N_i)\t {\delta} {^i^k} \t {\delta} {^j^l} = ... \nonumber \\
&...= \t {R}{_{\alpha} _{\beta} _{\mu} _{\nu}}\t {N}{^{\mu}_k}\t {N}{^{\nu}_l}\t {N}{^{\beta}_j}\t {N}{^{\alpha}_i}\t {\delta} {^i^k} \t {\delta} {^j^l}=\t {R} {_i_j_k_l}\t {\delta} {^i^k} \t {\delta} {^j^l}\nonumber \\
&...=\t {R}{_{\alpha} _{\beta} _{\mu} _{\nu}}\left(\t {N}{^{\mu}_k} \t {N}{^{\alpha}_i} \t {\delta} {^i^k}\right)\left(\t {N}{^{\nu}_l} \t {N}{^{\beta}_j} \t {\delta} {^j^l}\right).\nonumber 
\end{align}
Now considering the relation
(\ref{exp:delt_n_n})
we write
\begin{align}
\t {R} {_i_j_k_l} \t {\delta} {^i^k} \t {\delta} {^j^l}
&=\t {R}{_{\alpha} _{\beta} _{\mu} _{\nu}}\left(m^\mu \c{m}^\alpha + m^\alpha \c{m}^\mu \right)
\nonumber  \\
&\qquad \times
\left(m^\nu \c{m}^\beta + m^\beta \c{m}^\nu \right)\nonumber \\
&=\t {R}{_{\c{\b{m}}} _{\c{\b{m}}} _{\b{m}} _{\b{m}}}+\t {R}{_{\c{\b{m}}} _{\b{m}} _{\b{m}} _{\c{\b{m}}}}
\nonumber  \\
&\qquad \qquad \qquad
+\t {R}{_{\b{m}} _{\c{\b{m}}} _{\c{\b{m}}}  _{\b{m}}}
+\t {R}{ _{\b{m}} _{\b{m}} _{\c{\b{m}}} _{\c{\b{m}}}}, \nonumber
\end{align}
and by considering the symmetries of $\t {R}{_\mu _\nu _\alpha _\beta}$ we obtain
\begin{align}
\t {R} {_i_j_k_l}\t {\delta} {^i^k} \t {\delta} {^j^l}
&=-2\t {R}{_{\c{\b{m}}} _{\b{m}} _{\c{\b{m}}} _{\b{m}}}.\nonumber
\end{align}
Now let us use the decomposition
(\ref{Riccidecomp}) and write
\begin{align}
\t {R} {_i_j_k_l}\t {\delta} {^i^k} \t {\delta} {^j^l}
&=-2\left(\t {C}{_{\c{\b{m}}} _{\b{m}} _{\c{\b{m}}} _{\b{m}}}+\t {Y}{_{\c{\b{m}}} _{\b{m}} _{\c{\b{m}}} _{\b{m}}}-\t {S}{_{\c{\b{m}}} _{\b{m}} _{\c{\b{m}}} _{\b{m}}}\right)\label{DecRm_mm_m},
\end{align} 
where 
\begin{align}
\t {C}{_{\c{\b{m}}} _{\b{m}} _{\c{\b{m}}} _{\b{m}}}
&=\Psi _2+\c{\Psi _2} \label{Cm_mm_m},\\
\t {Y}{_{\c{\b{m}}} _{\b{m}} _{\c{\b{m}}} _{\b{m}}}
&=\frac{1}{2}\left(\inner{\c{\b{m}},\c{\b{m}}}\t {R}{_{\b{m}}_{\b{m}}}
-\inner{\c{\b{m}},\b{m}} \t {R}{_{\c{\b{m}}}_{\b{m}}}
\right. \nonumber  \\
&\qquad  \qquad \left.
-\inner{\b{m},\c{\b{m}}}\t {R}{_{\b{m}}_{\c{\b{m}}}}+\inner{\b{m},\b{m}} \t {R}{_{\c{\b{m}}}_{\c{\b{m}}}}\right)\nonumber \\
&=-\t {R}{_{\b{m}}_{\c{\b{m}}}} \label{Ym_mm_m}, \\
\t {S}{_{\c{\b{m}}} _{\b{m}} _{\c{\b{m}}} _{\b{m}}}
&=\frac{R}{6}\left(\inner{\c{\b{m}},\c{\b{m}}} \inner{\b{m},\b{m}}-\inner{\c{\b{m}},\b{m}}\inner{\b{m},\c{\b{m}}}\right)\nonumber \\
&=-\frac{R}{6} \label{Sm_mm_m}.
\end{align}
Equation (\ref{Cm_mm_m}) follows from the fact that Weyl tensor is traceless. To see this, consider the following. For any pair of vectors $\{\b{v},\,\b{w}\}$ one can write
\begin{align}
\t {g}{^x^y}\t {C}{_x_{\b{v}}_{y}_{\b{w}}}&=0 \nonumber \\
&=-\t {C}{_{\b{l}}_{\b{v}}_{\b{n}}_{\b{w}}}-\t {C}{_{\b{n}}_{\b{v}}_{\b{l}}_{\b{w}}}+\t {C}{_{\b{m}}_{\b{v}}_{\c{\b{m}}}_{\b{w}}}+ \t {C}{_{\c{\b{m}}}_{\b{v}}_{\b{m}}_{\b{w}}}.
\end{align}
Now let us set $\b{v}=\b{m},\,\b{w}=\c{\b{m}}$, then we obtain
\begin{align}
0&=-\t {C}{_{\b{l}}_{\b{m}}_{\b{n}}_{\c{\b{m}}}}-\t {C}{_{\b{n}}_{\b{m}}_{\b{l}}_{\c{\b{m}}}}+\t {C}{_{\b{m}}_{\b{m}}_{\c{\b{m}}}_{\c{\b{m}}}}+ \t {C}{_{\c{\b{m}}}_{\b{m}}_{\b{m}}_{\c{\b{m}}}} \nonumber \\
&=\t {C}{_{\b{l}}_{\b{m}}_{\c{\b{m}}}_{\b{n}}}+\t {C}{_{\b{l}}_{\c{\b{m}}}_{\b{m}}_{\b{n}}}+0-\t {C}{_{\c{\b{m}}}_{\b{m}}_{\c{\b{m}}}_{\b{m}}}.
\end{align}
Then by using the definition given in (\ref{Psi2}) we find
\begin{align}
\t {C}{_{\c{\b{m}}} _{\b{m}} _{\c{\b{m}}} _{\b{m}}}&=\Psi _2+\c{\Psi _2}.
\end{align}
In order to rewrite Eq.~(\ref{Ym_mm_m}) in terms of the curvature scalars consider
\begin{align}
R=2\left(-\t {R}{_{\b{l}}_{\b{n}}}+\t {R}{_{\b{m}}_{\c{\b{m}}}} \right)\qquad and \qquad
\Phi _{11}=\frac{1}{4}\left(\t {R}{_{\b{l}}_{\b{n}}}+\t {R}{_{\b{m}}_{\c{\b{m}}}}\right).
\end{align}
Then we write
\begin{align}
\t {R}{_{\b{m}}_{\c{\b{m}}}}&=\frac{R+8\Phi _{11}}{4}.
\end{align}
Therefore, substitution of Eqs.~(\ref{Cm_mm_m}), (\ref{Ym_mm_m}) and (\ref{Sm_mm_m}) into the decomposition (\ref{DecRm_mm_m}) yields
\begin{align}
g(R(N_k, N_l)&N^l,N^k)= ... \nonumber \\
&...=-2\left[\left(\Psi _2+\c{\Psi _2}\right)-\left(\frac{R+8\Phi _{11}}{4}\right)+\frac{R}{6}\right]\nonumber \\
&...=-2\left(\Psi _2+\c{\Psi _2}-2\Lambda -2 \Phi _{11}\right).
\end{align}
\subsection{Boost invariance of quasilocal charges}\label{App:SpinBoost}
\subsubsection{Transformation of $\tilde{\nabla} _{\mathbb{T}}\mathcal{J}$ under type-III Lorentz transformations:}
Under a type-III Lorentz transformation, the null vectors $\b{l}$ and $\b{n}$ transform according to the relations (\ref{lIII}) and (\ref{nIII}) respectively. The transformed spin coefficients, $\gamma  ^\prime , \, \mu  ^\prime, \, \rho  ^\prime$ and $\varepsilon  ^\prime$ can be obtained via the relations (\ref{gammaIII}), (\ref{muIII}), (\ref{rhoIII})  and (\ref{varepsilonIII}) so that the transformation of the term $\tilde{\nabla} _{\mathbb{T}}\mathcal{J}$ in Eq.~(\ref{DeltaJNP}) follows as
\begin{align}
\tilde{\nabla} _{\mathbb{T}}\mathcal{J} ^\prime 
&= 2\left(D_{\b{n} ^\prime}\rho  ^\prime-D_{\b{l} ^\prime}\mu  ^\prime \right)
-2\left[\left(\varepsilon  ^\prime+\c{\varepsilon  ^\prime}\right)\mu  ^\prime+\left(\gamma  ^\prime+\c{\gamma} ^\prime\right)\rho  ^\prime\right]\nonumber \\
&=2\left[\frac{1}{a^2}D_{\b{n}}\left(a^2\rho \right)-a^2D_{\b{l}}\left(\frac{1}{a^2}\mu\right)\right]
\nonumber \\
& \qquad
-2\Big\{ a^2 \left[\varepsilon + D_{\b{l}} \left(\ln a +i\theta \right)\right]
\nonumber  \\
&\qquad  \qquad \qquad 
+a^2 \left[\c{\varepsilon} + D_{\b{l}} \left(\ln a -i\theta \right)\right]
\Big\}\frac{1}{a^2}\mu 
\nonumber \\
& \qquad
-2\Big\{ \frac{1}{a^2} \left[\gamma + D_{\b{n}} \left(\ln a +i\theta \right)\right]
 \nonumber  \\
&\qquad  \qquad \qquad 
+ \frac{1}{a^2} \left[\c{\gamma} + D_{\b{n}} \left(\ln a -i\theta \right)\right]
\Big\}a^2\rho \nonumber \\
&= 2\left(D_{\b{n}}\rho -D_{\b{l}}\mu \right)-2\left[\left(\varepsilon +\c{\varepsilon }\right)\mu +\left(\gamma +\c{\gamma}\right)\rho \right].
\end{align}
Therefore $\tilde{\nabla} _{\mathbb{T}}\mathcal{J}$ is invariant under a type-III Lorentz transformation.
\subsubsection{Transformation of $\tilde{\nabla} _{\mathbb{S}}\mathcal{K}$ under type-III Lorentz transformations:}
By using Eq.~(\ref{DeltaKNP}), the transformed $\tilde{\nabla} _{\mathbb{S}}\mathcal{K}$ can be written as
\begin{align}
\tilde{\nabla} _{\mathbb{S}}\mathcal{K} ^\prime
&=2\left(D_{\b{m} ^\prime}\pi  ^\prime- D_{\c{\b{m}} ^\prime}\tau  ^\prime\right)
\nonumber \\
& \qquad
- 2\left[\left(\c{\alpha} ^\prime-\beta  ^\prime\right)\pi  ^\prime+ \left(\alpha  ^\prime-\c{\beta} ^\prime \right)\c{\pi} ^\prime\right],
\end{align}
in which the transformations of the complex null vectors $\b{m}$ and $\c{\b{m}}$ are given in relations (\ref{mIII}) and (\ref{m_III}) respectively. Also, the transformed spin coefficients $\tau  ^\prime,\, \beta  ^\prime,\, \alpha  ^\prime$ and $\pi  ^\prime$, are obtained via the relations (\ref{tauIII}), (\ref{betaIII}), (\ref{alphaIII}) and (\ref{piIII}) so that we have
\begin{align}
\tilde{\nabla} _{\mathbb{S}}\mathcal{K} ^\prime
&=2\left[e^{2i\theta}D_{\b{m}}\left(e^{-2i\theta}\pi\right)-
e^{-2i\theta}D_{\c{\b{m}}}\left(e^{2i\theta}\tau \right)\right]
\nonumber \\
& \qquad
-2\Big\{e^{2i\theta} \left[\c{\alpha} + D_{\b{m}} \left(\ln a -i\theta \right)\right]
\nonumber  \\
&\qquad  \qquad \qquad
-e^{2i\theta} \left[\beta + D_{\b{m}} \left(\ln a +i\theta \right)\right]\Big \}e^{-2i\theta}\pi 
\nonumber  \\
&\qquad  \qquad \qquad
-2\Big\{e^{-2i\theta} \left[\alpha + D_{\c{\b{m}}} \left(\ln a +i\theta \right)\right]
\nonumber  \\
&\qquad  \qquad \qquad
-e^{-2i\theta} \left[\c{\beta} + D_{\c{\b{m}}} \left(\ln a -i\theta \right)\right]\Big \}e^{2i\theta}\c{\pi}.
\end{align}
Now by further imposing our null tetrad condition, $\tau +\c{\pi}=0$ on the above equation we obtain
\begin{align}
\tilde{\nabla} _{\mathbb{S}}\mathcal{K} ^\prime
&= 2\left[D_{\b{m}}\pi - D_{\c{\b{m}}}\tau \right] - 2\left[\left(\c{\alpha}-\beta \right)\pi + \left(\alpha -\c{\beta} \right)\c{\pi}\right].
\end{align}
Then, $\tilde{\nabla} _{\mathbb{S}}\mathcal{K}$ transforms invariantly under the spin-boost transformation of the null tetrad.
\subsubsection{Transformation of $\mathcal{J}^2$ under type-III Lorentz transformations:}
The transformation of $\mathcal{J}^2$ follows from the definition (\ref{J2NP}) plus the transformation relations (\ref{muIII}), (\ref{sigmaIII}), (\ref{lambdaIII}) and (\ref{rhoIII})  of the spin coefficients $\mu  ^\prime,\, \sigma  ^\prime,\, \lambda  ^\prime$ and $\rho  ^\prime$. Then we write
\begin{align}
\mathcal{J}^{2 \prime}
&= 4\mu  ^\prime\rho  ^\prime+ 2\left( \sigma  ^\prime\lambda  ^\prime+ \c{\sigma}  ^\prime\c{\lambda} ^\prime \right)\nonumber \\
&=4\left(a^{-2}\mu \right)\left(a^{2}\rho \right)
\nonumber  \\
&\qquad 
+2\left[\left(a^{2}e^{4i\theta}\sigma \right)\left(a^{-2}e^{-4i\theta}\lambda \right)
\right. \nonumber \\
& \left. \qquad \qquad \qquad \qquad
+\left(a^{2}e^{-4i\theta}\c{\sigma} \right)\left(a^{-2}e^{4i\theta}\c{\lambda} \right)\right]\nonumber \\
&= 4\mu \rho + 2\left( \sigma \lambda + \c{\sigma} \c{\lambda}\right).
\end{align}
Therefore $\mathcal{J}^2$ transforms invariantly under the spin-boost transformation of the null tetrad.
\subsubsection{Transformation of $\mathcal{K}^2$ under type-III Lorentz transformations:}
By using Eq.~(\ref{K2NP}) as for the definition of $\mathcal{K}^2$ and considering relations (\ref{nuIII}), (\ref{tauIII}), (\ref{kappaIII})  and (\ref{piIII}) for the transformations of spin coefficients $\nu  ^\prime,\, \tau  ^\prime,\, \kappa  ^\prime$ and $\pi  ^\prime$ we write
\begin{align}
\mathcal{K}^{2 \prime}
&=-2\left(\kappa  ^\prime\nu  ^\prime+ \c{\kappa} ^\prime \c{\nu} ^\prime\right) + 2\left(\pi  ^\prime\c{\pi}  ^\prime+ \tau  ^\prime\c{\tau}  ^\prime\right)\nonumber \\
&= -2\left[\left(a^{4}\,e^{2i\theta} \kappa \right) \left(a^{-4}\,e^{-2i\theta} \nu \right)+
\right. \nonumber \\
& \left. \qquad \qquad \qquad \qquad
\left(a^{4}\,e^{-2i\theta} \c{\kappa} \right) \left(a^{-4}\,e^{2i\theta} \c{\nu} \right)\right]
\nonumber \\
& \qquad
+2\left[\left(e^{-2i\theta}\pi \right)\left(e^{2i\theta}\c{\pi} \right)+
\left(e^{2i\theta}\tau \right)\left(e^{-2i\theta}\c{\tau} \right)\right]\nonumber \\
&=-2\left(\kappa \nu + \c{\kappa} \c{\nu}\right) + 2\left(\pi \c{\pi} + \tau \c{\tau} \right).
\end{align}
Thus $\mathcal{K}^2$ is also invariant under spin-boost transformations.
\subsubsection{Transformation of $\mathcal{R_{\,W}}$ under type-III Lorentz transformations:}
The Weyl scalar $\Psi _2$ transforms invariantly under spin-boost transformations according to the relation (\ref{Psi2III}). Moreover, the parameter $\Lambda =R/24$ is invariant under such a transformation since the Ricci scalar is unchanged. Therefore, following Eq.~(\ref{RWSimpleNPgen}), it is easy to see that
\begin{align}
\mathcal{R_{\,W}} ^\prime &=-2\left(\psi _2 ^\prime + \c{\psi}_2  ^\prime+ 4\Lambda ^\prime \right) =-2\left(\psi _2 + \c{\psi}_2 + 4\Lambda \right),
\end{align}
and $\mathcal{R_{\,W}}$ is invariant under spin-boost transformations.
\bibliographystyle{apsrev4-1}
\bibliography{references}
\end{document}